\documentclass[10pt]{amsart}
\usepackage{amsmath,amssymb}
\usepackage[dvips]{epsfig}

\newtheorem{theorem}{Theorem}[section]
\newtheorem{lemma}[theorem]{Lemma}
\newtheorem{corollary}[theorem]{Corollary}
\newtheorem{proposition}[theorem]{Proposition}
\allowdisplaybreaks
\numberwithin{equation}{section}

\newcommand{\PP}{{\mathbb P}}
\def\Im{\mathop{\rm Im}\nolimits}
\def\Re{\mathop{\rm Re}\nolimits}
\def\div{\mathop{\rm Div}\nolimits}
\def\Jac{\mathop{\rm Jac}\nolimits}
\def\diag{\mathop{\rm Diag}\nolimits}
\def\dim{\mathop{\rm Dim}\nolimits}

\begin{document}

\title[Geometry of the 3-monopole]
{Remarks on the complex geometry of the 3-monopole}
\author{H.W. Braden}
\address{School of Mathematics, Edinburgh University, Edinburgh.}
\email{hwb@ed.ac.uk}
\author{V.Z. Enolski}
\address{Department of Mathematics and Statistics, Concordia
University, Montreal.
\newline
On leave: Institute of Magnetism, National Academy of Sciences of
Ukraine.} \email{vze@ma.hw.ac.uk}
\begin{abstract}
We develop the Ercolani-Sinha construction of $SU(2)$ monopoles
and make this effective for (a five parameter family of centred)
charge 3 monopoles. In particular we show how to solve the
transcendental constraints arising on the spectral curve. For a
class of symmetric curves the transcendental constraints become a
number theoretic problem and a recently proven identity of
Ramanujan provides a solution. The Ercolani-Sinha construction
provides a gauge-transform of the Nahm data.
\end{abstract}

 \maketitle

\tableofcontents

\section{Introduction}
Magnetic monopoles, or the topological soliton solutions of
Yang-Mills-Higgs gauge theories in three space dimensions, have
been objects of fascination for over a quarter of a century. BPS
monopoles in particular have been the focus of much research (see
\cite{ms04} for a recent review). These monopoles arise as a limit
in which the Higgs potential is removed and satisfy a first order
Bogomolny equation
$$B_i=\frac{1}{2}\sum_{j,k=1}\sp3\epsilon_{ijk}F\sp{jk}=D_i\Phi$$
(together with certain boundary conditions, the remnant of the
Higgs potential). Here $F_{ij}$ is the field strength associated
to a gauge field $A$, and $\Phi$ is the Higgs field. These
equations may be viewed as a dimensional reduction of the four
dimensional self-dual equations upon setting all functions
independent of $x_4$ and identifying $\Phi=A_4$. Just as Ward's
twistor transform relates instanton solutions in $\mathbb{R}\sp4$
to certain holomorphic vector bundles over the twistor space
$\mathbb{CP}\sp3$, Hitchin showed \cite{hitchin82} that the
dimensional reduction leading to BPS monopoles could be made at
the twistor level as well. Mini-twistor space is a two dimensional
complex manifold isomorphic to T$\mathbb{P}\sp1$, and BPS
monopoles may be identified with certain bundles over this space.
In particular a curve $\mathcal{C}\subset$ T$\mathbb{P}\sp1$, the
spectral curve, arises in this construction and, subject to
certain nonsingularity conditions, Hitchin was able to prove all
monopoles could be obtained by this approach \cite{hitchin83}.
Nahm also gave a transform of the ADHM instanton construction to
produce BPS monopoles \cite{nahm82}. The resulting Nahm's
equations have Lax form and the corresponding spectral curve is
again $\mathcal{C}$. Many striking results are now known, yet,
disappointingly, explicit solutions are rather few. This paper is
directed towards constructing new solutions.

In a seminal paper, Ercolani and Sinha  \cite{es89} sought to
bring methods from integrable systems to bear upon the
construction of solutions to Nahm's equations for the gauge group
$SU(2)$. Integrable structures have long been associated with the
self-dual equations and BPS monopoles: Ercolani and Sinha showed
how one could solve (a gauge transform of) the Nahm equations in
terms of a Baker-Akhiezer function for the curve $\mathcal{C}$.
While conceptually simple, the Ercolani-Sinha construction is
remarkably challenging to implement, and they noted that although
their \lq{procedure, can in principle, be carried out for
arbitrary monopole number, however, there are obvious technical
difficulties in almost every step of the}\rq\ construction. Here
we follow the approach of Ercolani-Sinha for the particular case
of charge 3 $SU(2)$ monopoles.

An outline of our paper is as follows. In section 2 we recall
aspects of the Hitchin, Nahm and Ercolani-Sinha constructions and
then proceed to extend the latter in section 3. Here we shall
present new formulae and clarify the appearance of constant gauge
transformations in the Ercolani-Sinha construction. Further we
will highlight the ingredients needed to make effective the
construction and show how these reduce to evaluating quantities
intrinsic to the curve. As an illustration of our general theory
we consider the charge 2 monopole in section 4. Key to expressing
the Baker-Akhiezer function for a curve $\mathcal{C}$ is
determining Riemann's theta function built from the period matrix
of $\mathcal{C}$. The first hurdle in implementing the
Ercolani-Sinha construction is to analytically determine the
period matrix for $\mathcal{C}$ and then understand the theta
divisor. In section 5 we will introduce a class of (genus 4)
curves for which we can do this. They are of the form
\begin{equation}
\eta^3+\hat\chi(\zeta-\lambda_1)(\zeta-\lambda_2)(\zeta-\lambda_3)
(\zeta-\lambda_4)(\zeta-\lambda_5)(\zeta-\lambda_6)=0,
\label{welstein99}
\end{equation}
where $\lambda_i$, $i=1,\ldots,6$ are distinct complex numbers.
(For appropriate $\lambda_i$ this yields a charge 3 monopole.)
This class of curves was studied by Wellstein over one hundred
years ago \cite{wel99} and more recently by Matsumoto
\cite{matsu00}. Here we will introduce our homology basis and
define branch points in terms of $\theta$-constants following
\cite{matsu00}.

Corresponding to (some of) Hitchin's nonsingularity conditions
Ercolani and Sinha obtain restrictions on the allowed period
matrices for the spectral curve. Equivalent formulations of these
conditions were given in \cite{hmr99}. The Ercolani-Sinha
conditions are transcendental constraints and to solve these is
the next (perhaps \emph{the}) major hurdle to overcome in the
construction. In section 6 we do this for our curves. At this
stage we have replaced the constraints by relations between
various hypergeometric integrals. To simplify matters for the
present paper we next demand more symmetry and consider in section
7 the genus 4 curves
\begin{equation}
\eta^3+\chi(\zeta^6+b \zeta^3-1)=0\label{bren03}
\end{equation}
where $b$ is a certain real parameter. This restriction has the
effect of reducing the number of hypergeometric integrals to be
calculated to two. Interestingly the relations we demand of these
integrals are assertions of Ramanujan only recently proven. We
will denote curves of the form (\ref{bren03}) as \emph{symmetric
monopole} curves (though in fact they may not satisfy all of
Hitchin's nonsingularity conditions). The tetrahedrally symmetric
charge 3 monopole is of this form.

The curve (\ref{bren03}) covers a hyperelliptic curve of genus two
and two elliptic curves. We discuss these coverings. Using
Weierstrass-Poincar\'e reduction theory we are able to express the
theta function behaviour of these symmetric monopoles in terms of
elliptic functions and fairly comprehensive results may be
obtained. Finally, in section 8, we shall consider the curve
(\ref{bren03}) associated with tetrahedrally symmetric 3-monopole
when  the above parameter $b=2\sqrt{5}$. This genus 4 curve covers
4 elliptic curves and all entries to the period matrices are
expressible in terms of elliptic moduli. The analytical means
which we are using for our analysis involve Thomae-type formulae,
Weierstrass-Poincar\'e reduction theory, multivariable
hypergeometric function and higher hyperegeometric equalities of
Goursat. Our conclusions in section 9 will highlight various of
our results.

\part{General Considerations}

\section{Monopoles}

In this section we shall recall various features of the spectral
curve coming from Hitchin's and Nahm's construction and then
describe the Ercolan-Sinha construction based on this curve.

\subsection{Hitchin Data}

Using twistor methods Hitchin \cite{hitchin83} has shown that each
static $SU(2)$ Yang-Mills-Higgs monopole in the BPS limit with
magnetic charge $n$ is equivalent to a spectral curve of a
restricted form. If $\zeta$ is the inhomogeneous coordinate on the
Riemann sphere, and $(\zeta,\eta)$ are the standard local
coordinates on $T\PP\sp1$ (defined by
$(\zeta,\eta)\rightarrow\eta\frac{d}{d\zeta}$), the spectral curve
is an algebraic curve $\mathcal{C} \subset T\PP\sp1$ which has the
form
\begin{equation}
P(\eta,\zeta)=\eta^n+\eta^{n-1} a_1(\zeta)+\ldots+\eta^r
a_{n-r}(\zeta)+ \ldots+\eta\,
a_{n-1}(\zeta)+a_n(\zeta)=0.\label{spectcurve}
\end{equation}
Here $a_r(\zeta)$  (for $1\leq r\leq n$) is a polynomial in
$\zeta$ of maximum degree $2r$.

The Hitchin data constrains the curve $\mathcal{C}$ explicitly in
terms of the polynomial $P(\eta,\zeta)$ and implicitly in terms of
the behaviour of various line bundles on $\mathcal{C}$. If the
homogeneous coordinates of $ \PP\sp1$ are $[\zeta_0,\zeta_1]$ we
consider the standard covering of this by the open sets
$U_0=\{[\zeta_0,\zeta_1]\,|\,\zeta_0\ne0\}$ and
$U_1=\{[\zeta_0,\zeta_1]\,|\,\zeta_1\ne0\}$, with
$\zeta=\zeta_1/\zeta_0$ the usual coordinate on $U_0$. We will
denote by $\hat U_{0,1}$ the pre-images of these sets under the
projection map $\pi:T\PP\sp1\rightarrow\PP\sp1$. Let
 $L^{\lambda}$ denote the holomorphic line bundle on
$T\PP\sp1$ defined by the transition function
$g_{01}=\rm{exp}(-\lambda\eta/\zeta)$ on $\hat U_{0}\cap \hat
U_{1}$, and let $L^{\lambda}(m)\equiv
L^{\lambda}\otimes\pi\sp*\mathcal{O}(m)$ be similarly defined in
terms of the transition function
$g_{01}=\zeta^m\exp{(-\lambda\eta/\zeta)}$. A holomorphic section
of such line bundles is given in terms of holomorphic functions
$f_\alpha$ on $\hat U_\alpha$ satisfying
$f_\alpha=g_{\alpha\beta}f_\beta$. We denote line bundles on
$\mathcal{C}$ in the same way, where now we have holomorphic
functions $f_\alpha$ defined on $\mathcal{C}\cap\hat U_\alpha$.

The Hitchin data constrains the curve to satisfy:\\

A1. Reality conditions
\begin{equation}\label{spectcurvereal}
a_r(\zeta)=(-1)^r\zeta^{2r}\overline{a_r(-\frac{1}{\overline{\zeta}})}
.\end{equation} This is the requirement that $\mathcal{C}$ is real
with respect to the standard real structure on $T\PP\sp1$
\begin{equation}
\tau:(\zeta,\eta)\mapsto(-\frac{1}{\bar{\zeta}},
-\frac{\bar{\eta}}{\bar{\zeta}^2}).
\end{equation}
This is the anti-holomorphic involution defined by reversing the
orientation of the lines in ${\mathbb R}\sp3$. A consequence of
the reality condition is that we may parameterize $a_r(\zeta)$ as
follows,
\begin{equation}\label{arpar}
a_r(\zeta)=\sum_{k=0}\sp{2r}a_{r k}\,\zeta\sp{k}= \chi_r
\left[\prod_{l=1}\sp{r}\left(
\frac{\overline{\alpha}_l}{\alpha_l}\right)\sp{1/2}\right]
\prod_{k=1}\sp{r}(\zeta-\alpha_k)(\zeta+\frac{1}{\overline{\alpha}_k}),\qquad
\alpha_r\in \mathbb{C},\ \chi_r\in\mathbb{R}.
\end{equation}
Thus each $a_r(\zeta)$ contributes $2r+1$ (real) parameters.

A2. $L^2$ is trivial on $\mathcal{C}$ and $L(n-1)$ is real. The
triviality of $L^2$  on $\mathcal{C}$ means that there exists a
nowhere-vanishing holomorphic section. In terms of our open sets
$\hat U_{0,1}$ we then have two, nowhere-vanishing holomorphic
functions, $f_0$ on $\hat U_0\cap\mathcal{C}$ and $f_1$ on $\hat
U_1\cap\mathcal{C}$, such that on $\hat U_{0}\cap \hat
U_{1}\cap\mathcal{C}$
\begin{equation}
 f_{0}(\eta,\zeta)=\mathrm{exp} \left\{
-2\frac{\eta}{\zeta} \right\} f_1(\eta,\zeta). \label{triv3}
\end{equation}

A3. $H^0(\mathcal{C},L^{\lambda}(n-2))=0$ for $\lambda\in(0,2)$.\\

For a generic $n$-monopole the spectral curve is irreducible and
has genus $g_\mathcal{C}=(n-1)^2$. This may be calculated as
follows. For fixed $\zeta$ the $n$ roots of $P(\eta,\zeta)=0$
yield an $n$-fold covering of the Riemann sphere. The branch
points of this covering are given by
$$0= {\text{ Resultant}_\eta}(P(\eta,\zeta),\partial_\eta P(\eta,\zeta))=
\prod_{i=1}\sp{n}\partial_\eta P(\eta_i,\zeta), \quad{\text{
where}}\ P(\eta_i,\zeta)=0.$$ This expression is of degree
$n\times\text{deg}\,a_{n-1}=n(2n-2)$ in $\zeta$ and so by the
Riemann-Hurwitz theorem we have that
$$2 g_\mathcal{C}-2=2n(g_{\mathbb{P}\sp1}-1)+n(2n-2)=2(n-1)\sp2-2,$$
whence the genus as stated.

The $n=1$ monopole spectral curve is given by
$$\eta= (x_1+i x_2)-2 x_3 \zeta-(x_1-ix_2)\zeta\sp2,$$
where $x=(x_1,x_2,x_3)$ is any point in ${\mathbb R}\sp3$. In
general the three independent real coefficients of $a_1(\zeta)$
may be interpreted as the centre of the monopole in ${\mathbb
R}\sp3$. Strongly centred monopoles have the origin as center and
hence $a_1(\zeta)=0$. The group $SO(3)$ of rotations of ${\mathbb
R}\sp3$ induces an action on $T\PP\sp1$ via the corresponding
$PSU(2)$ transformations. If
$$\begin{pmatrix} p&q\\-\bar q&\bar p\end{pmatrix}\in PSU(2),
\qquad |p|\sp2+|q|\sp2=1,
$$
the transformation on $T\PP\sp1$ given by
$$\zeta\rightarrow\dfrac{\bar p\, \zeta-\bar q}{q\, \zeta+p},
\qquad \eta\rightarrow \dfrac{\eta}{(q\, \zeta+p)\sp2}
$$
corresponds to a rotation by $\theta$ around ${\mathbf n}\in
S\sp2$, where $n_1\sin{(\theta/2)}=\Im q$,
$n_2\sin{(\theta/2)}=-\Re q$, $n_3\sin{(\theta/2)}=- \Im q$, and
$\cos{(\theta/2)}=\Re p$. (Here the $\eta$ transformation is given
by the derivative of the $\zeta$ transformation.) The $SO(3)$
action commutes with the real structure $\tau$. Although a general
M\"obius transformation does not change the period matrix of a
curve $\mathcal{C}$ only the subgroup $PSU(2)< PSL(2,\mathbb{C})$
preserves the desired reality properties . We have that
$$\alpha_k\rightarrow {\tilde\alpha}_k \equiv\frac{p\alpha_k+{\bar
q}}{{\bar p}-\alpha_k q},\qquad \chi_r \rightarrow
{\tilde\chi}_r\equiv \chi_r \prod_{k=1}\sp{r}\left[\frac{({\bar
p}-\alpha_k q)({ p}-{\bar\alpha_k}{\bar q})({\bar\alpha_k}{\bar
p}+q)(\alpha_k{ p}+ {\bar
q})}{{\alpha_k}{\bar\alpha_k}}\right]\sp{1/2},$$ and
$$
a_r\rightarrow \frac{{\tilde a}_r}{(q\, \zeta+p)\sp{2r}}\equiv
\frac{{\tilde\chi}_r}{(q\, \zeta+p)\sp{2r}}
\left[\prod_{l=1}\sp{r}\left(
\frac{\overline{\tilde\alpha}_l}{\tilde\alpha_l}\right)\sp{1/2}\right]
\prod_{k=1}\sp{r}(\zeta-{\tilde\alpha}_k)(\zeta+\frac{1}{\overline{\tilde\alpha}_k}).
$$
In particular the form of the curve does not change: that is, if
$a_r=0$ then so also ${\tilde a}_r=0$. It is perhaps worth
emphasizing that the reality conditions are an extrinsic feature
of the curve (encoding the space-time aspect of the problem)
whereas the intrinsic properties of the curve are invariant under
birational transformations or the full M\"obius group. Such
extrinsic aspects are not a part of the usual integrable system
story.

\subsection{Nahm Data}
The Nahm construction of  charge $n$ $SU(2)$ monopoles is in terms
of $n\times n$ matrices $(T_1,T_2,T_3)$ depending
on a real parameter $s\in[0,2]$ and satisfying the following:\\

B1. Nahm's equation
\begin{equation}
\frac{dT_i}{ds}=\frac{1}{2}\sum_{j,k=1}^3\epsilon_{ijk}[T_j,T_k].
\label{nahm}
\end{equation}

B2. $T_i(s)$ is regular for $s\in(0,2)$ and has simple poles at
$s=0$ and $s=2$, the residues of which form the irreducible
$n$-dimensional representation of $su(2)$.\\

B3. $T_i(s)=-T_i^\dagger(s)$, \hskip 0.5cm $T_i(s)=T_i^t(2-s)$.\\

A caution is perhaps worth giving regarding the second of the
constraints B3. All that really is required is that the matrices
$T_i(s)$ are conjugate to the matrices $T_i^t(2-s)$. Many explicit
examples often take this for granted. Thinking of the Nahm
equations as a one-dimensional gauge theory then we still have
some gauge freedom left, associated with constant gauge
transformations. The spectral curve is gauge invariant, so if it
has the correct reality properties this guarantees that there is a
gauge in which the $s\rightarrow 2-s$ relation is explicit even if
we do not happen to be in that gauge at the moment.\footnote{We
thank Paul Sutcliffe for discussions on this point.}

The Nahm equations admit a Lax formulation. Upon setting
\begin{align*}
A_{-1}&=T_1+iT_2,\
  A_0   =-2iT_3,\
  A_{1} =T_1-iT_2,\\
  A&=A_{-1}\zeta\sp{-1}+A_0+A_1\zeta,\qquad
  M=\frac{1}{2}A_0+A_1\zeta,
\end{align*} then
\begin{equation}\label{nahmlax}
    \dfrac{d A}{ds}=[A,M],\ \text{or equivalently}\
    [\dfrac{d}{ds}+M,A]=0.
\end{equation}

 Nahm's equation (\ref{nahm}) describes linear flow on a
complex torus, which is the Jacobian of an algebraic curve. This
algebraic curve is in fact the monopole spectral curve
$\mathcal{C}$ and may be explicitly read off from the Lax equation
\begin{equation}
P(\eta,\zeta)=\mbox{det}(\eta+(T_1+iT_2)-2iT_3\zeta+(T_1-iT_2)\zeta^2)=0.
\label{lax}
\end{equation}
The regularity condition B2 for $s\in(0,2)$ is the manifestation
in the ADHMN approach of the condition A3 for spectral curves.

\subsection{The Ercolani-Sinha construction}
We shall now present a short overview of the Ercolani-Sinha
construction which expresses a gauge transform of the Nahm data in
terms of Baker-Akhiezer functions on the spectral curve
$\mathcal{C}$. An explicit representation of these functions will
be given after that. Extensions to the theory of Ercolani and
Sinha will be presented in the next subsection.

\subsubsection{Overview}
Let $z=s-1$. Then $z\in[-1,1]$ for $s\in[0,2]$ and we have
that\footnote{The matrices of Ercolani-Sinha and Nahm are related
by $T_i\sp{\text{ES}}(z)=T_i\sp{\text{Nahm}}(z+1)$, whence
$$T_i\sp{\text{ES}}(z)=T_i\sp{\text{Nahm}}(z+1)=T_i\sp{\text{Nahm}\,t}(2-[z+1])
=T_i\sp{\text{Nahm}\,t}(1-z)=T_i\sp{\text{ES}\,t}(-z).$$}
\begin{equation}\label{aprops}
    A_0(z)=A_0\sp\dagger(z),\qquad
    A_1(z)=-A_{-1}\sp\dagger(z),\qquad
    A_\alpha(z)=A_\alpha\sp{t}(-z),\ \alpha=1,2,3.
\end{equation}

Ercolani and Sinha begin by focussing attention on the
differential operator
$$\dfrac{d}{dz}+M(z)=\dfrac{d}{dz}+\frac{1}{2}A_0(z)+A_1(z)\zeta$$
related to the Lax equation (\ref{nahmlax}). The spectral theory
of this equation enables the integration of the Lax equation. The
$z$-dependence of the term $A_1(z)$ means that
$$ \left(\dfrac{d}{dz}+\frac{1}{2}A_0(z)\right)\varphi =-\zeta
A_1(z)\varphi$$ is not of standard eigenvalue form. By considering
the gauge transformation
$$ Q_\alpha(z)=C\sp{-1}(z)A_\alpha(z)C(z),\qquad
\varphi=C(z)\Phi,$$ they obtain the standard eigenvalue equation
\begin{equation}\label{esgtde}
\left(\dfrac{d}{dz}+Q_0(z)\right)\Phi =-\zeta Q_1(0)\Phi,
\end{equation}
if and only if $C(z)$ satisfies
\begin{equation}\label{esgt}(
  C(z)\sp{-1}C'(z)=\frac{1}{2}Q_0(z),\
  \text{equivalently}\  \left(\dfrac{d}{dz}+\frac{1}{2}Q_0(z)\right)C\sp{-1} =0.
\end{equation}
The gauge transform was chosen so that $Q_1(z)=A_1(0)=Q_1(0)$.
From (\ref{aprops}) we see that this is a symmetric matrix, and
(by an overall constant gauge transformation) we may assume this
is diagonal,
\begin{equation}\label{aq0form}
A_1(0)=Q_1(0)=\text{diag}(\rho_1,\ldots,\rho_n).
\end{equation}
We see from (\ref{esgtde}) that the $\rho_j$ here (which may be
assumed distinct) correspond  to the roots of
${P(\eta,\zeta)}/{\zeta\sp{2n}}$ near $\zeta=\infty$,
\begin{equation}\label{rhodef}\frac{P(\eta,\zeta)}{\zeta\sp{2n}}\sim \prod_{j=1}\sp{n}\left(
\frac{\eta}{\zeta\sp2}-\rho_j\right).\end{equation} As a
consequence we see that at $\infty_j$ we have
\begin{equation}\label{infj}
\frac{\eta}{\zeta}\sim \rho_j\,\zeta,\qquad
{d}\left(\frac{\eta}{\zeta}\right)\sim \rho_j\,
{d}\zeta=\left(-\frac{\rho_j}{t^2} +O(1)\right){d}t,
\end{equation}
where $t=1/\zeta$ is a local coordinate. From (\ref{arpar}) we see
that at $\zeta=0$ we also have that
\begin{equation}\label{rhobdef}
P(\eta,0)=\prod_{j=1}\sp{n}\left(
\eta+\overline{\rho_j}\right).\end{equation}

The spectral curves $0=\det(\eta-A)=\det(\eta-Q)$ agree being
related by a gauge transformation $Q=C\sp{-1}AC$. Ercolani and
Sinha now construct $Q_0(z)$ in terms of a Baker-Akhiezer function
on $\mathcal{C}$. Baker-Akhiezer functions are a slight extension
to the class of meromorphic functions that allow essential
singularities at a finite number of points; they have many
properties similar to those of meromorphic functions. While for a
meromorphic function one needs to prescribe $g_\mathcal{C}+1$
poles in the generic situation, a non-trivial Baker-Akhiezer
function exists with $g_\mathcal{C}$ arbitrarily prescribed poles
on a surface of genus $g_\mathcal{C}$. The key result is the
following theorem.
\begin{theorem}[Krichever, 1977]\label{bathm}
Let $\mathcal{C}$ be a smooth algebraic curve of genus
$g_\mathcal{C}$ with $n\geqslant 1$ punctures $P_j $, $j
=1,\ldots, n$.  Then for each set of $g_S+n-1$ points $\delta
_{1},\ldots ,\delta _{g_S+n-1}$ in general position, there exists
a unique function $\Psi _{j }\left( t,P\right) $ and local
coordinates $w_{j }(P)$ for which $w_{j }(P_{j })=0$, such that
\begin{enumerate}
\item  The function $\Psi _{j }$ of $P\in \mathcal{C}$ is meromorphic
outside the punctures and has at most simple poles at $\delta
_{s}$ (if all of them are distinct);

\item  In the neighbourhood of the puncture $P_{l }$ the
function $\Psi _{j }$ has the form
\begin{equation}
\Psi _{j }\left( z, P\right) =e^{z\,{w_{l }}^{-m}}\left( \delta
_{j l }+\sum\limits_{k=1}^{\infty }\alpha\sp{k}_{j l }\left(
z\right) w_{l }^{k}\right), \qquad w_{l }=w_{l }\left(
P\right),\qquad m\in\mathbb{N}\sp{+}. \label{baexp}
\end{equation}
\end{enumerate}
\end{theorem}
The integer $m\ge1$ in the theorem is arbitrary and in
applications is determined by a given flow. Let $\tilde{w}_{j
}(P)$ be any local coordinates on $\mathcal{C}$ such that $\tilde
w_{j }(P_{j })=0$. To a particular flow we associate the unique
meromorphic differential $ d\Omega ^{[m]}$ on $\mathcal{C}$,
holomorphic outside the punctures $P_{j }$, with form
\begin{equation}\label{floa}
d\Omega ^{[m]}=d\left( \tilde{w}_{j }^{-m}+0\left( \tilde{w}_{j
}\right) \right)
\end{equation}
near the puncture $P_{j }$, and normalized with vanishing
$\mathfrak{a}$-periods
\begin{equation}\label{flob}
\oint_{\mathfrak{a}_{k}}d\Omega ^{[m]}=0.
\end{equation}

We may utilise the Baker-Akhiezer function in the monopole setting
as follows. Let $\Phi_1,\ldots,\Phi_n$ be the columns of the
fundamental matrix solution $\Omega$ to (\ref{esgtde}), normalized
so that
\begin{equation}\label{normba}
\exp(\zeta A_1(0) z)\Omega\big\vert_\infty=\mathrm{Id}_n. \end{equation} In
view of (\ref{infj}) we consider a differential $\gamma_\infty$ of
the second kind such that
\begin{align}
&\gamma_{\infty}(P)=
\left(\frac{\rho_l}{t^2}+O(1)\right){d}t,\quad \text{as}\quad
P\rightarrow\infty_l,\label{gaminf}\\
&\oint\limits_{\mathfrak{a}_k} \gamma_{\infty}(P) = 0,\quad
\forall k=1,\ldots,g,\label{gamnorm}
\end{align}
and take as punctures $\{P_l=\infty_l\}$ ($l=1,\ldots,n$),  the
$n$ points on $\mathcal{C}$ which lie above the point
$\zeta=\infty$. Then
\begin{theorem}[Ercolani and Sinha, 1989]\label{esba}
The $j$-th component of $\Phi_l$ normalised by (\ref{normba}) is
given by the expansion (\ref{baexp}) of $\Psi_j$ at $\infty_l$.
Further the matrix $Q_0$ has vanishing diagonal entries and may be
reconstructed from
\begin{equation}\label{esq0}
  \left(Q_0\right)_{jl}=-(\rho_j-\rho_l)\,\alpha\sp{1}_{jl}
=-(\rho_j-\rho_l)\,\lim_{P\rightarrow\infty_l}\zeta \exp(\zeta
\rho_l z) \Psi_j(z,P).
\end{equation}
\end{theorem}

The steps involved to obtain Nahm data in the Ercolani-Sinha
construction, are as follows:
\begin{enumerate}
    \item From the asymptotic properties of the curve solve for
    $\rho_j$ ($j=1,\ldots,g_\mathcal{C}$). Then $A_1(0)=Q_1(0)={\text{
    diag}}(\rho_j)$.
    \item Determine $Q_0(z)$ from (\ref{esgtde}) in terms of the Baker-Akhiezer function
    (\ref{esq0}).
    \item Determine $C(z)$ from (\ref{esgt}). Then
    \begin{enumerate}
        \item $A_0(z)=C(z)Q_0C\sp{-1}(z)$.
        \item $A_1(z)=C(z)A_1(0)C\sp{-1}(z)$.
        \item $A_{-1}(z)=-A_1\sp\dagger(z)$.
    \end{enumerate}
    \item From $A(z)$ reconstruct
    $T_i\sp{\text{ES}}(z)=T_i\sp{\text{Nahm}}(z+1)$.
\end{enumerate}
The constraints (B2, B3) or (A2, A3) arise in the Ercolani-Sinha
construction as constraints on the Baker-Akhiezer functions in
step (2). Thus to implement the approach we need to be able to
concretely express Baker-Akhieser functions. We shall consider
these in more detail in the next subsection, first in general and
then in the monopole context. Before doing this however it will be
useful to make some remarks regarding the gauge ambiguities of the
solution. The matrices $A$, $M$ and $C$ were initially defined up
to constant gauge transformations. By choosing the form
(\ref{aq0form}) these were reduced to constant diagonal gauge
transformations, such preserving the normalization (\ref{normba}).
At this stage then our matrix $Q_0(z)$ defined in terms of the
Baker-Akhieser function is defined up to constant diagonal gauge
transformations, $Q_0(z)_{ij}\sim d_i Q_0(z)_{ij} d_j\sp{-1}$ for
$d_i\ne0$ ($i,j=1\dots n$).

\subsubsection{Baker-Akhieser functions}
The functions $\Psi_j$ of theorem \ref{bathm} may be written
explicitly in terms of $\theta$-functions and the Abel map $\boldsymbol{\phi}$.
Given the normalised differential $d\Omega ^{[m]}$ of the second
kind (\ref{floa}, \ref{flob}) define the vector $U^{[m]}$ with
coordinates
\begin{equation*}
U_{k}^{[m]}=\frac{1}{2\pi i}\oint_{\mathfrak{b}_{k}}d\Omega
^{[m]}.
\end{equation*}
Then the function $\Psi _{j }\left( z, P\right) $ may be expressed
as
\begin{equation}\label{constnbafn}
\Psi _{j }\left( z, P\right) =g_{j }(P) \frac{ \theta \left(
\boldsymbol{\phi} (P)-\boldsymbol{\mathcal{Z}}_{j
}+z\,U^{[m]}\right) \theta \left(\boldsymbol{\phi} \left( P_{j
}\right)-\boldsymbol{\mathcal{Z}}_{j } \right) } { \theta
\left(\boldsymbol{\phi} \left( P_{j
}\right)-\boldsymbol{\mathcal{Z}}_{j }+ z\,U^{[m]}\right) \theta
\left( \boldsymbol{\phi} (P)-\boldsymbol{\mathcal{Z}}_{j }\right)
} e^{z\,\int\limits_{P_0}^{P}d\Omega\sp{[m]}}.
\end{equation}
Here
\begin{align*}
\boldsymbol{\mathcal{Z}}_{j } &=\boldsymbol{\mathcal{Z}}_{T}+\boldsymbol{\phi} \left( P_{j }\right)\equiv
\boldsymbol{\phi}(\Delta_j)+\boldsymbol{K} ,\qquad \boldsymbol{\mathcal{Z}}_{T}
=\sum\limits_{s=1}^{g_\mathcal{C}+n-1}\boldsymbol{\phi} \left( \delta
_{s}\right) -\sum\limits_{j =1}^{n}\boldsymbol{\phi} \left( P_{j
}\right)+\boldsymbol{K} ,
\end{align*}
where $\boldsymbol{K}$ is the vector of Riemann constants (with
base point $P_0$). Our conventions for theta functions are given
in the Appendix. By Abel's theorem $\boldsymbol{\mathcal{Z}}_{j }$ is
equivalent to a an effective divisor $\Delta_j$ of degree $g$. The
function $g_j(P)$ is the unique meromorphic function with
$$g_j(P_l)=\delta_{jl}$$
and for $n\ge2$ having poles from
$\{\delta_1,\ldots,\delta_{g_\mathcal{C}+n-1}\}$. For the case
$n=1$ we have $g_j(P)=1$. Again, this function may be explicitly
constructed. Set
$$ g_j(P)=\frac{f_j(P)}{f_j(P_j)},$$
where
$$
f_j(P)=\theta(\boldsymbol{\phi}(P)-\boldsymbol{\mathcal{Z}}_j)\,
\frac{\prod_{l\ne j}
\theta(\boldsymbol{\phi}(P)-\boldsymbol{R}_l)}
     {\prod_{k=1}\sp{n} \theta(\boldsymbol{\phi}(P)-\boldsymbol{S}_k)}
$$
and
\begin{align*}
\boldsymbol{R}_j&
= \sum\limits_{s=1}^{g_\mathcal{C}-1}\boldsymbol{\phi} \left( \delta
_{s}\right) +\boldsymbol{\phi} \left( P_{j }\right)+\boldsymbol{K},\qquad
\boldsymbol{S}_j=\sum\limits_{s=1}^{g_\mathcal{C}-1}\boldsymbol{\phi}
\left(\delta_{s}\right) +\boldsymbol{\phi} \left( \delta_{g_\mathcal{C}-1+j
}\right)+\boldsymbol{K}.
\end{align*}
Observe that for $n\ge2$ the factors
$\theta(\boldsymbol{\phi}(P)-\boldsymbol{\mathcal{Z}}_j)$ cancel between the term involving
$g_j(P)$ and the theta function in the denominator of
(\ref{constnbafn}), and so no extraneous poles are added.

Now the function $\Psi _{j }\left( z, P\right) $, which depends on
the choice of base point of the Abel map, has the requisite
properties of theorem \ref{bathm} aside from that of
normalization. Set
\begin{equation}
\nu_{j}\equiv\nu_{j}(P_0)=\lim_{P\rightarrow \infty_j}
\left[\int\limits_{P_0}^P {d}\Omega\sp{[m]}- \frac{1}{{ \tilde
w}\sp{m}(P)} \right].\label{banorm1}
\end{equation}
Thus for the local coordinate $\tilde w$ the Baker-Akhieser
function differs from the normalization of (\ref{baexp}) by the
exponential $\exp(z\nu_j)$. For this local coordinate the function
$\exp(-z\nu_j)\,\Psi_j$ has the desired normalization. Alternately
we may make a change of local coordinates
$$
w=\frac{ \tilde w}{1+\frac{1}{m}{ \tilde w}\sp{m}\nu(P_0)}
$$
for which
$$
\frac{1}{w\sp{m}}=\frac{(1+\frac{1}{m}{ \tilde
w}\sp{m}c(P_0))\sp{m}} {{ \tilde w}\sp{m}} =\frac{1}{{ \tilde
w}\sp{m}}+\nu(P_0)+O({ \tilde w}\sp{m})
$$
and we have a local coordinate for which the Baker-Akhieser
function has the desired expansion.

Let us conclude with some discussion of the divisor $\delta\equiv
\sum_{s=1}^{g_\mathcal{C}+n-1} \delta _{s}$ explaining what is
meant by saying that it is ``in general position". We may
interpret the meromorphic functions $g_j(P)$ as follows. Let $
L_\delta$ denote the line bundle on $\mathcal{C}$ determined by
the divisor $\delta$ and denote by $s_ \delta$ a (nonzero)
meromorphic section of this line bundle. (We shall further
identify $ L_\delta$ in the monopole setting in due course.) Then
the divisor of $g_j(P)$ is
\begin{equation}\label{divgj}
  \textrm{Div}\,g_j(P)=\Delta_j+P_1+\ldots+\hat P_j+\ldots+P_n-\delta.
\end{equation}
Thus $g_j(P)s_ \delta$ yields a holomorphic section of $
L_\delta$. Now by Riemann-Roch
$$ \dim H\sp0\mathcal{(C},\mathcal{O}(
L_\delta))=\textrm{deg} L_\delta +1-g_\mathcal{C}+\dim
H\sp1\mathcal{(C},\mathcal{O}( L_\delta))=n+\dim
H\sp1\mathcal{(C},\mathcal{O}( L_\delta))\ge n,$$ and $ L_\delta$
has precisely $n$ holomorphic sections when $\dim
H\sp1\mathcal{(C},\mathcal{O}( L_\delta))=0$. This latter
constraint means $\delta$ is a nonspecial divisor. This is a
condition on the divisor. Now consideration of the short exact
sequence
$$0\rightarrow \mathcal{O}(L)\xrightarrow{s_p} \mathcal{O}(LL_p)\rightarrow
\mathcal{O}_p(LL_p)\rightarrow0$$ and the corresponding long exact
sequence
$$0\rightarrow H\sp0\mathcal{(C},\mathcal{O}(L))\rightarrow
H\sp0\mathcal{(C},\mathcal{O}(L L_p))
\rightarrow\mathbb{C}\rightarrow H\sp1\mathcal{(C},\mathcal{O}(
L))\rightarrow H\sp1\mathcal{(C},\mathcal{O}( LL_p))\rightarrow0$$
shows us that either of two possibilities arise,
\begin{align*}
\textrm{(a)}\ & \dim H\sp0\mathcal{(C},\mathcal{O}(L L_p))= \dim
H\sp0(\mathcal{C},\mathcal{O}(L))+1, \ \textrm{and} \ \dim
H\sp1(\mathcal{C},\mathcal{O}( L))=\dim
H\sp1(\mathcal{C},\mathcal{O}( LL_p)),\\
\textrm{(b)}\ & \dim H\sp0(\mathcal{C},\mathcal{O}(L L_p))= \dim
H\sp0(\mathcal{C},\mathcal{O}(L)), \ \textrm{and} \ \dim
H\sp1(\mathcal{C},\mathcal{O}( L))=\dim
H\sp1(\mathcal{C},\mathcal{O}( LL_p))+1.
\end{align*}
In particular, if $\dim H\sp1(\mathcal{C},\mathcal{O}( L))=0$ then
$\dim H\sp0\mathcal{(C},\mathcal{O}(L L_p))= \dim
H\sp0(\mathcal{C},\mathcal{O}(L))+1$. (When the divisor of $LL_p$
is effective setting (a) is the generic situation, true for
general $p$.) Using these results together with Riemann-Roch we
find that (for each $j=1,\ldots,n$)
\begin{align*}
&\dim
H\sp0(\mathcal{C},\mathcal{O}(L_{\delta-\sum_{k=1}\sp{n}P_k}))=
\dim H\sp0(\mathcal{C},\mathcal{O}(L_{\Delta_j-P_j}))=0\nonumber\\
\Longleftrightarrow&\dim
H\sp1(\mathcal{C},\mathcal{O}(L_{\delta-\sum_{k=1}\sp{n}P_k}))=
\dim H\sp1(\mathcal{C},\mathcal{O}(L_{\Delta_j-P_j}))=0,\nonumber\\
\Longrightarrow& \begin{cases} \dim
H\sp0(\mathcal{C},\mathcal{O}(L_{\delta+P_j-\sum_{k=1}\sp{n}P_k}))=
\dim H\sp0(\mathcal{C},\mathcal{O}(L_{\Delta_j}))=1,  \\
\dim
H\sp1(\mathcal{C},\mathcal{O}(L_{\delta+P_j-\sum_{k=1}\sp{n}P_k}))=
\dim H\sp1(\mathcal{C},\mathcal{O}(L_{\Delta_j}))=0,
\end{cases}\\
\Longrightarrow& \dim
H\sp0(\mathcal{C},\mathcal{O}(L_{\delta}))=n,\quad \dim
H\sp1(\mathcal{C},\mathcal{O}(L_{\delta}))=0.\nonumber
\end{align*}
The condition $\dim
H\sp0(\mathcal{C},\mathcal{O}(L_{\Delta_j}))=1$ says that the
divisors $\Delta_j$ are nonspecial and we have used this in our
construction  to assert the uniqueness of the functions $g_j(P)$
and correspondingly that of the Baker-Akhieser functions. Actually
our requirement that $g_j(P_j)=1$ means that we can say more here.
Our analysis of the long exact sequence shows that either $\dim
H\sp0(\mathcal{C},\mathcal{O}(L_{\Delta_j-P_j}))=0$ or $1$. If the
latter then the divisor $\Delta_j-P_j$ is equivalent to an
effective divisor, $\Delta_j-P_j\sim_l
\sum_{k=1}\sp{g_\mathcal{C}-1} Q_k$, whence $\Delta_j\sim_l P_j+
\sum_{k=1}\sp{g_\mathcal{C}-1} Q_k$. But then $g_j(P_j)=0$, a
contradiction. Thus $\dim
H\sp0(\mathcal{C},\mathcal{O}(L_{\Delta_j-P_j}))=0$ and we have
established the necessary and sufficient condition for the
construction of the Baker-Akhieser function (for each
$j=1,\ldots,n$),
\begin{equation} \label{divdeltaa} \dim
H\sp0(\mathcal{C},\mathcal{O}(L_{\delta-\sum_{k=1}\sp{n}P_k}))=0
\Longleftrightarrow \dim
H\sp0(\mathcal{C},\mathcal{O}(L_{\delta+P_j-\sum_{k=1}\sp{n}P_k}))=1.
\end{equation}
Condition (\ref{divdeltaa}) says that the degree $g_\mathcal{C}-1$
divisor $\delta-\sum_{k=1}\sp{n}P_k$ is noneffective. In
particular this means that $\delta-\sum_{k=1}\sp{n}P_k\in
\textrm{Jac}\sp{g_\mathcal{C}-1}(\mathcal{C})\setminus \Theta$.
Here the theta divisor $\Theta$  is precisely the image (up to a
shift by the vector of Riemann constants) by the Abel map of
degree $g_\mathcal{C}-1$ effective divisors in the Jacobian
$\textrm{Jac}\sp{g_\mathcal{C}-1}(\mathcal{C})$. Now $\Theta$ is
of codimension one in the Jacobian, and so (\ref{divdeltaa}) will
hold for generic divisors. This is what we mean by $\delta$ being
``in general position". Finally let us remark that just as the
functions $g_j(P)=\Psi_j(0,P)$ yield sections of a line bundle
$L_\delta$, then similarly the functions $\Psi_j(z,P)$ yield
sections of a line bundle which we will denote $L_\delta\sp{z}$,
but now the transition functions in the vicinity of $P_l$ involve
the exponential term $\exp(z w_l\sp{-i})$.

\subsubsection{The Ercolani-Sinha constraints}
It follows from the last paragraph that, upon setting
\begin{equation}
\nu_{j}=\lim_{P\rightarrow \infty_j}
\left[\int\limits_{P_0}^P\gamma_{\infty}(P)+\frac{\eta}{\zeta}
\right],\label{etacondition}
\end{equation}
we may write the Baker-Akhieser function of theorem \ref{esba} as
\begin{equation}\label{newbafn}
\Psi _{j }\left( z, P\right) =g_{j }(P) \frac{ \theta \left( \boldsymbol{\phi}
(P)-\boldsymbol{\mathcal{Z}}_{j }
+z\,U\right) \theta \left(\boldsymbol{\phi} \left( P_{j
}\right)-\boldsymbol{\mathcal{Z}}_{j } \right) }
{ \theta \left(\boldsymbol{\phi} \left(
P_{j }\right)-\boldsymbol{\mathcal{Z}}_{j }+ z\,U\right)
\theta \left( \boldsymbol{\phi}
(P)-\boldsymbol{\mathcal{Z}}_{j }\right) }
e^{z\,\int\limits_{P_0}^{P}\gamma_{\infty}-z\,\nu_j}
\end{equation}
for a suitably generic divisor $\delta=\sum_{s=1}^{n(n-1)}\delta
_{s}$. Thus from (\ref{esq0}) we obtain the matrix $Q_0$ of
(\ref{esgtde}) as
\begin{equation}\label{esq0f}
\left(Q_0\right)_{jl}=-(\rho_j-\rho_l)\,c_{jl}\,e\sp{z[\nu_l-\nu_j]}\,
\frac{ \theta \left( \boldsymbol{\phi} (P_l)-\boldsymbol{\mathcal{Z}}_{j }
+z\,U\right)
\theta \left(\boldsymbol{\phi} \left( P_{j }\right)
-\boldsymbol{\mathcal{Z}}_{j } \right) }
{ \theta \left(\boldsymbol{\phi} \left( P_{j }\right)
-\boldsymbol{\mathcal{Z}}_{j }+
z\,U\right) \theta \left( \boldsymbol{\phi} (P_l)
-\boldsymbol{\mathcal{Z}}_{j }\right) },
\end{equation}
where
\begin{equation}\label{cijdef}
c_{jl}=\lim_{P\rightarrow\infty_l}\zeta\, g_j(P),\qquad
P=(\zeta,\eta)\in \mathcal{C}.
\end{equation}
We note that the constants $c_{ij}$ appearing in this solution
depend on the divisor $\delta$ through the functions $g_j(P)$. A
puzzle is what this dependence corresponds to in the physical
setting. In due course we shall show that this corresponds to a
gauge choice and give a simple form for these constants.

At this stage we have not imposed the Hitchin constraints A2, A3
on our curve $\mathcal{C}$. First let us identify  the line
bundles $L_\delta\sp{z}$ in the construction. From
(\ref{etacondition}, \ref{newbafn}) we have that
$\Psi_j\sim\exp(-z\eta/\zeta)\vert_{\infty_j}$ and so leads to
transition functions of this form in the neighbourhood of
$\zeta=\infty$. Now from (\ref{divgj}) the line bundle
$L_\delta\sp{z=0}=L_\delta$  has divisor (for each $j$)
\begin{equation*}
\delta \sim_l
  \Delta_j+\infty_1+\ldots+\hat\infty_j+\ldots+\infty_n,
\end{equation*}
and consequently has a zero of order $n-1$ above infinity. In
terms of the local coordinate $\tilde \zeta=1/\zeta$ this
corresponds to a section $s_1=\sum_{k=0}\sp{n-1}\mu_k
\tilde\zeta\sp{k}$ with transition function $g_{01}=z\sp{n-1}$.
(Here $s_0(P)=\Psi_j(z,P)$ on $\hat
U_0\cap\mathcal{C}\cap\setminus \{\delta\}$ and $s_1(P)=1$ on
$\hat U_1\cap\mathcal{C}$, while for patches $ V_{j+1}\subset \hat
U_0\cap\hat U_1\cap\mathcal{C}$, $\delta_j\in V_{j+1}$ and with
$V_j\cap V_k=\emptyset$ ($j\ne k$) we have for $P\in V_{j+1}$ that
$s_{j+1}(P)=w_j$.) Thus we may identify our bundle
$L_\delta\sp{z}$ with the bundle $L\sp{z+1}(n-1)$ of Hitchin.
Further the bundle $L_{\delta-\sum_{k=1}\sp{n}
P_k}=L_{\delta}\otimes \pi\sp* \mathcal{O}(-1)\equiv
L_{\delta}(-1)$ is then identified with Hitchin's $L\sp{1}(n-2)$.
Condition A3 for $\lambda=1$ is then the constraint
(\ref{divdeltaa}), $$L_\delta(-1)\in
\textrm{Jac}\sp{g_\mathcal{C}-1}(\mathcal{C})\setminus \Theta.$$
This condition means that the (push-forward) rank $n$ vector
bundle $E=\pi_* L_\delta$ on $\mathbb{P}\sp1$ is holomorphically
trivial. We shall now discuss the constraints A2 and A3.

\noindent{\textbf{A2}} With regards to A2 Ercolani and Sinha show
that the functions $g_j(P)$ form a basis of the holomorphic
sections of $L(n-1)$ and as a consequence $L(n-1)$ is real. Then
they consider the logarithmic derivative of (\ref{triv3})
representing the triviality of $L^2$ on $\mathcal{C}$,
\begin{equation}
\mathrm{d}\mathrm{log}\,f_{0}={d}\left(
-2\frac{\eta}{\zeta}\right) +\mathrm{d}\mathrm{log}
f_{1}\label{derivative}.
\end{equation}
(Hurtubise considered a similar construction in the $n=2$ case
\cite{hurt}.) Now in order to avoid essential singularities in
$f_{0,1}$ we have from (\ref{rhodef}, \ref{rhobdef}) that
\begin{align}
\mathrm{d}\mathrm{log}\,f_{1}(P)
&=\left(-\frac{2\eta_j(0)}{\zeta^2}
+O(1)\right){d}\zeta=\left(\frac{2{\bar \rho}_j(0)}{\zeta^2}
+O(1)\right){d}\zeta,\quad \text{at}\quad P\rightarrow 0_j,\\
\mathrm{d}\mathrm{log}\,f_{0}(P) &=\left(\frac{2\rho_j}{t^2}
+O(1)\right){d}t,\quad \text{at}\quad P\rightarrow
\infty_j.\label{foex}
\end{align}
Because $f_{0}$  is a function on $U_0=\mathcal{C}\setminus
\{P_j\}_{j=1}\sp{n}$, then
\begin{equation}
\mathrm{exp}\oint\limits_{\lambda}\mathrm{d}\mathrm{log}\,f_{0}=1,
\end{equation}
for all cycles $\lambda$ from $H_1(\mathbb{Z},\mathcal{C})$. A
similar result follows for $f_1$ and upon noting
(\ref{derivative}) we may define
\begin{align}
m_j =
-\frac{1}{2\pi\imath}\oint\limits_{\mathfrak{a}_j}\,\mathrm{d}\mathrm{log}
\,
f_{0}=-\frac{1}{2\pi\imath}\oint\limits_{\mathfrak{a}_j}\mathrm{d}
\mathrm{log} \, f_{1}\label{denotem},\\
n_j =
\frac{1}{2\pi\imath}\oint\limits_{\mathfrak{b}_j}\mathrm{d}\mathrm{log}
\, f_{0}=\frac{1}{2\pi\imath}\oint\limits_{\mathfrak{b}_j}
\mathrm{d}\mathrm{log} \, f_{1}.\label{denoten}
\end{align}
Further, in view of (\ref{gaminf}) and (\ref{foex}), we may write
\begin{equation}\label{gamfo}
\gamma_\infty=\frac{1}{2}\,\mathrm{d}\mathrm{log} \, f_{0}+\imath
\pi\,\sum_{j=1}^g m_j\int\limits_{P_0}^P v_j,
\end{equation}
where $v_j$ are canonically $\mathfrak{a}$-normalized holomorphic
differentials. Integrating $\gamma_\infty$ around
$\mathfrak{b}$-cycles leads to the Ercolani-Sinha constraints
\begin{equation*}
\oint\limits_{\mathfrak{b}_k}\gamma_\infty=\imath\pi  n_k
+\imath\pi\sum_{l=1}^g\tau_{kl} m_l,\label{winding}
\end{equation*}
which are necessary and sufficient conditions for $L\sp2$ to be
trivial when restricted to $ \mathcal{C}$. Thus the winding vector
$U$ appearing in the Baker-Akhieser function (\ref{newbafn}) takes
the form
\begin{equation}\label{ESC1}
\boldsymbol{U}=\frac12 \boldsymbol{n}+\frac12\tau\boldsymbol{m}.
\end{equation}
Therefore the vector $2\boldsymbol{U}\in\Lambda $, the period
lattice for the curve $\mathcal{C}$, and so the ``winding-vector"
vector $\mathcal{}$ is a half-period. Note that $\boldsymbol{U}\ne
0$ or otherwise $\gamma_\infty$ would be holomorphic contrary to
our choice.

Using the bilinear relations (and that
$\oint_{\mathfrak{a}_l}\gamma_{\infty}(P)=0$) we have that
\begin{align}
U_k&=\frac{1}{2\pi\imath}\oint_{\mathfrak{b}_k}\gamma_{\infty}(P)\nonumber\\
&=\frac{1}{2\pi\imath}\sum_{l=1}\sp{g}
\left(\oint_{\mathfrak{a}_l} v_k(P)
\oint_{\mathfrak{b}_l}\gamma_{\infty}(P')- \oint_{\mathfrak{b}_l}
v_k(P) \oint_{\mathfrak{a}_l}\gamma_{\infty}(P')
\right) \nonumber\\
 &=\frac{1}{2\pi\imath}\oint_{\partial
\Gamma}\gamma_{\infty}(P)
\int_{P_0}^P v_k(P') \\
&=\sum_{i=1}^n\mathrm{Res}_{P\rightarrow\infty_i}
\gamma_{\infty}(P)\int_{P_0}^P v_k(P') \nonumber\\
&=\sum_{i=1}^n \rho_j\,V\sp{(j)}_k. \nonumber
 \end{align}
Here we have defined ``winding vectors'' $\boldsymbol{T}^{(i)}$,
$\boldsymbol{V}^{(i)}$, $\boldsymbol{W}^{(i)}$, as the
coefficients of the expansion in the vicinity of $\infty_i$ of
\begin{equation}\label{TVW}
\left.\int\limits_{P_0}^P\boldsymbol{v} \right|_{P\rightarrow
\infty_i} =\boldsymbol{T}^{(i)}+ t\boldsymbol{V}^{(i)}
+\frac{t^2}{2}\boldsymbol{W}^{(i)}+\ldots,
\end{equation}
and so $  \boldsymbol{T}^{(i)}
=\int\limits_{P_0}^{\infty_i}\boldsymbol{v} $. More generally, for
any holomorphic differential $\Omega$
$$\sum_{l=1}\sp{g} U_l\oint_{\mathfrak{a}_l}\Omega=
\frac{1}{2\pi\imath}\sum_{l=1}\sp{g}\left(
\oint_{\mathfrak{a}_l}\Omega
\oint_{\mathfrak{b}_l}\gamma_{\infty}(P)-
\oint_{\mathfrak{b}_l}\Omega
\oint_{\mathfrak{a}_l}\gamma_{\infty}(P)
\right)=\sum_{i=1}^n\mathrm{Res}_{P\rightarrow\infty_i}
\gamma_{\infty}(P)\int_{P_0}^P \Omega.
$$
Houghton, Manton and Ram\~ao utilise this expression to express a
dual form of the Ercolani-Sinha constraints (\ref{ESC1}). Define
the 1-cycle
\begin{equation}\label{HMRcurve}
  \mathfrak{c}=\sum_{l=1}\sp{g}(n_l \mathfrak{a}_l +m_l \mathfrak{b}_l ).
\end{equation}
Then (upon recalling that $\tau_{lk}\oint_{\mathfrak{a}_k}\Omega
=\oint_{\mathfrak{b}_l}\Omega$, where $\tau$ is the period matrix)
we have the equivalent constraint:
\begin{equation}\label{ESC2}
\oint_{\mathfrak{c}}\Omega=2\sum_{i=1}^n\mathrm{Res}_{P\rightarrow\infty_i}
\gamma_{\infty}(P)\int_{P_0}^P \Omega.
\end{equation}
The right-hand side of this equation is readily evaluated. We may
express an arbitrary holomorphic differential $\Omega$ as,
\begin{align}
\Omega &=\frac{\beta_0\eta^{n-2}+\beta_1(\zeta)\eta^{n
-3}+\ldots+\beta_{n-2}(\zeta)}{\frac{\partial\mathcal{P}}{\partial
\eta}}\,d\zeta \label{holdefes} \\
&=\frac{\beta_0(\eta/\zeta^2)^{n-2}+\tilde\beta_1(1/\zeta)
(\eta/\zeta^2)^{n-3}+\ldots+\tilde\beta_{n-2}(1/\zeta)}{\sum_{i=1}\sp{n}
\prod_{\substack{j=1\\ j\ne
i}}\sp{n}\left(\eta/\zeta^2-\mu_j(1/\zeta)\right)}\,\frac{d\zeta}{\zeta^2},
\nonumber
\end{align}
where $\beta_j(\zeta)\equiv \zeta^{2j}\tilde \beta_j (1/\zeta)$ is
a polynomial of degree at most $2j$ in $\zeta$. Thus using
(\ref{rhodef}) we obtain
$$
\sum_{i=1}^n\mathrm{Res}_{P\rightarrow\infty_i}
\gamma_{\infty}(P)\int_{P_0}^P \Omega
=-\sum_{i=1}^n\frac{\beta_0\rho_i^{n-1}
+\tilde\beta_1(0)\rho_i^{n-2}+\ldots+\tilde\beta_{n-2}(0)\rho_i}
{\prod_{j\neq i}^n (\rho_i-\rho_j) }=-\beta_0,
$$
upon using Lagrange interpolation%
. At this stage we have from the
condition A2,
\begin{lemma}[Ercolani-Sinha Constraints] The following are equivalent:
\begin{enumerate} \item $L\sp2$ is trivial on $\mathcal{C}$.

\item There exists a 1-cycle
$\mathfrak{c}=\boldsymbol{n}\cdot{\mathfrak{a}}+
\boldsymbol{m}\cdot{\mathfrak{b}}$ such that for every holomorphic
differential $\Omega$ (\ref{holdefes}),
\begin{equation}
\oint\limits_{\mathfrak{c}}\Omega=-2\beta_0,\label{HMREScond}
\end{equation}
\item $2\boldsymbol{U}\in \Lambda\Longleftrightarrow$
\begin{equation}\label{EScond}
\boldsymbol{U}=\frac{1}{2\pi\imath}\left(\oint_{\mathfrak{b}_1}\gamma_{\infty},
\ldots,\oint_{\mathfrak{b}_g}\gamma_{\infty}\right)\sp{T}= \frac12
\boldsymbol{n}+\frac12\tau\boldsymbol{m} .
\end{equation}
\end{enumerate}
\end{lemma}
Here (2) is the dual form of the Ercolani-Sinha constraints given
by Houghton, Manton and Ram\~ao. Their 1-cycle generalises a
similar constraint arising in the work of Corrigan and Goddard
\cite{CorGod}. The only difference between (3) and that of
Ercolani-Sinha Theorem II.2 is in the form of $\boldsymbol{U}$ in
which we disagree. We also know that $\boldsymbol{U}\ne 0$.

The Ercolani-Sinha constraints impose $g$ conditions on the period
matrix of our curve. We have seen that the coefficients $a_r(\zeta)$
each give $2r+1$ (real) parameters, thus the moduli space of charge
$n$ centred $SU(2)$ monopoles is
$$\sum_{r=2}\sp{n}(2r+1)-g=(n+3)(n-1)-(n-1)\sp2=4(n-1)$$
(real) dimensional.

The 1-cycle appearing in the work of Houghton, Manton and Ram\~ao
further satisfies
\begin{corollary}[Houghton, Manton and Ram\~ao,
 2000]\label{HMRinvc} $\tau_*\mathfrak{c}=-\mathfrak{c}$.
\end{corollary}
This result is the dual of Hitchin's remark \cite[p164]{hitchin83}
that the triviality of $L\sp2$ together with the antiholomorphic
isomorphism $L\cong L\sp*$ yields an imaginary lattice point with
respect to $H\sp1(\mathcal{C},\mathbb{Z})\subset
H\sp1(\mathcal{C},\mathcal{O})$.

The Picard group of degree zero line bundles on $\mathcal{C}$ may
be identified with the (principally polarized) Jacobian of
$\mathcal{C}$ via the Abel map. We may identify the origin with
the trivial bundle. The degree of the trivial bundle
$L\sp2=L\sp{z=1}_\delta\otimes \mathcal{O}(-n+1)$ is zero. Thus
$$0=\boldsymbol{\phi}\left(\textrm{Div}(L\sp2)\right)=
\boldsymbol{\phi}\left(\textrm{Div}(L\sp{z=1}_\delta\otimes
\mathcal{O}(-n+1))\right)=\boldsymbol{\phi}\left(\textrm{Div}(L\sp{z=1}_\delta)
-(n-1)\sum_{k=1}\sp{n}\infty_k\right).
$$
Further, consideration of the sections $\Psi_j(1,P)$ associated to
$L\sp{z=1}_\delta$ gives us that
$$\boldsymbol{\phi}\left(\textrm{Div}(L\sp{z=1}_\delta)\right)=
\boldsymbol{\phi}\left(
\delta\right)-\boldsymbol{U}.$$ Together
with (\ref{divgj}) these yield that the winding vector
$\boldsymbol{U}$ may be expressed in terms of the degree zero
divisor (for each $j=1,\ldots,n$)
\begin{equation}\label{divU}
\boldsymbol{U}=\boldsymbol{\phi}\left(\Delta_j-\infty_j-(n-2)
\sum_{k=1}\sp{n}\infty_k\right).
\end{equation}

\noindent{\textbf{A3}} The full condition A3 is that
$L_{\delta}\sp{z}(-1)\in
\textrm{Jac}\sp{g_\mathcal{C}-1}(\mathcal{C})\setminus \Theta$ for
$z\in(-1,1)$. This constraint must be checked using knowledge of
the $\Theta$ divisor. The exact sequence
$\mathcal{O}(L\sp{s})\hookrightarrow \mathcal{O}(L\sp{s}(n-2))$
given by multiplication by a section of
$\pi\sp*\mathcal{O}(n-2)\vert_\mathcal{C}$ does however give us
the necessary condition
\begin{equation}\label{h0lz}
H\sp0\left(\mathcal{C},\mathcal{O}(L\sp{s}(n-2))\right)=0
\Longrightarrow
H\sp0\left(\mathcal{C},\mathcal{O}(L\sp{s})\right)=0,\qquad
s\in(0,2).
\end{equation}
If $L\sp{s}$ were trivial we would have a section, contradicting
this vanishing result. The same treatment given to the triviality
of $L\sp2$ shows that if $L\sp{s}$ were trivial then
$s\,\boldsymbol{U}\in \Lambda$. Therefore (\ref{h0lz}) shows that
$s\,\boldsymbol{U}\not\in \Lambda$ for $s\in(0,2)$. Thus
$2\boldsymbol{U}$ is a primitive vector in $\Lambda$ and we obtain
the final part of the Ercolani-Sinha constraints,
\begin{equation}\label{ESprim}
2\boldsymbol{U}\ \text{ is a primitive vector in} \ \Lambda
\Longleftrightarrow \mathfrak{c}\ \text{ is primitive in}\ H_1(
\mathcal{C},\mathbb{Z}).
\end{equation}

The Ercolani-Sinha constraints (\ref{EScond}) or (\ref{HMREScond})
place $g$ transcendental constraints on the spectral curve $
\mathcal{C}$ and a major difficulty in implementing this
construction has been in solving these, even in simple examples.
Beyond these constraints several further constants have appeared
in the construction (\ref{newbafn}) of the Baker-Akhieser
function. To make the the Ercolani-Sinha construction effective
these need to be calculated and to this we now turn.

\section{Extensions to the Ercolani-Sinha Theory} We shall now
both simplify and extend the formulae of Ercolani-Sinha. In
particular the construction thus far has depended on the divisor
$\delta$ through the constants $c_{ij}$ in (\ref{esq0f}). We shall
show that this divisor encodes a gauge choice, and show how the
constants $c_{ij}$ may be chosen in a particularly simple form
independent of $\delta$. Our form for the matrix $Q_0(z)$ (given
in (\ref{ourq0}) below) is wholly in terms of quantities intrinsic
to the curve. We will highlight the ingredients needed to
calculate $Q_0$ and conclude by showing how the fundamental
bi-differential may be employed in the construction.

\subsection{Vanishing and symmetry properties}
In addition to the Ercolani-Sinha vector, which from (\ref{divU})
may be written,
\begin{equation*}
\boldsymbol{U}=\boldsymbol{\phi}\left(\Delta_j-\infty_j-(n-2)
\sum_{k=1}\sp{n}\infty_k\right),
\end{equation*}
a further vector plays a special role in the monopole
construction. Set $$\widetilde{\boldsymbol{K}}=
\boldsymbol{K}+\boldsymbol{\phi}\left((n-2) \sum_{k=1}\sp{n}\infty_k\right).$$
Here $\boldsymbol{K}$ is the vector of Riemann constants; our
conventions regarding this are given in the Appendix. Let us
observe several points about this vector. First is that
$\widetilde{\boldsymbol{K}}$ is independent of the choice of base
point of the Abel map, for
\begin{align*}
\widetilde{\boldsymbol{K}}_P&= \boldsymbol{K}_P+\boldsymbol{\phi}_P\left((n-2)
\sum_{k=1}\sp{n}\infty_k\right) =\boldsymbol{K}_Q+(g-1)\boldsymbol{\phi}_{ Q}({
P})+\boldsymbol{\phi}_P\left((n-2) \sum_{k=1}\sp{n}\infty_k\right)
=\widetilde{\boldsymbol{K}}_Q,
\end{align*}
using the fact that $g-1=(n-1)^2-1=n(n-2)$. The same fact shows us
that $\widetilde{\boldsymbol{K}}\in
\boldsymbol{K}+\boldsymbol{\phi}\left(\mathcal{C}\sp{g-1}\right)$ and so
secondly,
\begin{equation}
\theta(\widetilde{\boldsymbol{K}})=0.
\end{equation}
We have already established that $\div \pi\sp*\mathcal{O}(1)=
\sum_{k=1}\sp{n}\infty_k$ whence $\div \pi\sp*\mathcal{O}(2n-4)=
2(n-2)\sum_{k=1}\sp{n}\infty_k$. Now Hitchin has established that
$$K_\mathcal{C}=\pi\sp*\mathcal{O}(2n-4)$$
utilising the adjunction formula. Thus
$$2\widetilde{\boldsymbol{K}} =2\widetilde{\boldsymbol{K}}+\boldsymbol{\phi}\left(2(n-2)
\sum_{k=1}\sp{n}\infty_k\right)=2\widetilde{\boldsymbol{K}}+\boldsymbol{\phi}(\div(K_\mathcal{C}))=0$$
upon using (\ref{KKC}). Thus we have thirdly,
\begin{equation}\label{tildeK2}
2\widetilde{\boldsymbol{K}}\in \Lambda.
\end{equation}
Finally Riemann's vanishing theorem for a degree $g-1$ line
bundle,
$${\rm multiplicity}_L \, \theta =\dim
H\sp0(\mathcal{C},\mathcal{O}(L)),$$ together with the fact that
each of the $n-1$ sections of $\mathcal{O}(n-2)$ on $ \PP\sp1$
yield sections of the pull-back, gives us that
$${\rm multiplicity}_{\tilde K}\, \theta=\dim
H\sp0(\mathcal{C},\pi\sp*\mathcal{O}(n-2))\ge n-1.$$ Thus for
$n\ge3$ we have fourthly, that
\begin{equation}
\widetilde{\boldsymbol{K}}\in \Theta_{\rm singular}.
\end{equation}
Indeed, from \cite[Prop. 4.5]{hitchin83} we find that the index of
speciality of $(n-2)\sum_{k=1}\sp{n}\infty_k$ is
\begin{equation}\label{indspec}
s\equiv i\left((n-2)\sum_{k=1}\sp{n}\infty_k\right)
=\dim H\sp0(\mathcal{C},\pi\sp*\mathcal{O}(n-2))=\begin{cases}
\frac14
n^2& \text{if $n$ is even},\\
\frac14 (n-1)^2& \text{if $n$ is odd}. \end{cases}
\end{equation}
This means that all partial derivatives of $\theta$ of order $s-1$
or less  vanish at the point $\widetilde{\boldsymbol{K}}$.
 The point $\widetilde{\boldsymbol{K}}$ is the distinguished point
Hitchin uses to identify degree $g-1$ line bundles with
$\Jac(\mathcal{C})$. Finally we remark that $(n-2)
\sum_{k=1}\sp{n}\infty_k$ and $ \Delta_j-\infty_j$ (for each $j$)
are theta characteristics (see the Appendix).

Using the point $\widetilde{\boldsymbol{K}}$ we may express the
functions (\ref{newbafn}) and (\ref{esq0f}) in the form
\begin{equation}\label{newbafnch}
\Psi _{j }\left( z, P\right) =g_{j }(P)\, \frac{
\theta_{\frac{\boldsymbol{m}}{2},\frac{\boldsymbol{n}}{2}} \left(
\boldsymbol{\phi} (P)-\boldsymbol{\phi}(\infty_{j
})+z\,\boldsymbol{U}-\widetilde{\boldsymbol{K}}\right)
\theta_{\frac{\boldsymbol{m}}{2},\frac{\boldsymbol{n}}{2}}
\left(-\widetilde{\boldsymbol{K}} \right) } {
\theta_{\frac{\boldsymbol{m}}{2},\frac{\boldsymbol{n}}{2}}
\left(\boldsymbol{\phi} (P)-\boldsymbol{\phi}(\infty_{j })
-\widetilde{\boldsymbol{K}}\right)
\theta_{\frac{\boldsymbol{m}}{2},\frac{\boldsymbol{n}}{2}} \left(
z\,\boldsymbol{U}-\widetilde{\boldsymbol{K}}\right) }
e^{z\,\int\limits_{P_0}^{P}\gamma_{\infty}-z\,\nu_j}
\end{equation}
and
\begin{equation}\label{esq0fch}
  \left(Q_0(z)\right)_{jl}=-(\rho_j-\rho_l)\,c_{jl}\,e\sp{z[\nu_l-\nu_j]}\,
\frac{ \theta_{\frac{\boldsymbol{m}}{2},\frac{\boldsymbol{n}}{2}}
\left( \boldsymbol{\phi} (\infty_l)-\boldsymbol{\phi}(\infty_{j
})+z\,\boldsymbol{U}-\widetilde{\boldsymbol{K}}\right)
\theta_{\frac{\boldsymbol{m}}{2},\frac{\boldsymbol{n}}{2}}
\left(-\widetilde{\boldsymbol{K}} \right) } {
\theta_{\frac{\boldsymbol{m}}{2},\frac{\boldsymbol{n}}{2}}
\left(\boldsymbol{\phi} (\infty_l)-\boldsymbol{\phi}(\infty_{j })
-\widetilde{\boldsymbol{K}}\right)
\theta_{\frac{\boldsymbol{m}}{2},\frac{\boldsymbol{n}}{2}} \left(
z\,\boldsymbol{U}-\widetilde{\boldsymbol{K}}\right) } ,
\end{equation}
where $c_{jl}$ has been defined in (\ref{cijdef}).

The matrix $Q_0$ is to satisfy $Q_0(z)=Q_0(-z)\sp{T}$. We find
that
$$Q_0(0)=Q_0(0)\sp{T}\Longleftrightarrow
c_{jl}=-c_{lj}.$$ This, together with (\ref{tildeK2}), yields that
$Q_0(z)=Q_0(-z)\sp{T}$. Thus we must establish that
$c_{jl}=-c_{lj}$.

Before doing this, let us consider the behaviour of
(\ref{esq0fch}). We require $z=0$ to be a regular point. This is
equivalent to our requirement that $\dim
H\sp0(\mathcal{C},\mathcal{O}(L_{\Delta_j-\infty_j}))=0$, for that
means
\begin{equation}\label{regcond}0\ne
\theta\left(
\boldsymbol{\phi}\left(\Delta_j-\infty_j\right)+\boldsymbol{K}\right)=
\theta\left(\boldsymbol{U} +\boldsymbol{\phi}\left((n-2)
\sum_{k=1}\sp{n}\infty_k\right)+\boldsymbol{K} \right)=
\theta\left(\boldsymbol{U}+\widetilde{\boldsymbol{K}} \right),
\end{equation}
and consequently that
$\boldsymbol{U}\pm\widetilde{\boldsymbol{K}}$ is a non-singular
even theta characteristic. Therefore we have the requirement of
the Ercolani-Sinha vector

\begin{lemma}\label{ueventheta}$\boldsymbol{U}\pm
\widetilde{\boldsymbol{K}}$ is a non-singular even theta characteristic.
\end{lemma}
Further the Nahm construction requires (\ref{esq0fch}) to have a
simple pole at $z=\pm1$. Because $2 \widetilde{\boldsymbol{K}} \in
\Lambda$ and $2 \boldsymbol{U} \in \Lambda$ this means we wish the
order of the vanishing of $\theta(\widetilde{\boldsymbol{K}})$ in
the direction $\boldsymbol{U}$ to be one more than the order of
the vanishing of $\theta(\boldsymbol{\phi} (\infty_l)-\boldsymbol{\phi}(\infty_{j
})+\widetilde{\boldsymbol{K}} )$ in the direction $\boldsymbol{U}$
(for each $j\ne l$). In principle the order of vanishing can be
higher than that given by the index of speciality and Riemann's
vanishing theorem, for these only provide the minimal order to
which all derivatives vanish and there may be some directions
yielding higher order vanishing. The desired vanishing of
(\ref{esq0fch}) may be deduced from the following property of
theta functions
$$\frac{\theta(\boldsymbol{\phi}(Q)-\boldsymbol{\phi}(P)+\boldsymbol{e})
\theta(\boldsymbol{\phi}(Q)-\boldsymbol{\phi}(P)-\boldsymbol{e})}{
\theta\sp2(\boldsymbol{e})
{E}(P,Q)\sp2}=\Omega_{\mathbf{B}}(P,Q)+\sum_{i,k=1}\sp{g}
\frac{\partial\sp2
\ln\theta(e)}{\partial z_i\partial z_k} v_i(P)v_k(Q)
$$
valid for all $P$, $Q\in\mathcal{C}$ and $e\in\mathbb{C}\sp{g}$
(see \cite[2.12]{fa73}). Here
$E(P,Q)=\mathcal{E}(P,Q)/\sqrt{dx(P)dx(Q)}$ is the prime form, and
$\Omega_{\mathbf{B}}(P,Q)$ a symmetric differential on
$\mathcal{C}\times\mathcal{C}$ with poles only on the diagonal.
Using that $\widetilde{\boldsymbol{K}}$ is a half period,
$2\widetilde{\boldsymbol{K}}=\boldsymbol{p}+\tau \boldsymbol{q}$
for $\boldsymbol{p}$,
$\boldsymbol{q}\in\mathbb{Z}\sp{g}$, we may use this identity to
obtain an expression of the form,
$$ F(s)F(-s)=G(s)e(s),$$
where
\begin{align*}F(s)&=\frac{\theta(\boldsymbol{\phi}(\infty_l)
-\boldsymbol{\phi}(\infty_j)+s\boldsymbol{U} -
\widetilde{\boldsymbol{K}})}{E(\infty_j,\infty_l)},\\
G(s)&=\theta\sp2(s\boldsymbol{U} -
\tilde{K})\Omega_{\mathbf{B}}(\infty_j,\infty_l)\\
&\qquad +\sum_{i,k=1}\sp{g}\left( \frac{\partial\sp2
\theta(s\boldsymbol{U} - \widetilde{\boldsymbol{K}})}{\partial
z_i\partial z_k}\theta(s\boldsymbol{U} -
\widetilde{\boldsymbol{K}})- \frac{\partial \theta(s\boldsymbol{U}
- \widetilde{\boldsymbol{K}})}{\partial z_i}\frac{\partial
\theta(s\boldsymbol{U} - \widetilde{\boldsymbol{K}})}{\partial
z_k}\right)
v_i(\infty_j)v_k(\infty_l),\\
 e(s)&=\exp(i\pi[q\cdot\tau\cdot
q+2q\cdot(\boldsymbol{\phi}(\infty_l)-\boldsymbol{\phi}(\infty_j)+s\boldsymbol{U}
-\widetilde{\boldsymbol{K}})]).
\end{align*}
Comparison of the Taylor series in $s$ shows that if the order of
vanishing of $\theta(s\boldsymbol{U} -
\widetilde{\boldsymbol{K}})$ is $m$ and that of
$\theta(\boldsymbol{\phi}(\infty_l)-\boldsymbol{\phi}(\infty_j)+s\boldsymbol{U} -
\widetilde{\boldsymbol{K}})$ is $k$ then $2k=2m-2$. Therefore the
order of vanishing of the denominator of (\ref{esq0fch}) is one
more than the numerator, and consequently that we have a simple
pole at $z=\pm1$. For $n\ge3$ the divisor
$\infty_j-\infty_l+(n-2)\sum_{k=1}\sp{n}\infty_k$ is special.

\subsection{The matrix $Q_0(z)$}
It remains to discuss the constants $c_{jl}$. Thus far in the
construction these depend on the divisor $\delta$. In this
subsection we now establish the following:
\begin{theorem} The matrix $Q_0(z)$ (which has poles of first order at
$z=\pm1$) may be written
\begin{equation}\label{ourq0}
Q_{0}(z)_{jl}  = \epsilon_{jl}\,\frac{(\rho_{j}-
\rho_{l})}{\mathcal{E}(\infty_j,\infty_l)}\,e\sp{i\pi\boldsymbol{\tilde
q}\cdot(\boldsymbol{\phi}(\infty_l)-\boldsymbol{\phi}(\infty_j))}\,
    \,\frac{\theta(\boldsymbol{\phi}(\infty_{l}) -\boldsymbol{\phi}(
    \infty_{j}) + [z+1]\boldsymbol{U} - \widetilde{\boldsymbol{K}})}{\theta(
    [z+1]\boldsymbol{U}
    - \widetilde{\boldsymbol{K}})}\,e^{z(\nu_{l} - \nu_{j})}.
\end{equation}
Here $E(P,Q)=\mathcal{E}(P,Q)/\sqrt{dx(P)dx(Q)}$ is the
Schottky-Klein prime form, $\boldsymbol{U} -
\widetilde{\boldsymbol{K}}=\frac12\boldsymbol{\tilde
p}+\frac12\tau\boldsymbol{\tilde q}$ ($\boldsymbol{p}$,
$\boldsymbol{q}\in\mathbb{Z}\sp{g}$) is a non-singular even theta
characteristic, and $\epsilon_{jl}=\epsilon_{lj}=\pm1$ is
determined (for $j<l$)
 by
$\epsilon_{jl}=\epsilon_{jj+1}\epsilon_{j+1j+2}\dots\epsilon_{l-1l}$.
The $n-1$ signs $\epsilon_{jj+1}=\pm1$ are arbitrary.

\end{theorem}

We remark that the prime form may be defined in terms of theta
functions with odd non-singular characteristic, the prime form
itself being independent of this choice. The construction
(\ref{ourq0}) depends on this choice via
${\mathcal{E}(\infty_j,\infty_l)}$, but any two choices lead to
$Q_0(z)$ differing by a constant diagonal gauge transformation,
the ambiguity noted earlier. We also note that the contours
implicit in $\boldsymbol{\phi}(\infty_l)$ are taken to be the same for each
term in (\ref{ourq0}), including that implicit in the limit
(\ref{etacondition}). A change to this contour leads to a constant
gauge transformation by a diagonal matrix with entries $\pm1$
which explains the signs $\epsilon_{jl}$ appearing in this
theorem.

\begin{proof} It will be convenient to introduce the following shorthand
for a recurring combination of functions appearing in this work.
For any divisor $\mathcal{A}$ set
$$<P-\mathcal{A}>\equiv\theta(\boldsymbol{\phi}(P)-\boldsymbol{\phi}(\mathcal{A})-{\boldsymbol{K}}).$$

The function $g_{j}(P)$ has been specified by
\begin{align*}
    g_{j}(P_{j})  = 1,  \qquad
    \div g_{j}(P) & = \Delta_{j} - P_{j} + \sum_{k=1}\sp{n} P_{k} -
    \delta.
\end{align*}
We may express $g_{j}(P)$ in several ways. First,
$$g_{j}(P)  = \frac{f_{j}(P)}{f_{j}(P_{j})},$$
where
$$f_{j}(P)  = \frac{<P - \Delta_{j}>}{<P - \tilde{\delta} -
    \delta_{g-1+j}>} \prod_{t \neq j} \frac{<P - \tilde{\delta} -
    P_{t}>}{<P- \tilde{\delta} - \delta_{g-1+t}>}$$
and $$
    \tilde{\delta} = \sum^{g-1}_{k=1} \delta_{k}, \qquad \delta
    =\sum_{k=1}\sp{g+n-1}
    \delta_{k}.$$
This was the form presented when we discussed the Baker-Akhiezer
function in general. In the case of the monopole we have
\begin{equation*}
    g = (n-1)^{2} \geq n - 1\ \text{for }\ n \geq 2,
\end{equation*}
and a second, more economical, representation exists.  Now we
could take
\begin{align*}
    f_{j}(P) &= \frac{ <P - \sum_{k \neq
    j}P_{k} - \sum^{g - (n-1)}_{s=1} \delta_{s}>\ <P - \Delta_{j}> }{<P -
    \sum^{n-1}_{t=1} \delta_{t+g-(n-1)} - \sum^{g-(n-1)}_{s=1}
    \delta_{s}> \quad<P - \sum^{n-1}_{t=1} \delta_{t+g} -
    \sum^{g-(n-1)}_{s=1} \delta_{s}>}\\
    & = \frac{<P - \sum_{k\neq j} P_{k} - \hat{\delta}><P
    - \Delta_{j}>}{<P - \delta^{(1)} - \hat{\delta}><P -
    \delta^{(2)} - \hat{\delta}>},
\end{align*}
where
$$
  \hat{\delta} = \sum^{g - (n-1)}_{k=1} \delta_{k},
    \qquad \delta^{(1)} = \sum^{n-1}_{j=1} \delta_{g - (n-1) + j},
    \qquad \delta^{(2)} = \sum^{n-1}_{j=1} \delta_{g+j}.$$
Then $\delta  = \hat{\delta} + \delta^{(1)} +
    \delta^{(2)}$.

We have
\begin{align*}
    \Psi_{j}(P) & = g_{j}(P)\, \frac{<P_{j} - \Delta_{j}>}{<P -
    \Delta_{j}>} \,\frac{\theta(\boldsymbol{\phi}(P) - \boldsymbol{\phi}(P_{j}) + (z+1)\boldsymbol{U} -
    \widetilde{\boldsymbol{K}})}{\theta((z+1)\boldsymbol{U} -
    \widetilde{\boldsymbol{K}})}e^{z[\int^{P}_{P_{0}}
    \gamma_{\infty}-\nu_{j}]} ,\\
    Q_{0}(z)_{jl} & = -(\rho_{j}- \rho_{l})\,
    \hat{c}_{jl}\,\frac{\theta(\boldsymbol{\phi}(P_{l} -
    P_{j}) + (z+1)\boldsymbol{U} - \widetilde{\boldsymbol{K}})}{\theta(
    (z+1)\boldsymbol{U}
    - \widetilde{\boldsymbol{K}})}\,e^{z(\nu_{l} - \nu_{j})}, \\
    \hat{c}_{jl} & = \lim_{P \rightarrow P_{l}} \zeta g_{j}(P)
    \frac{<P_{j} - \Delta_{j}>}{<P - \Delta_{j}>} \equiv c_{jl}
    \frac{<P_{j} - \Delta_{j}>}{<P_{l} - \Delta_{j}>}=
    c_{jl}
    \frac{\theta(\boldsymbol{U}  - \widetilde{\boldsymbol{K}})}{\theta(\boldsymbol{\phi}(P_{l} -
    P_{j}) +\boldsymbol{U} - \widetilde{\boldsymbol{K}})}
\end{align*}
where $c_{jl}  = \lim_{P \rightarrow P_{l}} \zeta
    g_{j}(P)$  is Ercolani-Sinha's constant.
We note that by breaking up the grouping of the theta functions in
our expression for $Q_0$ we need to be a little more careful
regarding its quasi-periodicity properties. If $2(\boldsymbol{U} -
\widetilde{\boldsymbol{K}})=\tilde{\boldsymbol{p}}+\tau\tilde{\boldsymbol{ q}}$
($\boldsymbol{p}$,
$\boldsymbol{q}\in\mathbb{Z}\sp{g}$) then

\begin{align*}{c}_{jl} = -{c}_{lj} &\Longleftrightarrow
\hat{c}_{jl} = -\hat{c}_{lj}\exp(2\pi i[\tilde{\boldsymbol{ q}}
\cdot(\boldsymbol{\phi}(P_{l} -
P_{j}) +\boldsymbol{U} - \widetilde{\boldsymbol{K}})
-\frac12 \tilde{\boldsymbol{ q}}\cdot\tau\tilde{\boldsymbol{q}}])\\
&\Longleftrightarrow\hat{c}_{jl}=-\hat{c}_{lj}\exp(2\pi i\tilde
{\boldsymbol{q}}\cdot\boldsymbol{\phi}(P_{l} -P_{j}) ).
\end{align*}
Here we have used that $\boldsymbol{U} -
\widetilde{\boldsymbol{K}}$ is an even theta characteristic.

Using the second representation we wish to evaluate
\begin{align*}
    \hat{c}_{jl} & = \lim_{P \rightarrow P_{l}} \zeta
    \frac{f_{j}(P)}{f_{j}(P_{j})} \, \frac{<P_{j} -
    \Delta_{j}>}{<P-\Delta_{j}>} \\
    & = \lim_{P \rightarrow P_{l}} \zeta \frac{<P - \sum_{k \neq j}
    P_{k} - \hat{\delta}>}{<P - \delta^{(1)} - \hat{\delta}><P -
    \delta^{(2)} - \hat{\delta}>} \cdot \frac{<P_{j} - \delta^{(1)} -
    \hat{\delta}><P_{j} - \delta^{(2)} - \hat{\delta}>}{<P_{j} -
    \sum_{k \neq j} P_{k} - \hat{\delta}>}.
\end{align*}
Now we use \cite[2.17]{fa73}
\begin{align*}
    <P - \sum^{g}_{i=1} x_{i}> & = c\,
    \frac{\det(v_{i}(x_{j}))}{\prod_{i<j}E(x_{i}, x_{j})} \cdot
    \frac{\sigma(P)}{\prod^{g}_{i=1}
    \sigma(x_{i})}\prod^{g}_{i=1}E(x_{i},P)
\end{align*}
with
    $$\sigma(P)  = \exp\left( -
\sum^{g}_{s=1} \int_{\mathfrak{a}_{s}} v_{s}(y) \ln E(y,P)\right).
$$
Taking $\{ x^{(0)}_{i}\}  = \{P_{k}\}_{k \neq j}\cup
\hat{\delta}$, $
    \{x^{(1)}_{i}\} = {\delta^{(1)}\cup \hat{\delta}} $ and $
    \{x^{(2)}_{i}\}  = {\delta^{(2)}\cup \hat{\delta}}$ this
    yields
\begin{align*}
    \hat{c}_{jl} & = \lim_{P \rightarrow P_{l}} \zeta
    \frac{\sigma(P)}{\sigma(P_{j})} \prod^{g}_{i=1}
    \frac{E(x_{i}^{(0)},P)}{E(x_{i}^{(0)},P_{j})} \, \left [ \frac{
    \sigma(P_{j})}{\sigma(P)}\right ] ^{2} \, \prod^{g}_{i=1}
    \left[\frac{E(x_{i}^{(1)}, P_{j})}{E(x^{(1)}_{i},P)} \,
    \frac{E(x_{i}^{(2)}, P_{j})}{E(x^{(2)}_{i},P)} \right]\\
    & = \lim_{P \rightarrow P_{l}} \zeta
    \frac{\sigma(P_{j})}{\sigma(P)} \cdot \left [\prod_{k\ne j}
    \frac{E(P_{k },P)}{E(P_{k},P_{j})} \,
    \prod_{\substack{ {\delta_i\in\delta^{(1)}}\\
    {{\delta_r\in\hat\delta},\,{\delta_s\in\delta^{(2)}}}}}
    \frac{
    E(\delta_i, P_{j})E(\delta_r, P_{j}) E(\delta_s, P_{j})}{E(\delta_i,P)
    E(\delta_r,P) E(\delta_s,P)} \right] \\
    & = \lim_{P \rightarrow P_{l}} \zeta
    \frac{\sigma(P_{j})}{\sigma(P)} \cdot
    \frac{E(P_{l},P)}{E(P_{l},P_{j})} \cdot \prod_{k \neq l,j}
    \frac{E(P_{k},P)}{E(P_{k},P_{j})} \cdot \prod^{g+n-1}_{i=1}
    \frac{E(\delta_{i},P_{j})}{E(\delta_{i},P_{l})} \\
    & = \left[\lim_{P \rightarrow P_{l}} \zeta
    \frac{E(P_{l},P)}{E(P_{l},P_{j})} \right]
    \frac{\sigma(P_{j})}{\sigma(P_l)} \prod_{k \neq l,j}
\frac{E(P_{k},P_{l})}{E(P_{k},P_{j})}
\cdot \prod^{g+n-1}_{i=1} \frac{E(\delta_{i},P_{j})}{E(\delta_{i},P_{l})}.
\end{align*}

The prime form $E(P,Q)$ that appears here is a differential of
weight $(-\frac12,-\frac12)$ on $\mathcal{C}\times\mathcal{C}$. If
$(\boldsymbol{a},\boldsymbol{b})$ is a non-singular odd theta
characteristic then we may write
\begin{equation}\label{primeform}
E(P,Q)=\frac{\theta_{\boldsymbol{a},\boldsymbol{b}}(\boldsymbol{\phi}(P)-\boldsymbol{\phi}(Q))}
{h_{\boldsymbol{a},\boldsymbol{b}}(P)h_{\boldsymbol{a},\boldsymbol{b}}(Q)},
\qquad h_{\boldsymbol{a},\boldsymbol{b}}\sp2(P)=\sum_{r=1}\sp{g}
\frac{\partial\theta_{\boldsymbol{a},\boldsymbol{b}}}{\partial
z_r}(0)\,v_r(P).
\end{equation}
The prime form is independent of the choice of $\alpha$. We remark
that in our expressions above the half-differentials
$h_{\boldsymbol{a},\boldsymbol{b}}$ cancel exactly between
numerator and denominator upon noting that the $P_k$ each are
pre-images of the same point $P\in \mathbb{P}\sp1$. It will  be
convenient to write
\begin{equation}\label{primeformr}E(P,Q)=\frac{\mathcal{E}(P,Q)}{\sqrt{dx(P)dx(Q)}}.
\end{equation}

Now $\lim_{P \rightarrow \infty_{l}} \zeta\,
\mathcal{E}(\infty_{l},P) = 1$ and we have
\begin{equation*}
    \hat{c}_{jl} \,\mathcal{E}(P_{l}, P_{j}) =
    \frac{\sigma(P_{j})}{\sigma(P_{l})} \prod_{k \neq l,j}
    \frac{E(P_{k},P_{l})}{E(P_{k},P_{j})} \prod^{g+n-1}_{i=1}
    \frac{E(\delta_{i}, P_{j})}{E(\delta_{i},P_{l})}.
\end{equation*}
Using the anti-symmetry of the prime form,$E(P_{l}, P_{j}) =
-E(P_{j}, P_{l})$, then
\begin{equation*}{c}_{jl} = -{c}_{lj} \Longleftrightarrow
    \left [{\sigma(P_{l})}\frac{\prod^{g+n-1}_{i=1} E(\delta_{i}, P_{l})}
    { \prod_{k \neq l,j} E(P_{k}, P_{l})}
     \,e\sp{i\pi\tilde q\cdot\boldsymbol{\phi}(P_l)}\right]^{2} = \left [
    {\sigma(P_{j})}\frac{\prod^{g+n-1}_{i=1} E(\delta_{i},P_{j})}
    { \prod_{k \neq l, j} E(P_{k}, P_{j})}
    \,e\sp{i\pi\tilde q\cdot\boldsymbol{\phi}(P_j)}\right ]^{2}.
\end{equation*}
But this is to be true for all $j\ne l$, thus
\begin{equation*}{c}_{jl} = -{c}_{lj} \Longleftrightarrow
     \left [
    {\sigma(P_{j})}\frac{\prod^{g+n-1}_{i=1} E(\delta_{i},P_{j})}
    { \prod_{k \neq j} E(P_{k}, P_{j})}
    \,e\sp{i\pi\tilde q\cdot\boldsymbol{\phi}(P_j)}
    \right ]^{2}= c^2
\end{equation*}
for some constant $c$, independent of $j$, whence we require
\begin{equation}\label{concjl}
    \frac{\sigma(P_{j})}{\sigma(P_{l})} \prod_{k \neq l,j}
    \frac{E(P_{k},P_{l})}{E(P_{k},P_{j})} \prod^{g+n-1}_{i=1}
    \frac{E(\delta_{i}, P_{j})}{E(\delta_{i},P_{l})}=\epsilon_{jl}
     \,e\sp{i\pi\tilde q\cdot(\boldsymbol{\phi}(P_l)-\boldsymbol{\phi}(P_j))}=\pm
     \,e\sp{i\pi\tilde q\cdot(\boldsymbol{\phi}(P_l)-\boldsymbol{\phi}(P_j))}.
\end{equation}
Because $\epsilon_{jl}$ is of the form $s_j/s_l$ (where $s_j=\pm
c$) then there are $n-1$ constraints here, which may be specified
by choosing $\epsilon_{j\,j+1}$. There are thus $2\sp{n-1}$
choices of signs for the $c_{jl}$. Thus
\begin{equation}
\hat{c}_{jl}\,\mathcal{E} (P_{l}, P_{j}) = \epsilon_{jl}
     \,e\sp{i\pi\tilde q\cdot(\boldsymbol{\phi}(P_l)-\boldsymbol{\phi}(P_j))}.
\end{equation}

Lets consider the various constraints involved. Having determined
$\boldsymbol{U}$ then, as $\boldsymbol{\phi}\left(P_{j}\right)$ are specified
once a choice of the Abel map has been made, we have that
$$\boldsymbol{\phi}\left( \Delta_{j}\right)=\boldsymbol{\phi}\left(P_{j}+(n-2)\sum_{k=1}\sp{n} P_{k}\right)
+\boldsymbol{U}.$$ This then determines the degree $g$ effective
divisor $\Delta_{j}$. Thus the degree $g+n-1$ divisor $\delta$ is
constrained by
$$\delta  \sim \Delta_{j} - P_{j} +\sum_{k=1}\sp{n} P_{k}, \quad
\boldsymbol{\phi}\left(\delta\right) = \boldsymbol{\phi}\left(\Delta_{j} - P_{j}
+\sum_{k=1}\sp{n} P_{k}\right)=\boldsymbol{U}+(n-1)
\boldsymbol{\phi}\left(\sum_{k=1}\sp{n} P_{k}\right).$$

This yields a further $g$ constraints on the divisor $\delta$ in
addition to the $n-1$ constraints of (\ref{concjl}). Thus we have
$g+n-1$ constraints on the degree $g+n-1$ nonspecial divisor
$\delta$. We remark that the $n-1$ constraints of (\ref{concjl})
which are of the form $s_j/s_l$ correspond to the constant
diagonal gauge freedom that exists for the matrix $Q_0(z)$. Our
solving the constraints (\ref{concjl}) is equivalent to choosing a
gauge.

For the moment let us suppose we may find a divisor $\delta$
satisfying the required constraints. If this is the case, then
bringing the above results together establishes the theorem, and
that we have
$$
Q_{0}(z)_{jl}  = \epsilon_{jl}\,\frac{(\rho_{j}-
\rho_{l})}{\mathcal{E}(P_j,P_l)}\,e\sp{i\pi\tilde
q\cdot(\boldsymbol{\phi}(P_l)-\boldsymbol{\phi}(P_j))}
    \,\frac{\theta(\boldsymbol{\phi}(P_{l} -
    P_{j}) + [z+1]\boldsymbol{U} - \widetilde{\boldsymbol{K}})}{\theta(
    [z+1]\boldsymbol{U}
    - \widetilde{\boldsymbol{K}})}\,e^{z(\nu_{l} - \nu_{j})}
,$$
where $\epsilon_{jl}=\epsilon_{lj}=\pm1$ is determined (for $j<l$)
 by
$\epsilon_{jl}=\epsilon_{jj+1}\epsilon_{j+1j+2}\dots\epsilon_{l-1l}$
and the $n-1$ signs $\epsilon_{jj+1}=\pm1$ are arbitrary.

\noindent\emph{Alternate Calculation} Our proof used the second
parameterization of the functions $f_j(P)$. The same constraints
arise if we use our first parameterization,
\begin{align*}
    g_{j}(P)  = \frac{f_{j}(P)}{f_{j}(P_{j})}, \qquad
    f_{j}(P) & = \frac{<P - \Delta_{j}>}{<P_{j} - \tilde{\delta}
    - \delta_{g - 1 + j}>} \prod_{t \neq j} \frac{<P - P_{t} -
    \tilde{\delta}>}{<P - \delta_{g - 1 + t} - \tilde{\delta}>},
\end{align*}
where $\tilde{\delta} = \sum^{g-1}_{k=1} \delta_{k}$. Let
$\{\tilde{x}^{(l)}\}  = \{P_{l}, \tilde{\delta} \}$,
$\tilde{y}^{(l)} = \{\delta_{g-1+l}, \tilde{\delta} \}$. Then
\begin{align*}
     \frac{<P - P_{t} - \tilde{\delta}>}{<P_{j} - P_{t} -
    \tilde{\delta}>}& \frac{<P_{j} - \delta_{g-1+t} -
    \tilde{\delta}>}{<P - \delta_{g-1+t} - \tilde{\delta}>} \\
    & = \frac{\sigma(P)}{\sigma(P_{j})} \cdot \left(\prod_{k\ne t}
    \frac{E(\tilde{x}^{(k)},P)}{E(\tilde{x}^{(k)},P_{j})} \cdot
    \frac{E(\tilde{y}^{(k)},P_{j})}{E(\tilde{y}^{(k)},
    P)}\right)\cdot \frac{\sigma(P_{j})}{\sigma(P)} \\
    & = \frac{E(P_{t}, P)}{E(P_{t}, P_{j})} \cdot
    \frac{E(\delta_{g-1+t}, P_{j})}{E(\delta_{g-1+t}, P)}
.\end{align*}
We again wish to evaluate
\begin{align*}
    \hat{c}_{jl} & = \lim_{P \rightarrow P_{l}} \zeta
    \frac{f_{j}(P)}{f_{j}(P_{j})} \cdot \frac{<P_{j}-
    \Delta_{j}>}{<P - \Delta_{j}>} \\
    & = \lim_{P \rightarrow P_{l}} \zeta \frac{<P_{j} -
    \tilde{\delta} - \delta_{g - 1 + j}>}{<P - \tilde{\delta} -
    \delta_{g - 1 + j}>} \prod_{t \neq j} \frac{E(P_{t},
    P)}{E(P_{t},P_{j})} \cdot \frac{E(\delta_{g-1+t},
    P_{j})}{E(\delta_{g-1+t}, P)} \\
    & = \lim_{P \rightarrow P_{l}} \zeta
    \frac{\sigma(P_{j})}{\sigma(P)} \cdot \left( \prod^{g-1}_{k=1}
    \frac{E(\delta_{k}, P_{j})}{E(\delta_{k}, P)}\right)
    \frac{E(\delta_{g-1+j}, P_{j})}{E(\delta_{g-1+j},P)} \prod_{t
    \neq j} \frac{E(P_{t}, P) E(\delta_{g-1+t}, P_{j})}{E(P_{t},
    P_{j})E(\delta_{g-1+t}, P)},
\end{align*}
Therefore
$$  \hat{c}_{jl}\, E(P_{l}, P_{j}) = \left[ \lim_{P
    \rightarrow P_{l}} \zeta E(P_{l}, P)\right]
     \frac{\sigma(P_{j})}{\sigma(P_{l})} \cdot \prod_{k \neq l,j}
    \frac{E(P_{k}, P_{l})}{E(P_{k}, P_{j})} \cdot
    \prod^{g+n-1}_{i=1} \frac{E(\delta_{i}, P_{j})}{E(\delta_{i},
    P_{l})},
 $$
which is the same expression as our previous method of
calculating.
\end{proof}

We have established our new expression for $Q_0(z)$ once the
following is established:

\begin{proposition} Given $n$ (generic) constants $\alpha_j\ne0$  there
exists a degree
$g+n-1$ nonspecial divisor $\delta=\sum_{s=1}\sp{g+n-1}\delta_s$
such that \begin{equation}\label{ajdelta}  \boldsymbol{\phi}\left(\delta -
(n-1)\sum_{k=1}\sp{n} \infty_{k}\right)=\boldsymbol{U}
\end{equation}
and satisfying the $n-1$ (equivalent) constraints
\begin{align*}
\text{C}&\qquad
\alpha_1\prod_{s=1}\sp{g+n-1}
\theta_{\boldsymbol{a},\boldsymbol{b}}
\left(\int\sp{\delta_s}_{\infty_1}\boldsymbol{v}\right)=
\alpha_2\prod_{s=1}\sp{g+n-1}
\theta_{\boldsymbol{a},\boldsymbol{b}}
\left(\int\sp{\delta_s}_{\infty_2}\boldsymbol{v}\right)=\dots=
\alpha_n\prod_{s=1}\sp{g+n-1}
\theta_{\boldsymbol{a},\boldsymbol{b}}
\left(\int\sp{\delta_s}_{\infty_n}\boldsymbol{v}\right),\\
\text{C}\sp\prime&\qquad \prod_{i=1}\sp{g+n-1}
\frac{E(\delta_i,\infty_s)}{E(\delta_i,\infty_n)}=\tilde\alpha_s,\qquad
s=1,\ldots,n-1,\\
\text{C}\sp{\prime\prime}&\qquad
\sum_{i=1}\sp{g+n-1}\int_{P_0}\sp{\delta_i}\omega_{\infty_s,\infty_n}(P)=\hat
\alpha_s,\qquad s=1,\ldots,n-1.
\end{align*}
\end{proposition}
Here $\tilde\alpha_s=\alpha_n/\alpha_s$ and $\hat
\alpha_s=\ln\tilde
\alpha_s-(g+n-1)\ln\left[{E(P_0,\infty_s)}/{E(P_0,\infty_n)}\right]$,
$\boldsymbol{\phi}$ is the Abel map, $\boldsymbol{v}$ the $\mathfrak
a$-normalized holomorphic differentials,
$({\boldsymbol{a},\boldsymbol{b}})$ is a non-singular odd (half)
theta characteristic and $\boldsymbol{U}$ a further specified
theta (half) characteristic. The equivalence of the constraints C
and C$\sp{\prime}$ follows upon writing the prime form in terms of
theta functions (\ref{primeform}) and observing that  the
half-differentials $h_{\boldsymbol{a},\boldsymbol{b}}$ in the
definition of the primeform cancel exactly between numerator and
denominator upon noting that the $\infty_k$ each are pre-images of
the same point $\infty\in \mathbb{P}\sp1$. The final equivalence
with C$\sp{\prime\prime}$ is obtained upon using the expression
\begin{equation}\label{d3kpf}
\omega_{\mathcal{P},\mathcal{Q}}(P)=d\;\mathrm{ln}\;\prod_{j=1}^m
\frac{E(P,P_j)}{E(P,Q_j)},
\end{equation}
where $\mathcal{P}=P_1+\ldots,P_m$ and
$\mathcal{Q}=Q_1+\ldots+Q_m$ are divisors of the same degree $m$. We
have for example that
$$\omega_{\infty_s,\infty_n}(P)=d\ln\frac{E(P,\infty_s)}{E(P,\infty_n)}$$
and so
$$\int_{P_0}\sp{\delta_i}\omega_{\infty_s,\infty_n}(P)=
\ln\frac{E(\delta_i,\infty_s)}{E(\delta_i,\infty_n)}
-\ln\frac{E(P_0,\infty_s)}{E(P_0,\infty_n)}.
$$

The proposition then is an extension of the usual Abel-Jacobi
inversion: (\ref{ajdelta}) is the usual Abel-Jacobi map and
$\text{C}\sp{\prime\prime}$ adds to the usual holomorphic
differentials abelian differentials of the third kind. Such
generalized Abel-Jacobi maps frequently arise when considering
integrable systems. Clebsch and Gordan \cite{clebgor} considered
the situation when the holomorphic differentials are supplemented
by $n$ abelian differentials $\omega_{X_i Y_i}$ of the third kind
for  distinct pairs $(X_i,Y_i)$. (This is enough to solve the
$n=2$ genus 1 case.) More recently Fedorov \cite{fed99} has
developed this theory. Though our form $\text{C}\sp{\prime\prime}$
in which there is a point common to each of the abelian
differentials appears new, the approach of \cite{fed99} can deal
with this case and a proof of the proposition that will appear
elsewhere with Fedorov.

\subsection{Ingredients for the construction}
It is perhaps worthwhile recording the elements needed to make
effective the construction $Q_0$, the second step in the
Ercolani-Sinha construction of the Nahm data, the roots $\rho_j$
of (\ref{rhodef}) having been calculated in the first step.

The whole construction is predicated on the theta functions built
from the spectral curve. Thus we need
\begin{enumerate}
    \item To construct the period matrix $\tau$ associated to
    $\mathcal{C}$.
    \item To determine the half-period $\widetilde{\boldsymbol{K}}$.
    \item To determine the Ercolani-Sinha vector $\boldsymbol{U}$.
    \item For normalised holomorphic differentials $\boldsymbol{v}$ to
    calculate $\int\limits_{\infty_i}^{\infty_j}
   \boldsymbol{v}=\boldsymbol{\phi}(\infty_j)-\boldsymbol{\phi}(\infty_i)$.
   \item To determine $E(\infty_j,\infty_l)$.
   \item To determine $\gamma_{\infty}(P)$ and
   $\nu_i=\lim_{P\to\infty_i}\left(\int_{P_0}\sp{P}\gamma_{\infty}(P')+\frac{\eta}
{\zeta}(P)\right)$.
\end{enumerate}

\subsection{The fundamental bi-differential} We shall now describe how to
calculate the meromorphic differential $\gamma_\infty$ and the
constants appearing in the Ercolani-Sinha construction. This will
be in terms of the fundamental bi-differential.

Let $\boldsymbol{v}$ be the vector of $\mathfrak{a}$-normalised
holomorphic differentials with expansion (\ref{TVW}). Introduce
directional derivatives along the vectors fields
$\boldsymbol{V}^{(i)}$, $\boldsymbol{W}^{(i)}$ etc,
\begin{align*}
\partial_{\boldsymbol{V}}f(\boldsymbol{v})&=\sum_{k=1}^gV_k
\frac{\partial}{\partial v_k} f(\boldsymbol{v}),
&\partial_{\boldsymbol{V},\boldsymbol{W}}f(\boldsymbol{v})
=\sum_{k=1}^g\sum_{l=1}^gV_kW_l \frac{\partial^2}{\partial
v_k\partial v_l} f(\boldsymbol{v}).
\end{align*}
Recall (see \cite{fa73}) that the fundamental bi-differential
$\Omega_{\mathbf{B}}$ is the symmetric 2-differential of the
second kind on $\mathcal{C}\times\mathcal{C}$ is defined by,
\begin{equation}\label{asymberg}
\Omega_{\mathbf{B}}(P,Q) =\frac{\partial^2}{\partial x\partial y }
\mathrm{ln}\,\theta_{\boldsymbol{a},\boldsymbol{b}}
\left(\int_{Q}^P\boldsymbol{v}\right)\,dx
dy=\left(\frac1{(x(P)-y(Q))\sp2}+\text{nonsingular}\right)\,dx dy,
\end{equation}
where $P$, $Q$ are two different points of the curve $\mathcal{C}$
with local coordinates $x(P)$, $y(Q)$, and
$[{\boldsymbol{a},\boldsymbol{b}}]$ is an odd, nonsingular,
half-integer theta characteristic. The kernel has a second order
pole along the diagonal. For fixed $Q\in\mathcal{C}$ we have that
\begin{equation}\label{perberg}\oint_{\mathfrak{a}_j}\Omega_{\mathbf{B}}(P,Q)
=0,\qquad \oint_{\mathfrak{b}_j}\Omega_{\mathbf{B}}(P,Q) =2\pi
i\,v_j(Q).\end{equation}

Consider for each infinity $\infty_i$ ($i=1,\ldots,n$) the
differential of the second kind,
\[\Omega_{\mathbf{B}}^{(i)}(P)=\left.
\frac{\Omega_{\mathbf{B}}(P,Q)}{{d} t(Q)} \right|_{Q=\infty_i}
=-\sum_{k,l=1}\sp{g} \frac{\partial\sp2}{
\partial z\sp{k}\partial z\sp{l}}\mathrm{ln}\, \theta_{\boldsymbol{a},\boldsymbol{b}}
\left(\int_{\infty_i}^P\boldsymbol{v}\right)\,v_k(P)\,\left.\frac{v_l(Q)}{dt(Q)}
\right|_{Q=\infty_i} ,\] where $t(Q)$ is a local coordinate in the
vicinity of $Q$. Then noting the expansion (\ref{TVW}) we have
${v_l(Q)}/{dt(Q)}\rightarrow V_l\sp{(i)}$ as
$Q\rightarrow\infty_i$. Using (\ref{perberg}) we see that
\begin{equation}
\oint\limits_{\mathfrak{a}_l} \Omega_{\mathbf{B}}^{(i)}(P)
=0,\quad \oint\limits_{\mathfrak{b}_l}
\Omega_{\mathbf{B}}^{(i)}(P) =2\imath\pi V_l^{(i)}\quad
l=1,\ldots,g.\label{periods}
  \end{equation}
The quantities $\int_{P_0}\sp{P}\Omega_{\mathbf{B}}^{(i)}(P)$ are
then Abelian integrals of the second kind with unique pole of the
first order at $\infty_i$. Further, from (\ref{asymberg}), we know
that if $t(P)$ is a local coordinate in the vicinity of $\infty_i$
that $\Omega_{\mathbf{B}}^{(i)}(P)=(1/t^2+\mathrm{regular})dt$.
This together with (\ref{periods}) shows that
\begin{equation}\label{gammares}
\gamma_{\infty}(P)=\sum_{i=1}^n
\rho_i\,\Omega_{\mathbf{B}}^{(i)}(P)=-\frac{\partial\ \ \ }
{\partial x(P)} \sum_{i=1}^n
\rho_i\,\partial_{\boldsymbol{V}^{(i)}}\,\mathrm{ln}\,
\theta_{\boldsymbol{a},\boldsymbol{b}}
\left(\int\limits_{\infty_i}^P  \boldsymbol{v} \right){d}x(P)
\end{equation}
and that
$\boldsymbol{U}=\sum_{i=1}\sp{g}\rho_i\boldsymbol{V}\sp{(i)}$. Now
\begin{align*}
\int\limits_{P_0}^P\Omega_{\mathbf{B}}^{(i)}(P')&=
\int\limits_{P_0}^P-\frac{\partial\ \ \ } {\partial x(P')}
\partial_{\boldsymbol{V}^{(i)}}\,\mathrm{ln}\,
\theta_{\boldsymbol{a},\boldsymbol{b}}
\left(\int\limits_{\infty_i}^{P'}  \boldsymbol{v} \right){d}x(P')
=-\partial_{\boldsymbol{V}^{(i)}}\,\mathrm{ln}\left[\frac{
\theta_{\boldsymbol{a},\boldsymbol{b}}
\left(\int\limits_{\infty_i}^P \boldsymbol{v}\right)}{
\theta_{\boldsymbol{a},\boldsymbol{b}}
\left(\int\limits_{\infty_i}^{P_0} \boldsymbol{v}\right)}\right].
\end{align*}
Combining these with (\ref{gammares}) yields
\begin{align*}
\nu_i&=\lim_{P\to\infty_i}\left(\int_{P_0}\sp{P}\gamma_{\infty}(P')+\frac{\eta}
{\zeta}(P)\right)\\ &=
\lim_{P\to\infty_i}\rho_i\left[\frac1{t(P)}-
\partial_{\boldsymbol{V}^{(i)}}\mathrm{ln}\;
\theta_{\boldsymbol{a},\boldsymbol{b}}
\left(\int\limits_{\infty_i}^{P}
   \boldsymbol{v}\right)\right]\\
&\qquad -\sum_{j\ne i}
\rho_j\,\partial_{\boldsymbol{V}^{(j)}}\mathrm{ln}\;
\theta_{\boldsymbol{a},\boldsymbol{b}}
\left(\int\limits_{\infty_j}^{\infty_i}
   \boldsymbol{v}\right)+\mathrm{Res}_{\infty_i}\frac{\eta}
{\zeta\sp2}+c(P_0),
\end{align*}
where $c(P_0)=\sum_{i=1}\sp{g}
\rho_i\,\partial_{\boldsymbol{V}^{(i)}}\mathrm{ln}\;
\theta_{\boldsymbol{a},\boldsymbol{b}}
\left(\int\limits_{\infty_i}^{P_0}
   \boldsymbol{v}\right)$ is a constant independent of the index
   $i$.
The first limit may be calculated directly. Assuming for the sake
of exposition a first order vanishing we find in terms of the
local coordinate $t$ that
$$\lim_{P\to\infty_i}\partial_{\boldsymbol{V}^{(i)}}\mathrm{ln}\;
\theta_{\boldsymbol{a},\boldsymbol{b}}
\left(\int\limits_{\infty_i}^{P}
   \boldsymbol{v}\right)
   =
   \frac{1}{t}+
\frac{1}{2}\frac{\partial^{2}_{\boldsymbol{V}^{(i)},
\boldsymbol{V}^{(i)}}
\theta_{\boldsymbol{a},\boldsymbol{b}}(\boldsymbol{0}) }
{\partial_{\boldsymbol{V}^{(i)}}
   \theta_{\boldsymbol{a},\boldsymbol{b}}(\boldsymbol{0})}
-\frac{1}{2}\frac{\partial_{\boldsymbol{W}^{(i)}}
\theta_{\boldsymbol{a},\boldsymbol{b}}(\boldsymbol{0}) }
{\partial_{\boldsymbol{V}^{(i)}}
   \theta_{\boldsymbol{a},\boldsymbol{b}}(\boldsymbol{0})}+\ldots
   $$
Now because $[\boldsymbol{a},\boldsymbol{b}]$ is an odd theta
characteristic then $\partial^{2}_{\boldsymbol{V}^{(i)},
\boldsymbol{V}^{(i)}}
\theta_{\boldsymbol{a},\boldsymbol{b}}(\boldsymbol{0}) =0$.
Therefore
\begin{equation}\label{nuiberg}
\nu_i= \frac{\rho_i}{2}\frac{\partial_{\boldsymbol{W}^{(i)}}
\theta_{\boldsymbol{a},\boldsymbol{b}}(\boldsymbol{0}) }
{\partial_{\boldsymbol{V}^{(i)}}
   \theta_{\boldsymbol{a},\boldsymbol{b}}(\boldsymbol{0})}
    -\sum_{j\ne i}
\rho_j\,\partial_{\boldsymbol{V}^{(j)}}\mathrm{ln}\;
\theta_{\boldsymbol{a},\boldsymbol{b}}
\left(\int\limits_{\infty_j}^{\infty_i}
   \boldsymbol{v}\right)+\mathrm{Res}_{\infty_i}\frac{\eta}
{\zeta\sp2}+c(P_0).
\end{equation}

Because we are only interested in calculating the differences
$\nu_i-\nu_j$ there exists a further representation making use of
normalised differentials of the third kind. Let
$\omega_{\infty_i,\infty_j}$ be the meromorphic  differential of
the third kind, with simple pole of residue $+1$ at $\infty_i$ and
$-1$ at $\infty_j$ with vanishing ${\mathfrak{a}}$-periods. Now
consider the integral
\[\mathfrak{I}= \int_{\partial \Gamma }   \int_{P_0}^{P}\left[
 \gamma_{\infty}(P')
+d\left(\frac{\eta}{\zeta}\right)(P') \right]
 \omega_{\infty_i,\infty_j}(P)
 \]
taken over the boundary  $\partial \Gamma$ of the fundamental
domain and compute it in two ways: as sum of residues and as a
contour integral. Because of the normalisation of the
differentials $\gamma_{\infty}$ and $\omega_{\infty_i,\infty_j}$
the contour integral vanishes. Therefore the sum of residues
vanishes too, which upon using
\[ \mathrm{Res}_{P=0_k  }\frac{\eta}{\zeta}(P)=\eta(0_k),  \]
leads to the equality
\begin{equation}
\nu_i-\nu_j=-\sum_{k=1}^n\eta(0_k)\omega_{\infty_i,\infty_j}(0_k).
\label{nuijtk}
\end{equation}
Using (\ref{d3kpf}) this formula can be written in terms of the
prime form and we obtain the formula
\begin{equation}
\nu_i-\nu_j=-\sum_{k=1}^n\eta(0_k) \,
\left.\frac{\partial}{\partial z} \mathrm{ln}\,\frac
{\mathcal{E}(P,\infty_i)}{\mathcal{E}(P,\infty_j)}\right|_{P=0_k}
=-\sum_{k=1}^n\eta(0_k) \, \left.\frac{\partial}{\partial z}
\mathrm{ln}\,\frac
{\theta_{\boldsymbol{a},\boldsymbol{b}}\left(\int_{\infty_i}^P\boldsymbol{v}
\right)}
{\theta_{\boldsymbol{a},\boldsymbol{b}}\left(\int^P_{\infty_j}\boldsymbol{v}\right)}\right|_{P=0_k}
\label{nuij1}
\end{equation}
We remark that more explicit results we can be found upon
utilising the Klein-Weierstrass realisation of third kind
differentials \cite{bak95}.

With the general construction now at hand we shall turn to some
explicit examples. The case of $n=2$ has been treated by several
authors and by way of illustration we too treat this example using
the formulae just described. Our formulae, though different in
detail, lead to the known results. Going beyond these results we
consider the case of $n=3$ in the second part of this work.

\section{An illustration: the charge 2 monopole}

We shall consider the well-studied case of $n=2$ to enable
comparison with other authors. Our first step will be to assemble
the ingredients for the construction, noted above, and so to
determine the matrix $Q_0(z)$. For completeness we will also
perform the remaining steps needed to reconstruct the Nahm data.

We will work with the (centred) spectral curve in the form chosen
by Ercolani-Sinha,
\begin{align}0&=
\eta^2+\frac{\kappa^2}{4}(\zeta^4+2(k^2-k'^2)\zeta\sp2+1)
\label{esellform}\\
&=\eta^2+\frac{\kappa^2}{4}(\zeta-k'-ik)
(\zeta-k'+ik)(\zeta+k'-ik)(\zeta+k'+ik)
\end{align}
where $k'\sp2=1-k^2$. With $k'=\cos\alpha$, $k=\sin\alpha$, then
the roots may be written as $\pm e\sp{\pm i\alpha}$ and these lie
on the unit circle. We may take $0\le\alpha\le\pi/4$. We choose
cuts between $-k'+ik=-e\sp{- i\alpha}$ and $k'+ik=e\sp{ i\alpha}$
as well as $-k'-ik$ and $k'-ik$. Let $\mathfrak{b}$ encircle
$-k'+ik$ and $k'+ik$ with $\mathfrak{a}$ encircling $k'+ik$ and
$-k'+ik$ on the two sheets as on the diagram. We take as our
assignment of sheets ($j=1$, $2$, with analytic continuation from
$\zeta=0$ avoiding the cuts) to be
$$\eta_j
=(-1)\sp{j}\,i\frac{\kappa}{2}\sqrt{\zeta^4+2(k^2-k'^2)\zeta\sp2+1}.$$
Then, upon using the substitutions $\zeta=e\sp{i\theta}$ and
$k\sin u=\sin\theta$ on sheet $1$,
$$\frac{d\zeta}{\eta}=i\frac{2}{\kappa}\frac{d\zeta}{\sqrt{(\zeta\sp2-
e\sp{2 i\alpha})(\zeta\sp2- e\sp{-2 i\alpha})}} =
\frac{-1}{k\kappa}\frac{d\theta}{\sqrt{1-\frac{1}{k^2}\sin\sp2\theta}}=
\frac{-1}{\kappa}\frac{du}{\sqrt{1-{k^2}\sin\sp2 u}}.$$ Thus
$$\oint_\mathfrak{a}\frac{d\zeta}{\eta}=
\frac{-2}{\kappa}\int_\alpha\sp{-\alpha}
\frac{d\theta}{\sqrt{{k^2}-\sin\sp2\theta}}=
\frac{4}{\kappa}\int_0\sp{\pi/2}\frac{du}{\sqrt{1-{k^2}\sin\sp2
u}}=\frac{4}{\kappa}\,\mathbf{K}(k),$$ where
$\mathbf{K}=\mathbf{K}(k)$ is the complete elliptic integral of
the first kind. Similarly (with $\zeta=\exp i(w+\pi/2)$)
$$\oint_\mathfrak{b}\frac{d\zeta}{\eta}=
\frac{2i}{\kappa}\int_{\alpha-\pi/2}\sp{\pi/2-\alpha}
\frac{dw}{\sqrt{{k'^2}-\sin\sp2 w}}=
\frac{4i}{\kappa}\int_0\sp{\pi/2}\frac{du}{\sqrt{1-{k'^2}\sin\sp2
u}}=\frac{4i}{\kappa}\,\mathbf{K}'(k).$$ For the curve
(\ref{esellform}) and this choice of homology basis the normalized
holomorphic differential is then $\boldsymbol{v}={ \kappa
d\zeta}/{(4\mathbf{K}\eta)}$. Comparison with (\ref{HMREScond})
shows that the Ercolani-Sinha constraint is satisfied for $\kappa=
\mathbf{K}(k)$ and $\mathfrak{c}=-\mathfrak{a}$. Thus
$\boldsymbol{U}=-1/2$. The period matrix for the curve is then
$\tau=i\mathbf{K}'/\mathbf{K}$. Symmetry now enables us to
evaluate various Abel-maps (with base point $P_0=k'+ik$):
$$\begin{matrix} \boldsymbol{\phi}(\infty_1)=\dfrac{1+\tau}{4}=-\boldsymbol{\phi}(\infty_2),&&
\boldsymbol{\phi}(0_1)=\dfrac{1-\tau}{4}=-\boldsymbol{\phi}(0_2).
\end{matrix}
$$
From (\ref{rhodef}) and our assignment of sheets we have that
$$\rho_1=-\frac{i}{2}\,\mathbf{K},\qquad
\rho_2=\frac{i}{2}\,\mathbf{K}.$$

Many features of this example can be determined without
calculation. For example, $\widetilde{\boldsymbol{K}}$ is a theta
(half-) characteristic such that $\theta$ vanishes to order
$s=\frac14 2^2=1$ at $\widetilde{\boldsymbol{K}}$. This identifies
$\tilde K$ as the unique odd theta characteristic
$\widetilde{\boldsymbol{K}}=(1+\tau)/2$. Further
$\tau\sp*(d\zeta/\eta)=-\overline{(d\zeta/\eta)}$. The property
$\tau_*(\mathfrak{c})=-\mathfrak{c}$ then fixes
$\mathfrak{c}=\pm\mathfrak{a}$ and consequently
$\boldsymbol{U}=\pm1/2$, the relevant sign being selected by
(\ref{HMREScond}). The non-singular even theta characteristic
$\boldsymbol{U}+\widetilde{\boldsymbol{K}}=\tau/2$. Then
$$\boldsymbol{U} - \widetilde{\boldsymbol{K}}=-1-\tau/2=\frac12\tilde
p+\frac12\tau\tilde q\Rightarrow \tilde p=-2,\ \tilde q=-1.$$
Substitution of the quantities collected thus far into
(\ref{ourq0}) yields
\begin{align}\label{q0charge2}
Q_{0}(z)_{12}
=\epsilon_{12}\,\frac{-i\mathbf{K}}{\mathcal{E}(\infty_1,\infty_2)}\,e\sp{i\pi
(1+\tau)/2}\,
    \,\frac{\theta(-[z+1]/2 - 1-\tau)}{\theta(-
    [z+1]/2
    - (1+\tau)/2)}\,e^{z(\nu_{2} - \nu_{1})}.
\end{align}
Upon using the identification of $\theta(z)$ with the Jacobi theta
function $\theta_3(z)$ and the periodicities of the Jacobi theta
functions $\theta_*(z)$ then
\begin{align*}
\theta(-[z+1]/2 - 1-\tau)&=-e\sp{-i\pi(z+\tau)}\,\theta_4(z/2),\\
\theta(- [z+1]/2 -
(1+\tau)/2)&=e\sp{-i\pi(z/2+\tau/4)}\,\theta_2(z/2),
\end{align*}
giving
$$Q_{0}(z)_{12}=-
\epsilon_{12}\,\frac{\mathbf{K}}{\mathcal{E}(\infty_1,\infty_2)}\,e\sp{-i\pi
\tau/4} \,\frac{\theta_4(z/2)}{\theta_2(z/2)} \,e^{z(\nu_{2} -
\nu_{1}-i\pi/2)}.
$$

\subsubsection*{The prime form}
Let us now evaluate the prime form. We have from (\ref{primeform})
that
$$E(P,Q)=\frac{\theta\left[\begin{matrix}1/2\\1/2\end{matrix}
\right](\boldsymbol{\phi}(P)-\boldsymbol{\phi}(Q))}{h(P)h(Q)}, \qquad
h\sp2(P)=\frac{\partial\theta\left[
\begin{matrix}1/2\\1/2\end{matrix}\right](0)}{\partial
z}\,v(P).
$$
If $\zeta=1/t$ is a local parameter at $\infty_j$ then
$v(\infty_j)=d\zeta/4\eta|_{\infty_j}=-dt/(4 \rho_j)$. Identifying
$\theta\left[
\begin{matrix}1/2\\1/2\end{matrix}\right](z)$ with the Jacobi theta
function $-\theta_1(z)$ then gives
\begin{align*}
\mathcal{E}(\infty_1,\infty_2)&= E(\infty_1,\infty_2)dt=-
4\sqrt{\rho_1\rho_2} \,
\frac{\theta_1\left(\int_{\infty_2}\sp{\infty_1}\boldsymbol{v}\right)}
{\theta_1'}=
-2\mathbf{K}\frac{\theta_1\left(\int_{\infty_2}\sp{\infty_1}\boldsymbol{v}\right)}
{\theta_1'}\\
&=-2\mathbf{K}\,e\sp{-i\pi\tau/4}\,\frac{\theta_3}{\theta_1'}.
\end{align*}

Upon noting that $\theta_1'= \pi \theta_2 \theta_3 \theta_4$, we
then have that \begin{equation} \label{gen1q0} Q_{0}(z)_{12}=
\epsilon_{12}\,\frac{\pi\theta_2 \theta_4 }{2}
\,\frac{\theta_4(z/2)}{\theta_2(z/2)} \,e^{z(\nu_{2} -
\nu_{1}-i\pi/2)}.
\end{equation}

\subsubsection*{Alternate calculation}
Let us check this calculation proceeding from the form
(\ref{esq0fch}). Then
\begin{align*}
Q_{0}(z)_{12}&=- (\rho_{1}- \rho_{2})\,
    {c}_{12}\,\frac{\theta(\boldsymbol{\phi}(\infty_2)-\boldsymbol{\phi}(\infty_1)+
(z+1)\boldsymbol{U} -
\widetilde{\boldsymbol{K}})\theta(\boldsymbol{U} -
\widetilde{\boldsymbol{K}})}{\theta(\boldsymbol{\phi}(\infty_2)-\boldsymbol{\phi}(\infty_1)+\boldsymbol{U}
- \widetilde{\boldsymbol{K}})\theta(
    (z+1)\boldsymbol{U}
    - \widetilde{\boldsymbol{K}})}\,e^{z(\nu_{2} - \nu_{1})}\\
&=c_{12}\,
i\mathbf{K}\,\frac{\theta_2\,\theta_4(z/2)}{\theta_4\,\theta_2(z/2)}
\,e^{z(\nu_{2} - \nu_{1}-i\pi/2)}.
\end{align*}
where $c_{12}  = \lim_{P \rightarrow \infty_2} \zeta
    g_{1}(P)$  is Ercolani-Sinha's constant. To evaluate this we
need the function $g_1(P)=f_1(P)/f_1(\infty_1)$, with
$$f_1(P)=\frac{\theta(\boldsymbol{\phi}(P)-\boldsymbol{\phi}(\Delta_1)-{\boldsymbol{K}}) \,
\theta(\boldsymbol{\phi}(P)-\boldsymbol{\phi}(\infty_2)-{\boldsymbol{K}})}
{\theta(\boldsymbol{\phi}(P)-\boldsymbol{\phi}(\delta_1)-{\boldsymbol{K}}) \,
\theta(\boldsymbol{\phi}(P)-\boldsymbol{\phi}(\delta_2)-{\boldsymbol{K}})}$$ where here
$\delta_1+\delta_2\sim\Delta_1-\infty_1$ and
$\boldsymbol{\phi}(\Delta_1-\infty_1)=\boldsymbol{U}=-1/2$. Using the values of
the Abel map this last equality yields
$\boldsymbol{\phi}(\Delta_1)=(-1+\tau)/4$ and we may identify $\Delta_1=0_2$.
(Similarly $\Delta_2=0_1$.) With these properties then
$$g_1(P)=
\frac{\theta_1\left(\int_{\infty_2}\sp{P}\boldsymbol{v}\right)\,
\theta_1\left(\int_{0_2}\sp{P}\boldsymbol{v}\right)\,
\theta_1\left(\int_{\delta_1}\sp{\infty_1}\boldsymbol{v}\right)\,
\theta_1\left(\int_{\delta_2}\sp{\infty_1}\boldsymbol{v}\right)}{
\theta_1\left(\int_{\infty_2}\sp{\infty_1}\boldsymbol{v}\right)\,
\theta_1\left(\int_{0_2}\sp{\infty_1}\boldsymbol{v}\right)\,
\theta_1\left(\int_{\delta_1}\sp{P}\boldsymbol{v}\right)\,
\theta_1\left(\int_{\delta_2}\sp{P}\boldsymbol{v}\right) }.$$ This
function has poles in $\delta$ and vanishes at $\infty_2$ and
$\Delta_1=0_2$.\footnote{We disagree with the formulae of Ercolani
and Sinha at this stage: their function (IV.26a) does not have
poles where stated.} Now
$$\lim_{P \rightarrow \infty_2}\zeta(P)\,\theta_1\left(\int_{\infty_2}\sp{P}\boldsymbol{v}\right)
=-\frac{\theta_1'}{4\rho_2}
$$
and
$$c_{12}=\frac{\pi \theta_4\sp2}{2\mathbf{K}}\,
\frac{\theta_1\left(\int_{\delta_1}\sp{\infty_1}\boldsymbol{v}\right)\,
\theta_1\left(\int_{\delta_2}\sp{\infty_1}\boldsymbol{v}\right)}{
\theta_1\left(\int_{\delta_1}\sp{\infty_2}\boldsymbol{v}\right)\,
\theta_1\left(\int_{\delta_2}\sp{\infty_2}\boldsymbol{v}\right)
}.$$ A similar calculation shows also that
$$
c_{21}=\frac{\pi \theta_4\sp2}{2\mathbf{K}}\, \frac{
\theta_1\left(\int_{\delta_1}\sp{\infty_2}\boldsymbol{v}\right)\,
\theta_1\left(\int_{\delta_2}\sp{\infty_2}\boldsymbol{v}\right) }
{\theta_1\left(\int_{\delta_1}\sp{\infty_1}\boldsymbol{v}\right)\,
\theta_1\left(\int_{\delta_2}\sp{\infty_1}\boldsymbol{v}\right)} .
$$
Thus $$c_{12}=-c_{21}\Longleftrightarrow \frac{
\theta_1\left(\int_{\delta_1}\sp{\infty_2}\boldsymbol{v}\right)\,
\theta_1\left(\int_{\delta_2}\sp{\infty_2}\boldsymbol{v}\right) }
{\theta_1\left(\int_{\delta_1}\sp{\infty_1}\boldsymbol{v}\right)\,
\theta_1\left(\int_{\delta_2}\sp{\infty_1}\boldsymbol{v}\right)}=\pm
i ,$$ which again yields the solution (\ref{gen1q0}). To solve for
the divisor $\delta$ we note that
$\boldsymbol{\phi}(\delta_1)+\boldsymbol{\phi}(\delta_2)=-1/2$, whence we wish to solve for
$x=\boldsymbol{\phi}(\delta_1)$,
$$\frac{\theta_1\left(\frac{1+\tau}{4}-x\right)\,\theta_1
\left(\frac{3+\tau}{4} +x\right)}{
\theta_1\left(\frac{1+\tau}{4}+x\right)\,
\theta_1\left(\frac{3+\tau}{4}-x\right)}=\pm i.
$$
The left-hand side is an elliptic function (with periods $1$ and
$x$) in $x$ and so has solutions. For example
$x=\boldsymbol{\phi}(\delta_1)=3\tau/4$ and $1/2+\tau/4$ yield the plus sign,
while $x=\boldsymbol{\phi}(\delta_1)=1/2+3\tau/4$ and $\tau/4$ yield the minus
sign. The two solutions for a fixed sign correspond to the
interchange of $\delta_1$ and $\delta_2$, thus up to equivalence
there are the two solutions arising from the different choices of
sign.

\subsubsection*{The fundamental bi-differential} Finally let us calculate the
fundamental bi-differential and use our formulae to show that
$$\nu_2-\nu_1=\frac{i\pi}{2}.$$
The evenness of the curve (\ref{esellform}) means that for $P$
near $\infty_i$ we have
\begin{align*}
\int_{P_0}\sp{P}\boldsymbol{v}&=\int_{P_0}\sp{\infty_i}\boldsymbol{v}
+\int_{\infty_i}\sp{P_0}\boldsymbol{v}=
\int_{P_0}\sp{\infty_i}\boldsymbol{v}-\frac{t}{4\rho_i}+O(t^3),\\
\mathbf{V}\sp{(i)}&=-\frac{1}{4\rho_i},\ \mathbf{W}\sp{(i)}=0,\\
\mathrm{Res}_{\infty_1,\infty_2}\frac{\eta} {\zeta\sp2}&=0,\\
\Omega_{\mathbf{B}}^{(i)}(P)&=\frac{1}{4\rho_i}\partial_x\left[
\frac{\theta_1'\left(\int_{\infty_i}\sp{P}\boldsymbol{v}\right)}
{\theta_1\left(\int_{\infty_i}\sp{P}\boldsymbol{v}\right)}\right]\,dx(P).
\end{align*}
Further, $\theta_1(x)$ is an odd function and so $\theta''(0)=0$.
Thus
\begin{align*}
\nu_2-\nu_1=\frac14\,\partial_x\ln\left[
\frac{\theta_1\left(x+\int_{\infty_1}\sp{\infty_2}\boldsymbol{v}\right)}
{\theta_1\left(x+\int_{\infty_2}\sp{\infty_1}\boldsymbol{v}\right)}\right]_{x=0}
=\frac14\,\partial_x\ln\left[ \frac{\theta_1\left(
x-\frac{1+\tau}{2}\right)}
{\theta_1\left(x+\frac{1+\tau}{2}\right)}\right]_{x=0}=\frac{i\pi}{2},
\end{align*}
upon using $\theta_1(x+\frac{1+\tau}{2})=B(x)\,\theta_3(x)$, with
$B(x)=\exp-i\pi(x+\tau/4)$. The same result ensues from
(\ref{nuij1}),
\begin{align*}
\nu_1-\nu_2&=-\eta(0_1)\left.\frac{\partial}{\partial z}
\mathrm{ln}\, \frac{\theta_1\left( \int_{\infty_1}^P
\boldsymbol{v} \right)} {\theta_1\left( \int_{\infty_2}^P
\boldsymbol{v} \right)}\right|_{P=0_1}
-\eta(0_2)\left.\frac{\partial}{\partial z} \mathrm{ln}\,
\frac{\theta_1\left( \int_{\infty_1}^P \boldsymbol{v} \right)}
{\theta_1\left( \int_{\infty_2}^P \boldsymbol{v}
\right)}\right|_{P=0_2}\\
&=-\frac14\partial_x\; \mathrm{ln}\left.
\frac{\theta_1\left(x+\int_{\infty_1}^{0_1}\boldsymbol{v}  \right)
} {\theta_1\left(x+\int_{\infty_2}^{0_2}\boldsymbol{v}
\right)}\right|_{x=0} -\frac14\partial_x\; \mathrm{ln}\left.
\frac{\theta_1\left(x+\int_{\infty_2}^{0_1}\boldsymbol{v} \right)
} {\theta_1\left(x+\int_{\infty_1}^{0_2}\boldsymbol{v}
\right)}\right|_{x=0}\\
&=-\frac14\partial_x\;\mathrm{ln}\left.
\frac{\theta_1(x-\frac{\tau}{2})}{\theta_1(x+\frac{\tau}{2})}\right|_{x=0}+
\frac14\partial_x\;\mathrm{ln}\left.
\frac{\theta_1(x+\frac{1}{2})}{\theta_1(x-\frac{1}{2})}
\right|_{x=0}.
\end{align*}
Here we have used our expressions for $\boldsymbol{\phi}(\infty_{1,2})$ and
$\boldsymbol{\phi}(0_{1,2})$. Now upon using $\theta_1(x+\tau/2)=\imath
B(x)\theta_4(x)$,
$B(x)=\mathrm{exp}\left\{-\imath\pi(x+\tau/4)\right\}$ we again
conclude  that $ \nu_1-\nu_2=\frac{\imath\pi}{2}$. For
completeness we record that
\[\omega_{\infty_1,\infty_2}=-\frac{\imath \mathbf{K}(k)}{2}
 \frac{\zeta}{\eta}d\zeta +c\, \boldsymbol{v}(P), \]
where the normalisation constant is given by
\[ c=\frac{\imath \mathbf{K}(k)}{8} \oint_{\mathfrak{a}}
\frac{\zeta d\zeta}{\eta}=\frac{\imath\pi}{4}.
 \]
 One may simply work from this and (\ref{nuijtk}) to obtain the
 same result.

\subsubsection*{Determining the Nahm data}
At this stage we have established that
$$Q_{0}(z)_{12}=
\epsilon_{12}\,\frac{\pi\theta_2 \theta_4 }{2}
\,\frac{\theta_4(z/2)}{\theta_2(z/2)}=\epsilon_{12}\mathbf{K}
k'\,\frac{1}{\mathrm{cn}\,\mathbf{K}z},$$ and that we have the
matrix
$$Q_0(z)=\left(\begin{array}{cc}0&Q_0(z)_{12}\\
Q_{0}(z)_{12}&0\end{array} \right).$$ Without loss of generality
we may take $\epsilon_{12}=1$. We conclude by deriving the well
known elliptic solution of the Nahm equations given by
\begin{align}
T_j(z)=\frac{\sigma_j}{2\imath}\, f_j(z), \quad j=1,2,3,
\end{align}
where $\sigma_j$ are Pauli matrices and the functions $f_j(z)$ are
expressible in terms of Jacobian elliptic functions
\begin{equation}
\begin{split}\label{nahmsolution}
f_1(z)&=\mathbf{K}\,\frac{\mathrm{dn}\,\mathbf{K}z
}{\mathrm{cn}\,\mathbf{K}z}=\frac{\pi\theta_2 \theta_3 }{2}
\,\frac{\theta_3(z/2)}{\theta_2(z/2)},\quad f_2(z)=\mathbf{K}
k'\,\frac{\mathrm{sn}\,\mathbf{K}z
}{\mathrm{cn}\,\mathbf{K}z}=\frac{\pi\theta_3 \theta_4 }{2}
\,\frac{\theta_1(z/2)}{\theta_2(z/2)},\\
 f_3(z)&=\mathbf{K}k'\,\frac{1}{\mathrm{cn}\,\mathbf{K}z}=\frac{\pi\theta_2 \theta_4 }{2}
\,\frac{\theta_4(z/2)}{\theta_2(z/2)}.
\end{split}
\end{equation}
We shall further use
\begin{equation}
\int \frac{du}{\mathrm{cn}\,u } =\frac{1}{k'}\mathrm{ln}\frac{
\mathrm{dn}u + k' \mathrm{sn} u }{\mathrm{cn} u}.\label{primitive}
\end{equation}

Following theorem (\ref{esba}) we have outlined the steps involved
in determining the Nahm data one $Q_0(z)$ is known. First we find
the matrix $C(z)$, subject to initial condition $C(0)=\rm{Id}_2$
and satisfying the differential equation
\[  \frac{d C(z)}{dz}=\frac12\, C(z)Q_0(z) . \]
The solution of this satisfying our initial condition is simply
\[ C(z)=\left(\begin{array}{cc}
F(z)&G(z)\\
G(z)&F(z)
\end{array}    \right)   \]
with
\[F(z)=\mathrm{ch} \left(\frac12\int_0\sp{z} f_3(u)du\right),\quad G(z)=
\mathrm{sh} \left(\frac12\int_0\sp{z}  f_3(u)du\right).
\]
Therefore we have that
\begin{equation} A_0(z)=C(z)Q_0(z)C^{-1}(z)=\left(\begin{array}{cc}
0&1\\1&0\end{array}\right) Q_0(z)_{12} =\frac{1}{2\imath}\sigma_3
f_3(z).
 \end{equation}
Step (b) of our procedure then says that
\begin{equation}
\begin{split} A_1(z)&=C(z)A_1(0)C^{-1}(z)=C(z)\left(\begin{array}{cc}
\rho_1&0\\0&\rho_2\end{array}\right)C^{-1}(z)\\
&=\frac{\mathbf{K}}{2\imath} \left(\begin{array}{cc}
1+2G(z)^2&-2F(z)G(z)\\
2F(z)G(z)&-1-2G(z)^2\end{array} \right).
\end{split}
\end{equation}
Straightforward calculations (where we now employ
(\ref{primitive})) now give that
\begin{align*}
G(z)^2=\frac12 \dfrac{\mathrm{dn}\,\mathbf{K}z}{
\mathrm{cn}\,\mathbf{K}z }-\frac12,\qquad
F(z)G(z)=k'\dfrac{\mathrm{sn}\,\mathbf{K}z
}{\mathrm{cn}\,\mathbf{K}z }.
\end{align*}
Therefore
\begin{align*}
A_1(z)=\frac{\mathbf{K}}{2\imath}\left(
\begin{array}{cc} \dfrac{\mathrm{dn}\,\mathbf{K}z}{\mathrm{cn}\,\mathbf{K}z  }
&-k'\dfrac{\mathrm{sn}\,\mathbf{K}z}{\mathrm{cn}\,\mathbf{K}z  }\\\\
k'\dfrac{\mathrm{sn}\,\mathbf{K}z}{\mathrm{cn}\,\mathbf{K}z  }&
-\dfrac{\mathrm{dn}\,\mathbf{K}z}{\mathrm{cn}\,\mathbf{K}z  }
\end{array}
\right)
\end{align*}
and
\begin{align*}
A_{-1}(z)=-A^{\dagger}(z)=-\frac{\mathbf{K}}{2\imath}\left(
\begin{array}{cc} -\dfrac{\mathrm{dn}\,\mathbf{K}z}{\mathrm{cn}\,\mathbf{K}z  }
&-k'\dfrac{\mathrm{sn}\,\mathbf{K}z}{\mathrm{cn}\,\mathbf{K}z  }\\\\
k'\dfrac{\mathrm{sn}\,\mathbf{K}z}{\mathrm{cn}\,\mathbf{K}z  }&
\dfrac{\mathrm{dn}\,\mathbf{K}z}{\mathrm{cn}\,\mathbf{K}z  }
\end{array}
\right).
\end{align*}
The final step in obtaining the Nahm data is then
\begin{align*}
T_1(z)&=\frac12(A_1(z)+A_{-1}(z))=\frac{\mathbf{K}}{2\imath}\left(
\begin{array}{cc}
 \frac{\mathrm{dn}\,\mathbf{K}z }{\mathrm{cn}\,\mathbf{K}z }&0\\
 0&- \frac{\mathrm{dn}\,\mathbf{K}z }{\mathrm{cn}\,\mathbf{K}z } \end{array}  \right)
 =\frac{1}{2\imath}\sigma_1 f_1(z),\\
T_2(z)&=\frac{1}{2\imath}(A_{-1}(z)-A_{1}(z))
=\frac{\mathbf{K}}{2\imath}\left( \begin{array}{cc}0& -\imath
k'\frac{\mathrm{sn}\,\mathbf{K}z }{\mathrm{cn}\,\mathbf{K}z
}\\\imath k'
\frac{\mathrm{sn}\,\mathbf{K}z}{\mathrm{cn}\,\mathbf{K}z }&0
\end{array}  \right) =\frac{1}{2\imath}\sigma_2 f_2(z),
\end{align*}
and our procedure leads to the known solution
(\ref{nahmsolution}).

\part{Charge Three Monopole Constructions}

\section{The trigonal curve}
We shall now introduce the class of curves that will be the focus
of our attention. These are
\begin{equation}
\eta^3+\hat\chi(\zeta-\lambda_1)(\zeta-\lambda_2)(\zeta-\lambda_3)
(\zeta-\lambda_4)(\zeta-\lambda_5)(\zeta-\lambda_6)=0.
\label{cubic}
\end{equation}
For suitable $\lambda_i$ they correspond\footnote{Here
$\{\lambda_i\}_{i=1}\sp{6}=\{\alpha_j,-{1}/{{\overline\alpha}_j}\}_{j=1}\sp{3}$
and $\hat\chi=\chi_3 \left[\prod_{l=1}\sp{3}\left(
\frac{\overline{\alpha}_l}{\alpha_l}\right)\sp{1/2}\right]$.} to
centred charge three monopoles restricted by $a_2(\zeta)=0$. Thus
the eight dimensional moduli space of centred monopoles has been
reduced to three dimensions. The asymptotic behaviour of the curve
gives us
\begin{equation}
\rho_k= -\hat\chi\sp{\frac13}\,e\sp{2\imath k\pi/3}.
\label{curverho}
\end{equation}
For notational convenience we will study (\ref{cubic}) in the form
($w=-\hat\chi\sp{-\frac{1}{3}}\eta$, $z=\zeta$)
\begin{equation}
w^3=\prod_{i=1}^6(z-\lambda_i). \label{curvegena}
\end{equation}
The moduli space of such curves with an homology marking can be
regarded as the configuration space of six distinct points on
$\mathbb{P}\sp1$. This class of curves has been studied by
Picard \cite{picard83}, Wellstein
\cite{wel99}, Shiga \cite{shiga88} and more
recently by Matsumoto \cite{matsu00}; we shall recall some of their
results. To make concrete the $\theta$-functions arising in the
Ercolani-Sinha construction we need to have the period matrix for
the curve, the vector of Riemann constants, and to understand the
special divisors. We shall now make these things explicit, beginning
first with our choice of homology basis.

\subsection{The curve and homologies}
Let $\mathcal{C}$ denote the curve (\ref{curvegena}) of genus four
where the six points $\lambda_i\in\mathbb{C}$ are assumed distinct
and ordered according to the rule $\mathrm{arg}(\lambda_1)<
\mathrm{arg}(\lambda_2)<\ldots< \mathrm{arg}(\lambda_6)$. Let
$\mathcal{R}$ be the automorphism of $\mathcal{C}$ defined by
\begin{equation}\label{curvesym}
    \mathcal{R}:(z,w)\rightarrow (z,\rho w),\quad
\rho=\mathrm{exp}\{2\imath\pi/3\}.
\end{equation}

The bilinear transformation $(z,w)\leftrightarrow (Z,W)$
\begin{align}
\begin{split}
Z&= \frac{(\lambda_2-\lambda_1)(z-\lambda_4)   }
{(\lambda_2-\lambda_4)(z-\lambda_1)},\\
W&=
-\frac{w}{(z-\lambda_1)^2}
\left(\prod_{k=2}^6(\lambda_1-\lambda_k)\right)^{-\frac13}
\left(\frac{(\lambda_1-\lambda_4)(\lambda_1-\lambda_2)}
{\lambda_2-\lambda_4}\right)^{\frac53}\end{split}\label{rotation1}\end{align}
and its inverse
\begin{align}
\begin{split}
z&=\frac{Z\lambda_1(\lambda_2-\lambda_4)
+\lambda_4(\lambda_1-\lambda_2)}{Z(\lambda_2-\lambda_4)
-(\lambda_2-\lambda_1)}\\
w&=-\frac{W}{(Z(\lambda_2-\lambda_4)-(\lambda_2-\lambda_1))^2}
\left(\prod_{k=2}^6(\lambda_1-\lambda_k)\right)^{\frac13}
\, (\lambda_1-\lambda_2)^{\frac13}(\lambda_1-\lambda_4)^{\frac13}
(\lambda_2-\lambda_4)^{\frac53}\end{split} \label{rotation2}
\end{align}
leads to the following normalization of the curve (\ref{curvegen})
\begin{equation}
W^3=Z(Z-1)(Z-\Lambda_1)(Z-\Lambda_2)(Z-\Lambda_3),
\label{curvegenb}
\end{equation}
where
\begin{equation}\Lambda_i=\frac{\lambda_2-\lambda_1}{\lambda_2-\lambda_4}
\frac{\lambda_{2+j(i)}-\lambda_4}{\lambda_{2+j(i)}-\lambda_1}, \quad
i=1,2,3;\ j(1)=1,\ j(2)=3,\ j(3)=4.
\end{equation}

Fix the following lexicographical ordering of independent canonical
holomorphic differentials of $\mathcal{C}$,
\begin{equation} {d}u_1= \frac{{d} z}{w},\quad
                 {d}u_2= \frac{{d} z}{w^2},\quad
                  {d}u_3= \frac{z{d} z}{w^2},\quad
                  {d}u_4= \frac{z^2{d} z}{w^2}.
\label{diffbasis}
 \end{equation}

To construct the symplectic basis
$(\mathfrak{a}_1,\ldots,\mathfrak{a}_4;
\mathfrak{b}_1,\ldots,\mathfrak{b}_4)$ of
$H_1(\mathcal{C},\mathbb{Z})$ we introduce oriented paths
$\gamma_k(z_i,z_j)$ going  from $P_i=(z_i,w_i)$ to $P_j=(z_j,w_j)$
in the  $k$-th sheet. Define 1-cycles
$\mathfrak{a}_i,\mathfrak{b}_i$ on $\mathcal{C}$ as
follows\footnote{This is the basis from \cite{matsu00}; another
but equivalent basis can be found in \cite{wel99}.}
\begin{align}\begin{split}
\mathfrak{a}_1&=\gamma_1(\lambda_1,\lambda_2)+\gamma_2(\lambda_2,\lambda_1),
\qquad
\mathfrak{b}_1=\gamma_1(\lambda_2,\lambda_1)+\gamma_3(\lambda_1,\lambda_2),\\
\mathfrak{a}_2&=\gamma_1(\lambda_3,\lambda_4)+\gamma_2(\lambda_4,\lambda_3)
,\qquad
\mathfrak{b}_2=\gamma_1(\lambda_4,\lambda_3)+\gamma_3(\lambda_3,\lambda_4),\\
\mathfrak{a}_3&=\gamma_1(\lambda_5,\lambda_6)
+\gamma_2(\lambda_6,\lambda_5),\qquad
\mathfrak{b}_3=\gamma_1(\lambda_6,\lambda_5)
+\gamma_3(\lambda_5,\lambda_6),\\
\mathfrak{a}_4&=\gamma_3(\lambda_1,\lambda_2)
+\gamma_1(\lambda_2,\lambda_6)
+\gamma_3(\lambda_6,\lambda_5)+\gamma_2(\lambda_5,\lambda_1),
\\
\mathfrak{b}_4&
=\gamma_2(\lambda_2,\lambda_1)+\gamma_3(\lambda_6,\lambda_2)
+\gamma_2(\lambda_5,\lambda_6)+\gamma_1(\lambda_1,\lambda_5).
\end{split}\label{homology}
\end{align}
The $\mathfrak{a}$-cycles of the homology basis are given in Figure
1, with the $\mathfrak{b}$-cycles shifted by one sheet.  We have the
pairings $\mathfrak{a}_k\circ \mathfrak{a}_l=\mathfrak{b}_k\circ
\mathfrak{b}_l=0$, $\mathfrak{a}_k\circ
\mathfrak{b}_l=-\mathfrak{b}_k\circ \mathfrak{a}_l= \delta_{k,l}$
and therefore
$(\mathfrak{a}_1,\ldots,\mathfrak{a}_4;\mathfrak{b}_1,\ldots,\mathfrak{b}_4)
$ is a symplectic basis of $ H_1(\mathcal{C},\mathbb{Z})$. In the
homology basis introduced we have
\begin{equation}
\mathcal{R}(\mathfrak{b}_i)=\mathfrak{a_i},\quad i=1,2,3, \quad
\mathcal{R}(\mathfrak{b}_4)=-\mathfrak{a}_4.
\end{equation}
As $(1+\mathcal{R}+\mathcal{R}\sp2)\mathfrak{c}=0$ for any cycle
$\mathfrak{c}$ we have, for example, that
$\mathcal{R}(\mathfrak{a}_i)=-\mathfrak{a}_i-\mathcal{R}\sp2(\mathfrak{a}_i)=
-\mathfrak{a}_i-\mathfrak{b}_i$ for $i=1,2,3$ and
$\mathcal{R}(\mathfrak{a}_4)= -\mathfrak{a}_4+\mathfrak{b}_4$, so
completing the $\mathcal{R}$ action on the homology basis.



\begin{figure}
\centering
\begin{minipage}[l]{0.6\textwidth}
\includegraphics[width=6cm]{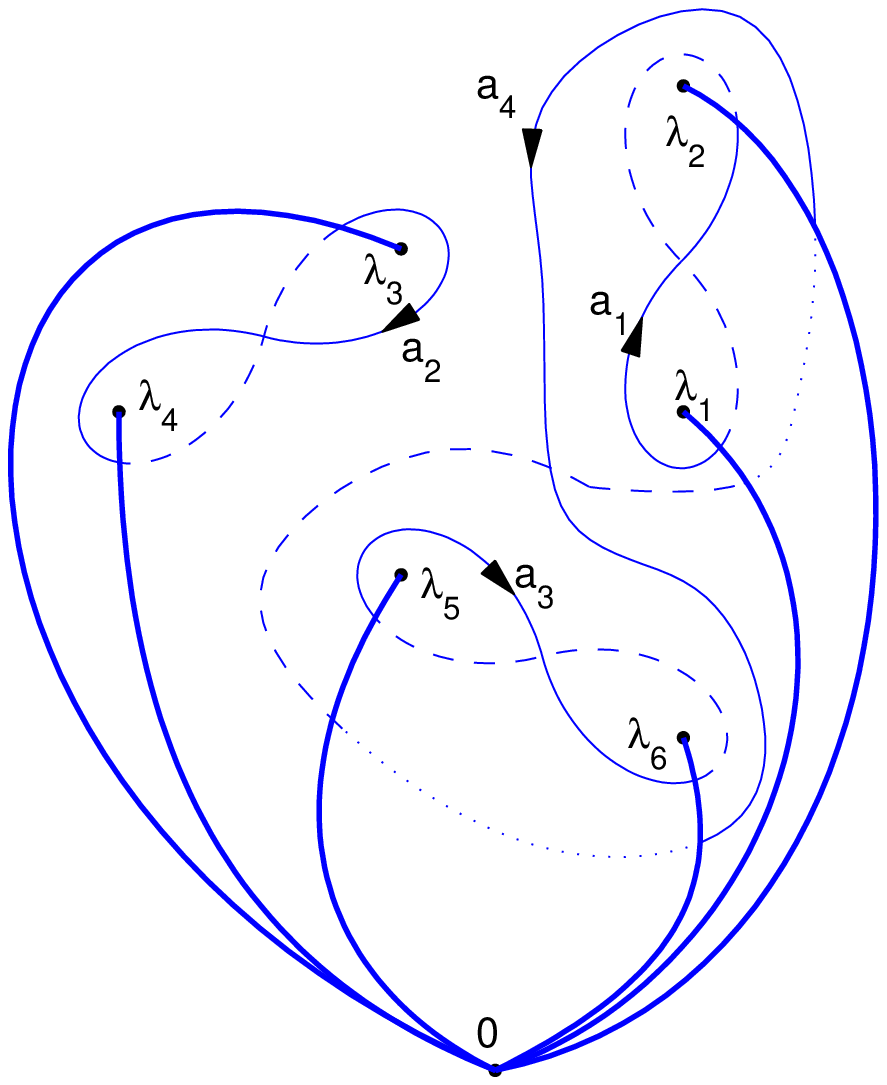} \caption{Homology basis: $\mathfrak{a}$-cycles }
\end{minipage}%
\begin{minipage}[l]{0.4\textwidth}
\includegraphics[width=3cm]{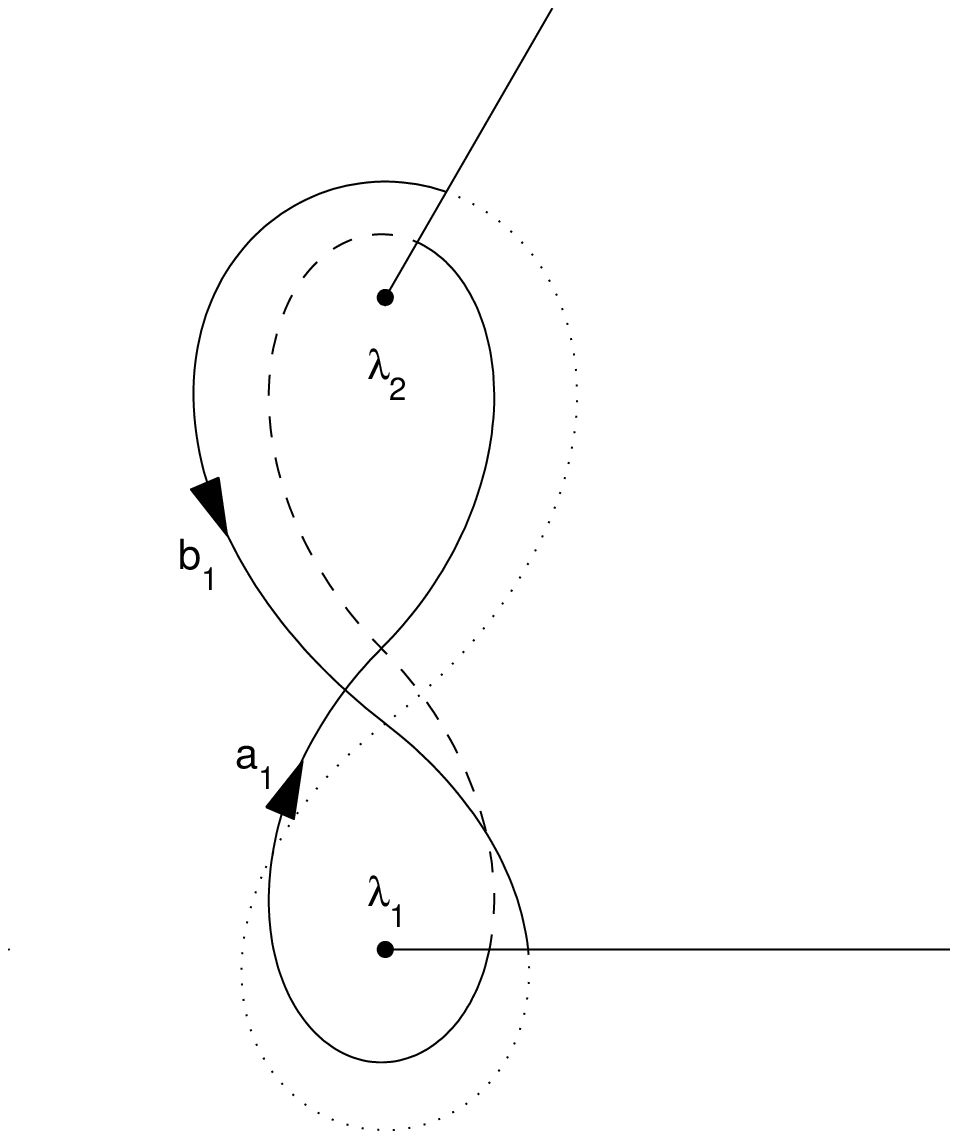} \caption{Cycles $\mathfrak{a}_1$ and
 $\mathfrak{b}_1$ }
\end{minipage}
\end{figure}

\subsection{The Riemann period matrix}
Denote vectors
\begin{align*}
\boldsymbol{x}&=(x_1,x_2,x_3,x_4)^T
=\left(
\oint_{\mathfrak{a}_1}{d}u_1,\ldots,\oint_{\mathfrak{a}_4}{d}u_1 \right)^T ,\\
\boldsymbol{b}&=(b_1,b_2,b_3,b_4)^T
=\left(
\oint_{\mathfrak{a}_1}{d}u_2,\ldots,\oint_{\mathfrak{a}_4}{d}u_2 \right)^T ,\\
\boldsymbol{c}&=(c_1,c_2,c_3,c_4)^T
=\left(
\oint_{\mathfrak{a}_1}{d}u_3,\ldots,\oint_{\mathfrak{a}_4}{d}u_3 \right)^T ,\\
\boldsymbol{d}&=(d_1,d_2,d_3,d_4)^T =\left(
\oint_{\mathfrak{a}_1}{d}u_4,\ldots,\oint_{\mathfrak{a}_4} {d}u_4
\right)^T .
\end{align*}
Crucial for us is the fact that the symmetry (\ref{curvesym})
allows us to relate the matrices of $\mathfrak{a}$ and
$\mathfrak{b}$-periods. For any contour $\Gamma$ and one form
$\omega$ we have that
$\oint\limits_{\mathcal{R}(\Gamma)}\omega=\oint\limits_{\Gamma}\mathcal{R}\sp*\omega$.
If $(\tilde z,\tilde w)=(z,\rho w)=R(z,w)$ then, for example,
$$\mathcal{R}\sp*\left(du_2\right)=\mathcal{R}\sp*\left(
\frac{d\tilde z}{{\tilde w}\sp2}\right)=\frac{d z}{{\tilde
w}\sp2}=\frac{d z}{\rho\sp2 {w}\sp2}=\rho\frac{d z}{{w}\sp2}$$
leading to
$$\oint\limits_{\mathfrak{a}_1} {d}u_2
=\oint\limits_{\mathcal{R}(\mathfrak{b}_1)} {d}u_2=
\oint\limits_{\mathfrak{b}_1}\mathcal{R}\sp*({d}u_2)=
\oint\limits_{\mathfrak{b}_1}\frac{dz}{\rho\sp{2}w\sp2}=
\rho\oint\limits_{\mathfrak{b}_1} {d}u_2.
$$
We find that
\begin{align}\begin{split}
\mathcal{A}&=\left(\mathcal{A}_{ki}\right)=\left(
  \oint\limits_{\mathfrak{a}_k} {d}u_i\right)_{i,k=1,\ldots,4}
=(\boldsymbol{x},\boldsymbol{b},\boldsymbol{c},\boldsymbol{d})
\\
 \mathcal{B}&=\left(\mathcal{B}_{ki}\right)=\left(
  \oint\limits_{\mathfrak{b}_k} {d}u_i\right)_{i,k=1,\ldots,4}
= (\rho H\boldsymbol{x},\rho^2 H\boldsymbol{b},\rho^2
H\boldsymbol{c}, \rho^2
H\boldsymbol{d})=H\mathcal{A}\Lambda,\end{split}\label{calba}
\end{align}
where $H=\mathrm{diag}(1,1,1,-1)$ and
$\Lambda=\mathrm{diag}(\rho,\rho\sp2,\rho\sp2,\rho\sp2)$. This
relationship between the $\mathfrak{a}$ and $\mathfrak{b}$-periods
leads to various simplifications of the Riemann identities,
$$\sum_i\left(
\oint\limits_{\mathfrak{a}_i} {d}u_k \oint\limits_{\mathfrak{b}_i}
{d}u_l - \oint\limits_{\mathfrak{b}_i} {d}u_k
\oint\limits_{\mathfrak{a}_i} {d}u_l \right)=0.
$$
For $k=1$ and $l=2,3,4$ we obtain (respectively) that
\begin{equation}
\boldsymbol{x}^TH\boldsymbol{b} =\boldsymbol{x}^TH\boldsymbol{c}
=\boldsymbol{x}^TH\boldsymbol{d}=0, \label{brr}
\end{equation}
relations we shall employ throughout the paper.

Given $\mathcal{A}$ and $\mathcal{B}$ we now construct the Riemann
period matrix which belongs to the Siegel upper half-space
$\mathbb{S}^4$ of degree 4. If one works with canonically
$\mathfrak{a}$-normalized differentials the period matrix (in our
conventions) is $\tau_\mathfrak{a}=\mathcal{B}\mathcal{A}^{-1}$
while for canonically $\mathfrak{b}$-normalized differentials it
is $\tau_{\mathfrak{b}}=\mathcal{A}\mathcal{B}^{-1}$. Clearly
$\tau_{\mathfrak{b}}=\tau_\mathfrak{a}\sp{-1}$ and we shall simply
denote the period matrix by $\tau$ if neither normalization is
necessary.

\begin{proposition}[Wellstein, 1899; Matsumoto, 2000]
\label{matsumoto1} Let $\mathcal{C}$ be the triple covering of
$\mathbb{P}^1$ with six distinct point $\lambda_1,
\ldots,\lambda_6$,
\begin{equation}
w^3=\prod_{i=1}^6(z-\lambda_i) .\label{curvegen}
\end{equation}
Then the Riemann period matrix is of the form
\begin{align}
\tau_{\mathfrak{b}}&=\rho\left(
H-(1-\rho)\frac{\boldsymbol{x}\boldsymbol{x}^T} {\boldsymbol{x}^T
H\boldsymbol{x}}   \right),\label{taumat}
\end{align}
where $H=\mathrm{diag}(1,1,1,-1)$. Then $\tau_{\mathfrak{b}}$ is
positive definite if and only if
\begin{align} \bar{\boldsymbol{x}}^T H \boldsymbol{x} <0.
\label{condition1}
\end{align}
\end{proposition}
Both Wellstein and Matsumoto give broadly similar proofs of
(\ref{taumat}) and we shall present another variant as we need to
use an identity established in the proof later in the text.

\begin{proof}From (\ref{brr}) we see that we have
$$\mathcal{A}\sp{T} H \boldsymbol{x}=(\Delta,0,0,0)\sp{T},\qquad
\Delta:={\boldsymbol{x}}^T H \boldsymbol{x}.$$ We know that
$\mathcal{A}$ is nonsingular and consequently $\boldsymbol{x}\ne0$
and $\Delta\ne0$. Now
$H\boldsymbol{x}=\mathcal{A}\sp{T\,-1}(\Delta,0,0,0)\sp{T}$ which
gives
\begin{equation}\label{invcalA}
    (H\boldsymbol{x})_{\mu}=\mathcal{A}\sp{-1}_{1\mu}\Delta.
\end{equation}
Now from (\ref{calba}) we see that
$$\mathcal{B}\mathcal{A}^{-1}=\rho\sp2H+(\rho-\rho\sp2)H
(\boldsymbol{x},0,0,0)\mathcal{A}^{-1}.$$ From (\ref{invcalA}) we
obtain
$$H(\boldsymbol{x},0,0,0)\mathcal{A}^{-1}=\frac{1}{\Delta}H
\boldsymbol{x}\boldsymbol{x}\sp{T}H$$ and therefore
$$\mathcal{B}\mathcal{A}^{-1}=\rho\sp2H+\frac{(\rho-\rho\sp2)}{\Delta}H
\boldsymbol{x}\boldsymbol{x}\sp{T}H.$$ Finally one sees that
$$\left[\rho\sp2H+\frac{(\rho-\rho\sp2)}{\Delta}H
\boldsymbol{x}\boldsymbol{x}\sp{T}H\right]\left[ \rho
H-\frac{(\rho-\rho\sp2)}{\Delta}
\boldsymbol{x}\boldsymbol{x}\sp{T}\right]=1,$$ whence the result
(\ref{taumat}) follows for
$\tau_{\mathfrak{b}}=\mathcal{A}\mathcal{B}^{-1}$. The remaining
constraint arises by requiring $\Im\tau$ to be positive definite.
We note that (\ref{condition1}) ensures that both
$\boldsymbol{x}\ne0$ and $\Delta\ne0$.
\end{proof}

The branch points can be expressed in terms of $\theta$-constants.
Following Matsumoto \cite{matsu00} we introduce the set of
characteristics
\begin{equation}  (\boldsymbol{a},\boldsymbol{b}),\qquad \boldsymbol{b}
=-\boldsymbol{a} H, \quad a_i\in \left\{ \frac16,\frac36, \frac
56\right\} \label{characteristics} \end{equation} and denote
$\theta_{\boldsymbol{a},-H\boldsymbol{b}}(\tau)
=\theta\{6\boldsymbol{a }\}(\tau)$ (see Appendix A for out theta
function conventions). The characteristics (\ref{characteristics})
are classified in \cite{matsu00} by the representations of the braid
group. Further, the period matrix determines the branch points as
follows.
\begin{proposition} [Diez 1991, Matsumoto 2000] \label{matsumoto2}
Let $\tau_{\mathfrak{b}}$ be the period matrix of
(\ref{curvegenb})
given in Proposition  \ref{matsumoto1}. Then
\begin{equation}
\Lambda_1=\left(\frac{\theta\{3,3,3,5\}} {\theta\{1,1,3,3\}
}\right)^3,\ \ \Lambda_2=-\left(\frac{\theta\{1,5,3,3\}}
{\theta\{1,1,5,5\} }\right)^3,\ \
 \Lambda_3
=-\left(\frac{\theta\{1,1,3,3\}} {\theta\{5,1,1,1\} }\right)^3.
\label{lambdas}
\end{equation}
\end{proposition}

These results have the following significance for our construction
of monopoles. First we observe that the period matrix is invariant
under $\mathbf{x}\rightarrow \lambda\mathbf{x}$. Thus to our
surface we may associate a point $[x_1:x_2:x_3:x_4]\in
\mathbb{B}\sp3=\{\mathbf{x}\in\mathbb{P}\sp3\,|\, \boldsymbol{x}^T
H \boldsymbol{x}<0\}\subset \mathbb{P}\sp3$ and from this point we
may obtain the normalized curve (\ref{curvegenb}). It is known
that a dense open subset of $\mathbb{B}\sp3$ arises in this way
from curves with distinct roots with the complement corresponding
to curves with multiple roots. Correspondingly, if we choose a
point $[x_1:x_2:x_3:x_4]\in \mathbb{B}\sp3$ we may construct a
period matrix and corresponding normalized curve.

We note that with $d\boldsymbol{u}=(du_1,\ldots,du_4)$ then
$$\tau\sp*(d\boldsymbol{u})=\overline{d\boldsymbol{u}}\cdot T,
\qquad T=\begin{pmatrix}
  -\kappa & 0& 0 & 0 \\
  0 & 0 & 0 & \kappa\sp2 \\
  0 & 0 & -\kappa\sp2 & 0 \\
  0 & \kappa\sp2 & 0 & 0
\end{pmatrix},\qquad
\kappa=\frac{\hat\chi\sp{\frac{1}{3}}}{\overline{\hat\chi}\sp{\frac{1}{3}}},
$$
and so we obtain
$$\oint_{\tau_*\mathfrak{c}}d\boldsymbol{u}=\oint_{\mathfrak{c}}
\tau\sp*(d\boldsymbol{u})=\oint_{\mathfrak{c}}\overline{d\boldsymbol{u}}\cdot
T=\overline{\left(
\oint_{\mathfrak{c}}{d\boldsymbol{u}}\right)}\cdot
T=\overline{\left(6\hat \chi\sp{\frac13}\left(
  1 \ 0\ 0 \ 0\right)\right)}\cdot
T=-6 \hat\chi\sp{\frac13}\left(
  1 \ 0\ 0 \ 0\right)=\oint_{-\mathfrak{c}}d\boldsymbol{u}$$
  and as a consequence corollary (\ref{HMRinvc}) of Houghton, Manton and
  Rom\~ao, $\tau_*\mathfrak{c}=-\mathfrak{c}$. More generally, let
  us write for an arbitrary cycle
  $$\gamma=\mathbf{p}\cdot \mathfrak{a}+\mathbf{q}\cdot \mathfrak{b}=
  \left(%
\begin{array}{cc}
  \mathbf{p} & \mathbf{q} \\
\end{array}%
\right)\left(%
\begin{array}{c}
  \mathfrak{a} \\
  \mathfrak{b} \\
\end{array}%
\right), \qquad \tau(\gamma)=  \left(%
\begin{array}{cc}
  \mathbf{p} & \mathbf{q} \\
\end{array}%
\right)\mathcal{M}\left(%
\begin{array}{c}
  \mathfrak{a} \\
  \mathfrak{b} \\
\end{array}%
\right).
$$
Then the equality
$$\oint_{\tau(\mathfrak{\gamma})}d\boldsymbol{u}=\oint_{\mathfrak{\gamma}}
\tau\sp*(d\boldsymbol{u})
=\oint_{\mathfrak{\gamma}}\overline{d\boldsymbol{u}}\cdot
T=\overline{\left(
\oint_{\mathfrak{\gamma}}{d\boldsymbol{u}}\right)}\cdot T$$ leads
to the equation
$$
 \left(%
\begin{array}{cc}
  \mathbf{p} & \mathbf{q} \\
\end{array}%
\right)\mathcal{M}\left(%
\begin{array}{c}
  \mathcal{A} \\
  \mathcal{B} \\
\end{array}%
\right) =\left(%
\begin{array}{cc}
  \mathbf{p} & \mathbf{q} \\
\end{array}%
\right)\overline{
\left(%
\begin{array}{c}
  \mathcal{A} \\
  \mathcal{B} \\
\end{array}%
\right)}\cdot T.
$$
We have then that the matrix $\mathcal{M}$ representing the
involution $\tau$ on homology and Ercolani-Sinha vector satisfy
\begin{equation}
\label{cplxMT} \mathcal{M}\sp2=\rm{Id},\quad
\mathcal{M}\left(%
\begin{array}{c}
  \mathcal{A} \\
  \mathcal{B} \\
\end{array}%
\right) =\overline{
\left(%
\begin{array}{c}
  \mathcal{A} \\
  \mathcal{B} \\
\end{array}%
\right)}\cdot T, \quad \mathbf{U}\mathcal{M}=
\left(%
\begin{array}{cc}
  \mathbf{n} & \mathbf{m} \\
\end{array}%
\right)\mathcal{M}=-\left(%
\begin{array}{cc}
  \mathbf{n} & \mathbf{m} \\
\end{array}%
\right).
\end{equation}
A calculation employing the algorithm of Tretkoff and Tretkoff
\cite{tret84} to describe the homology basis generators and
relations, together with some analytic continuation of the paths
associated to our chosen homology cycles (with the sheet
conventions described later in the text), yields that for our
curve
$$\mathcal{M}=\left( \begin {array}{cccccccc} 0&0&-1&0&0&-1&0&-1
\\\noalign{\medskip}0&0&1&1&-1&0&1&0\\\noalign{\medskip}-1&1&0&-1&0&1&0
&1\\\noalign{\medskip}0&-1&1&2&-1&0&1&0\\\noalign{\medskip}0&-1&1&1&0&0
&1&0\\\noalign{\medskip}-1&0&0&-1&0&0&-1&1\\\noalign{\medskip}1&0&0&0&
1&-1&0&-1\\\noalign{\medskip}1&-1&0&2&0&-1&1&-2\end {array}
\right).$$ The matrix $\mathcal{M}$ is not symplectic but
satisfies
$$\mathcal{M}J\mathcal{M}\sp{T}=-J,$$
where $J$ is the standard symplectic form. (The minus sign appears
here because of the reversal of orientation under the
antiholomorphic involution.)

\subsection{The vector $\widetilde{\boldsymbol{K}} $
and $\int\limits_{\infty_i}\sp{\infty_j}
\boldsymbol{v}$} We shall now describe the vector
$\widetilde{\boldsymbol{K}}$ and various related results,
including the quantity $\boldsymbol{\phi}(\infty_i)
-\boldsymbol{\phi}({\infty_j})$.

First let us record some elementary facts about our curve. For
ease in defining various divisors of the curve (\ref{curvegena})
let $\infty_{1,2,3}$ be the three points over infinity  and
$Q_i=(\lambda_i,0)$ ($i=1,\ldots 6$) be the branch points. Then
$$
\begin{array}{lll}
\div(z-\lambda_i)=\dfrac{Q_i\sp3}{\infty_1 \infty_2 \infty_3},&
\div(w)=\dfrac{\prod_{i=1}\sp6 Q_i}{(\infty_1 \infty_2
\infty_3)^2},&
\div(dz)=\dfrac{(\prod_{i=1}\sp6 Q_i)^2}{(\infty_1 \infty_2 \infty_3)^2},\\ \\
\div\left(\dfrac{dz}{w}\right)=\prod_{i=1}\sp6 Q_i,&
\div\left(\dfrac{dz}{w^2}\right)= (\infty_1 \infty_2 \infty_3)^2,&
\div\left(\dfrac{(z-\lambda_i)dz}{w^2}\right)= Q_i^3 \infty_1 \infty_2 \infty_3,\\
\\ \div\left(\dfrac{(z-\lambda_i)^2dz}{w^2}\right)=Q_i^6.
\end{array}
$$
Consideration of the function $(z-\lambda_i)/(z-\lambda_j)$ shows
that $3\int_{Q_j}\sp{Q_i}\boldsymbol{v}\in \Lambda$. The order of
vanishing of the differentials $d(z-\lambda_i)/w^2$,
$d(z-\lambda_i)/w$, $(z-\lambda_i)d(z-\lambda_i)/w^2$ and
$(z-\lambda_i)d(z-\lambda_i)/w^2$ at the point $Q_i$ are found to
be $0$, $1$, $3$ and $6$ respectively, which means that the gap
sequence at $Q_i$ is $1$, $2$, $4$ and $7$. From this we deduce
that the index of speciality of the divisor $Q_i^3$ is
$i(Q_i^3)=2$. Because the genus four curve $\mathcal{C}$ has the
function $w$ of degree $3$ then $\mathcal{C}$ is not
hyperelliptic. The function $1/(z-\lambda_i)$ has divisor
$\mathcal{U}/D$, with $\mathcal{U}=\infty_1 \infty_2 \infty_3$ and
$D=Q_i^3$ such that $D^2$ is canonical. This means that any other
function of degree $3$ on $\mathcal{C}$ is a fractional linear
transformation of $w$ and that $\Theta_{{\rm singular}}$ consists
of precisely one point which is of order $2$ in
$\Jac(\mathcal{C})$ \cite[III.8.7, VII.1.6]{fk80}. The vector of
Riemann constants $\boldsymbol{K}_{Q_i}$ is a point of order $2$
in $\Jac(\mathcal{C})$ because $Q_i\sp6$ is canonical
\cite[VI.3.6]{fk80}. Let us fix $Q_1$ to be our base point. Then
as $\boldsymbol{K}_{Q_1}=\boldsymbol{\phi}_{Q_1}(Q_1^3)+\boldsymbol{K}_{Q_1}$
we have that $\boldsymbol{K}_{Q_1}\in\Theta$. Because $i(Q_1^3)=2$
we may identify $\boldsymbol{K}_{Q_1}$ as the unique point in
$\Theta_{{\rm singular}}$. We may further identify
$\boldsymbol{K}_{Q_1}$ as the unique even theta characteristic
belonging to $\Theta$.

With $Q_1$ as our base point $\boldsymbol{\phi}\left(\sum_k \infty_k\right)$
corresponds to the image under the Abel map of the divisor of the
function $1/(z-\lambda_1)$, and so vanishes (modulo the period
lattice). Thus for our curve $\widetilde{\boldsymbol{K}}=
\boldsymbol{K}_{Q_1}+\boldsymbol{\phi}\left(\sum_k
\infty_k\right)=\boldsymbol{K}_{Q_1}=\Theta_{{\rm singular}}$ is
the unique even theta characteristic. The point
$\boldsymbol{K}_{Q_1}$ may be constructed several ways: directly,
using the formula (\ref{vecR}) of the Appendix (the evaluation of
the integrals of normalised holomorphic differentials between
branch points is described in Appendix B); by enumeration we may
find which of the $136$ even theta characteristics
$\left[\begin{matrix}\epsilon
\\ \epsilon'\end{matrix}\right]$ leads to the vanishing of $\theta
\left[\begin{matrix}\epsilon
\\ \epsilon'\end{matrix}\right](z;\tau)$; using a monodromy argument of Matsumoto
\cite{matsu00}. One finds that the relevant half period is
$\dfrac12\left[\begin{matrix}1&1&1&1\\1&1&1&1\end{matrix}\right]$.

The analysis of the previous paragraph, together with
(\ref{regcond}), tells us that $\boldsymbol{U}$ must also be an
even theta characteristic.

Again using that $\sum_k \infty_k\sim_l 0$ we have that
$\infty_i-\infty_j\sim_l 2\infty_i+\infty_k$ (with $i$, $j$, $k$
distinct) and so
$\theta(\boldsymbol{\phi}(\infty_j)-\boldsymbol{\phi}(\infty_i)-\widetilde{\boldsymbol{K}})=
\theta(\boldsymbol{\phi}(2\infty_i+\infty_k)+ \boldsymbol{K})=0$. One sees from
the above divisors (in particular $\div(dz/w^2)$) that $\dim
H\sp0( \mathcal{C}, L_{2\infty_i+\infty_k})=
i(2\infty_i+\infty_k)=1$. Thus
$\theta(w+\boldsymbol{\phi}(\infty_j)-\boldsymbol{\phi}(\infty_i)-\widetilde{\boldsymbol{K}}$
and $\theta(w-\widetilde{\boldsymbol{K}})$ have order of vanishing
differing by one for (generic) $w\rightarrow0$.

\subsection{Calculating $\nu_i-\nu_j$}
From the results of the previous section we see that
$$\div\left(\dfrac{z^4dz}{w^2}\right)=\frac{(0_10_20_3)\sp4}{(\infty_1\infty_2\infty_3)\sp2}.
$$
This has precisely the same divisor of poles as $\gamma_\infty$
and we will use this to represent $\gamma_\infty$. It is
convenient to introduce the (meromorphic) differential
$$dr_1(P)=\frac{z^4 dz}{3w^2},$$
the factor of three here being introduced to give the pairing
\[ \sum_{s=1}^3  \mathrm{Res}_{P=\infty_s}d{r}_1(P)
  \int_{P_0}^P d{u}_1(P')
=1.
\]
We may therefore write \begin{equation}\label{gpdr}
\gamma_{\infty}(P)=-3 dr_1(P)+\sum_{i=1}^4 c_i {v}_i(P).
\end{equation} The constants $c_i$ are found from the condition of
normalisation
\[ \oint_{\mathfrak{a}_k } \gamma_{\infty}(P)=0\quad \Longleftrightarrow
c_k=3\oint_{\mathfrak{a}_k} dr_1(P)\equiv 3y_k,\quad k=1,\ldots,4,
\] where we have defined the vector of
$\mathfrak{a}$-periods
$\boldsymbol{y}\sp{T}=\left(\oint_{\mathfrak{a}_1} dr_1(P),\ldots,
\oint_{\mathfrak{a}_4} dr_1(P)     \right)$. The vector of
$\mathfrak{b}$-periods of $dr_1$ is found to be $\rho^2 H
\boldsymbol{y}$. The pairing with $du_1$ then yields the Legendre
relation
\begin{equation}\label{legndre}
\boldsymbol{y}{\cdot } H\boldsymbol{x}=-\frac{2\pi}{\sqrt{3}}.
\end{equation}
Now the $\mathfrak{b}$-periods of the differential
$\gamma_{\infty}$ give the Ercolani-Sinha vector. Using
(\ref{gpdr}) we then obtain the equality
\begin{equation}
-3(\rho^2 H-\tau_{\mathfrak{a}})\boldsymbol{y}=\pi\imath
\boldsymbol{n}+\pi\imath \tau \boldsymbol{m}.\label{relation}
\end{equation}

Finally, using (\ref{gpdr}), we may write
\begin{equation}\label{nuijgpgr}
\nu_i-\nu_j=3\boldsymbol{y}\cdot\int_{\infty_j}\sp{\infty_i}
\boldsymbol{v}+\int_{\infty_j}\sp{\infty_i}\left[
d\left(\frac{w}{z}\right)-3dr_1\right].
\end{equation}

\section{Solving the Ercolani-Sinha constraints}
We shall now describe how to solve the Ercolani-Sinha constraints
for the spectral curve (\ref{cubic}). This reduces to constraints
just on the four periods $\boldsymbol{x}$.  Later we shall
restrict attention to the curves (\ref{bren03}), which has the
effect of reducing the number of integrals to be evaluated to two
and consequently simplifies our present analysis.

We shall work with the Ercolani-Sinha constraints in the form
(\ref{HMREScond}). Let the holomorphic differentials be ordered as
in (\ref{diffbasis}). Then there exist two integer 4-vectors
$\boldsymbol{n}, \boldsymbol{m}\in \mathbb{Z}^4$ and values of the
parameters $\lambda_1,\ldots,\lambda_6 $ and $\chi$ such that
\begin{equation}\boldsymbol{n}^T \mathcal{A}+\boldsymbol{m}^T\mathcal{B}
=\nu(1,0,0,0).\label{esconda}
 \end{equation}
Here $\nu$ depends on normalizations.  For us this will be
\begin{equation*}
 \nu = 
 6 \hat\chi ^{\frac{1}{3}} .
\end{equation*}
To see this observe that (\ref{HMREScond}) requires that
$$-2\delta_{1k} = \oint_{\boldsymbol{n}\cdot\mathfrak{a}+
\boldsymbol{m}\cdot\mathfrak{b}} \Omega^{(k)}\ \textrm{ for the
differentials}\ \Omega^{(1)} = \frac{\eta^{n-2} d \zeta}
{\frac{\partial P}{\partial \eta}} = \frac{d \zeta}{n \eta},\\
\Omega^{(2)} = \frac{\eta^{n-3} d \zeta} {\frac{\partial P}
{\partial \eta}} ,\dots.
$$ In the parameterisation (\ref{curvegena}) we are using we have that
\begin{equation*}
    x_{i} = {\oint_{\mathfrak{a}_i}}\frac{dz}{w} =
   {\oint_{\mathfrak{a}_i}} \frac{d\zeta}{-\hat\chi^{-\frac{1}{3}}\eta} =
    -3\hat\chi^{\frac{1}{3}}{\oint_{\mathfrak{a}_i}} \Omega^{(1)}.
\end{equation*}
We wish
\begin{align}
    -2 = \oint_{\boldsymbol{n}\cdot\mathfrak{a}+
\boldsymbol{m}\cdot\mathfrak{b}} \Omega^{(1)} & =
    \frac{-1}{3 \hat\chi^{\frac{1}{3}}}(\boldsymbol{n}.\boldsymbol{ x} + \rho
    \boldsymbol{m}.H.\boldsymbol{x}) \nonumber \intertext{and so}
    \boldsymbol{n}.\boldsymbol{ x} + \rho
    \boldsymbol{m}.H.\boldsymbol{x}& = \nu,\label{esonx}
    \end{align}
with the value of $\nu$ stated. Consideration of the other
differentials then yields (\ref{esconda}), transcendental
constraints on the curve $\mathcal{C}$. These constraints may be
solved using the following result.
\begin{proposition}\label{ourh2}
The Ercolani-Sinha constraints (\ref{esconda}) are satisfied for
the curve (\ref{cubic}) if and only if
\begin{align}\label{expxx}
    \boldsymbol{x} & = \xi(H\boldsymbol{n} + \rho^{2}\boldsymbol{m}),
    \intertext{where }\xi &   = \frac{\nu}{[\boldsymbol{n}.H
    \boldsymbol{n}-\boldsymbol{m}.\boldsymbol{n}+\boldsymbol{m}.H
    \boldsymbol{m}]} = \frac{6
\hat\chi^{\frac{1}{3}}}{[\boldsymbol{n}.H
    \boldsymbol{n}-\boldsymbol{m}.\boldsymbol{n}+\boldsymbol{m}.H
    \boldsymbol{m}]}.\label{expxxx}
\end{align}
\end{proposition}

\begin{proof}
Rewriting (\ref{esconda}) we have that
\begin{align*}
    \boldsymbol{n}^{T} + \boldsymbol{m}^{T}\mathcal{B}\mathcal{A}^{-1}
    & = \nu (1,0,0,0)\mathcal{A}^{-1} = \nu
    \mathcal{A}^{-1}_{1\mu}. \intertext{Upon using (\ref{invcalA})
    we obtain}
    \boldsymbol{n}^{T} + \boldsymbol{m}^{T}\mathcal{B}\mathcal{A}^{-1}
    & = \frac{\nu}{\Delta}
    \boldsymbol{x}^{T}H.
    \end{align*}
  Therefore \begin{align*}  \boldsymbol{x}& = \frac{\Delta}{\nu}
  (H\boldsymbol{n} + H(\mathcal{B}\mathcal{A}^{-1})^{T}\boldsymbol{m}) \\
    & = \frac{\Delta}{\nu}(H\boldsymbol{n} + \rho^{2}\boldsymbol{m}
     + (\frac{\rho -
    \rho^{2}}{\Delta})\boldsymbol{x}\,\boldsymbol{x}^{T}H\boldsymbol{m})\end{align*}
  upon using that the period matrix is symmetric and our earlier expression
  for $\mathcal{B}\mathcal{A}^{-1}$. Rearranging now gives us that
  \begin{equation}\label{expx}
    (1 + \frac{\rho^{2}-\rho}{\nu}\boldsymbol{x}^{T}H\boldsymbol{m})
    \boldsymbol{x}  =
    \frac{\Delta}{\nu}(H\boldsymbol{n} + \rho^{2}\boldsymbol{m})
\end{equation}
and so we have established (\ref{expxx}) where
\begin{align}\label{expxxxx}
    \xi & = \frac{\Delta}{\nu}(1 +
    \frac{\rho^{2}-\rho}{\nu}\,\boldsymbol{x}.H\boldsymbol{m})^{-1}.
\end{align}
There are several constraints. First, the Ercolani-Sinha condition
(\ref{esonx}) is that
$$
    [\boldsymbol{n}^{T} + \rho \boldsymbol{m}^{T}H]
    \xi[H\boldsymbol{n} + \rho^{2}\boldsymbol{m}] = \nu $$
    and consequently
\begin{equation}\label{esnu}
[\boldsymbol{n}.H\boldsymbol{n} - \boldsymbol{m}.\boldsymbol{n} +
\boldsymbol{m}.H\boldsymbol{m}]\xi  = \nu =
6\hat\chi^{\frac{1}{3}},
\end{equation}
thus establishing (\ref{expxxx}). We remark that if $\hat\chi$ is
real, then $\hat\chi^{\frac{1}{3}}$ may be chosen real and hence
 $\xi$ is real. We observe that (\ref{expxxx}) and (\ref{expxxxx}) are consistent
with
\begin{align*}
    \Delta = \boldsymbol{x}^{T}H\boldsymbol{x} & =
     \xi^{2}(\boldsymbol{n}^{T}H + \rho^{2}\boldsymbol{m}\sp{T})H
     (H\boldsymbol{n} + \rho^{2}\boldsymbol{m})
     = \xi^{2}[\boldsymbol{n}.H\boldsymbol{n} + 2\rho^{2}
    \boldsymbol{m}.\boldsymbol{n} + \rho
    \boldsymbol{m}.H\boldsymbol{m}].
\end{align*}
A further consistency check is given by  (\ref{relation}). Using
the form of the period matrix, the Legendre relation
(\ref{legndre}) and the proposition (with $\nu=-6$) we obtain
(\ref{relation}).
\end{proof}

At this stage we have reduced the Ercolani-Sinha constraints to
one of imposing the four constraints (\ref{expxx}) on the periods
$x_k$. In particular this means we must solve
\begin{equation}
\frac{x_1}{n_1+\rho^2m_1} =\frac{x_2}{n_2+\rho^2m_2}
=\frac{x_3}{n_3+\rho^2m_3}
=\frac{x_4}{-n_4+\rho^2m_4}=\xi,\label{nmconditions1}
\end{equation}
which means $x_i/x_j\in\mathbb{Q}[\rho]$. Further we have from the
conditions (\ref{condition1}) that
\begin{align}
\frac{\bar{\boldsymbol{x}}\sp{T} H
\boldsymbol{x}}{|\xi|\sp2}&=[\boldsymbol{n}.H\boldsymbol{n} -
\boldsymbol{m}.\boldsymbol{n} +
\boldsymbol{m}.H\boldsymbol{m}]
= \sum_{i=1}^3 (n_i^2-n_im_i+m_i^2)-n_4^2-m_4^2-m_4n_4<0.
\label{nmconditions}
\end{align}

Our result admits another interpretation. Thus far we have assumed
we have been given an appropriate curve and sought to satisfy the
Ercolani-Sinha constraints. Alternatively we may start with a
curve satisfying (most of) the Ercolani-Sinha constraints and seek
one satisfying the reality constraints (and any remaining
Ercolani-Sinha constraints). How does this progress? First note
that the period matrix (\ref{taumat}) for a curve satisfying
(\ref{nmconditions1}) is independent of $\xi$: it is determined
wholly in terms of the Ercolani-Sinha vector. Let us then start
with a primitive vector $\mathbf{U}=(\mathbf{n},\mathbf{m})$
satisfying the hyperboloid condition (\ref{nmconditions}) and
lemma \ref{ueventheta}. From this we construct a period matrix and
then, via Proposition \ref{matsumoto2}, a normalized curve
(\ref{curvegenb}). Now we must address whether the curve has the
correct reality properties. For this we must show that there
exists a M\"obius transformation of the set $S=\{0,1,\infty,
\Lambda_1,\Lambda_2,\Lambda_3\}$ to one of the form
$H=\{\alpha_j,-{1}/{{\overline\alpha}_j}\}_{j=1}\sp{3}$. We will
show below that this question may be answered, with the roots
$\alpha_i$ being determined up to an overall rotation. At this
stage we have (using the rotational freedom) a curve of the form
$$W^3=Z(Z-a)Z(Z+\frac1{a})(Z-w)(Z+\frac{1}{\overline{w}}),\qquad
a\in\mathbb{R},\ w\in\mathbb{C}.$$ To reconstruct a monopole curve
we need a normalization $\hat\chi=\chi_3 \left[
\frac{\overline{w}}{w}\right]\sp{1/2}$. This is encoded in $\xi$,
which has not appeared thus far. To calculate the normalization we
must calculate a period. Then using (\ref{nmconditions1}) and
(\ref{expxxx}) we determine $\hat\chi$. This is a constraint. For
a consistent monopole curve we require $$\arg(\xi)=\arg\left[
\frac{\overline{w}}{w}\right]\sp{1/6}.$$ Of course, to complete
the construction we need to check there are no roots of the theta
function in $[-1,1]$. Although the procedure outlined involves
several transcendental calculations it is numerically feasible and
gives a means of constructing putative monopole curves.

To conclude we state when there exists a M\"obius transformation
of the set $S=\{0,1,  \infty,   \Lambda_1,  \Lambda_2,  \Lambda_3\}$ to
one of the form
$H=\{\alpha_j,-{1}/{{\overline\alpha}_j}\}_{j=1}\sp{3}$. For
simplicity we give the case of distinct roots:
\begin{theorem}The roots $S=\{0,1,\infty,
\Lambda_1,\Lambda_2,\Lambda_3\}$ are M\"obius equivalent to
$H=\{\alpha_j,-{1}/{{\overline\alpha}_j}\}_{j=1}\sp{3}$ if and
only if \begin{enumerate} \item If just one of the roots, say
$\Lambda_1$, is real and
\begin{itemize}
\item $\Lambda_1<0$ then
$\Lambda_2\overline{\Lambda}_3=\Lambda_1$, \item $0<\Lambda_1<1$
then
$\frac{\Lambda_2}{\Lambda_2-1}\overline{\frac{\Lambda_3}{\Lambda_3-1}}=\frac{\Lambda_1}{\Lambda_1-1}$,
\item $1<\Lambda_1$ then
$(1-\Lambda_2)\overline{(1-\Lambda_3)}=1-\Lambda_1$.
\end{itemize}
If all three roots are real then, up to relabelling, one of the
above must hold.

\item All three roots are complex and, up to relabelling,
$$0<\Lambda_1\overline{\Lambda}_2\in\mathbb{R},\qquad
1<\frac{\Lambda_1}{\Lambda_2},\qquad
\Lambda_3=\Lambda_2\,\frac{1-\overline{\Lambda}_1}{1-\Lambda_2}.$$
\end{enumerate}
\end{theorem}

\section{Symmetric 3-monopoles}
In this section we shall consider the curve $\mathcal{C}$ specialized
to the form
\begin{equation}
\eta^3+\chi(\zeta^6+b\zeta^3-1)=0,\label{curve}
\end{equation}
where $b$ is a real parameter. In this case branch points are
\[ (\lambda_1,\lambda_2,\lambda_3,\lambda_4,\lambda_5,\lambda_6)
 =(\alpha,\;\rho^2 \beta,\; \rho \alpha,\; \beta,\; \rho^2 \alpha,\;
\rho \beta),   \] where $\alpha$ and $\beta$ are real,
\[ \alpha=\sqrt[3]{\frac{-b+\sqrt{b^2+4}}{2}}>0,\quad
\beta=\sqrt[3]{\frac{-b-\sqrt{b^2+4}}{2}}<0,\quad
\alpha^3\beta^3=-1. \] Here $\chi=\hat\chi$ is real and we choose
our branches so that $\hat\chi^{\frac{1}{3}}$ is also real.

The effect of choosing such a symmetric curve will be to reduce
the four period integrals $x_i$ to two independent integrals. The
tetrahedrally symmetric monopole is in the class (\ref{curve}). We
note that a general rotation will alter the form of $a_3(\zeta)$.
Thus the dimension of the moduli space is reduced from three by
the 3 degrees of freedom of the rotations yielding a discrete
space of solutions. We are seeking then a discrete family of
spectral curves.

We shall begin by calculating the period integrals, and then
imposing the Ercolani-Sinha constraints. We shall also consider
the geometry of the curves (\ref{curve}).

\subsection{The period integrals}
In terms of our Wellstein parameterization we are working with
$$
w^3=z^6+bz^3-1=(z^3-\alpha\sp3)(z^3+\frac{1}{\alpha\sp3})
$$
($1/\alpha\sp3=-\beta\sp3=(b+\sqrt{b^2+4})/{2}$). We choose the
first sheet so that
$w=\sqrt[3]{(z^3-\alpha\sp3)(z^3+{1}/{\alpha\sp3})}$ is negative
and real on the real $z$-axis between the branch points
$(-1/\alpha,\alpha)$.

Introduce integrals computed on the first sheet
\begin{align}\begin{split}
\mathcal{I}_1(\alpha)&=\int\limits_{0}^{\alpha}\frac{{d}z}{w}
=-\frac{2\pi\sqrt{3}\alpha}{9} {_2F_1}\left(\frac13,\frac13;1;-\alpha^6\right),\\
\mathcal{J}_1(\alpha)&=\int\limits_{0}^{\beta}\frac{{d}z}{w} =
\frac{2\pi\sqrt{3}}{9\alpha} {_2F_1}\left(\frac13,\frac13;1;
-\alpha^{-6}\right).\end{split} \label{integralsij}\end{align}
Here $_2F_1(a,b;c;z)$ is the standard Gauss hypergeometric
function and we have, for example, evaluated the first integral
using the substitution $z=\alpha t\sp{1/3}$ and our specification
of the first sheet. We also have that
\[ \int\limits_{0}^{\rho^k\alpha}\frac{{d}z}{w}=\rho^k\mathcal{I}_1(\alpha),
\quad
\int\limits_{0}^{\rho^k\beta}\frac{{d}z}{w}=\rho^k\mathcal{J}_1(\alpha),\quad
k=1,2. \]

Our aim is to express the periods for our homology basis
(\ref{homology}) in terms of the integrals $\mathcal{I}_1(\alpha)$
and $\mathcal{J}_1(\alpha)$. Consider for example
\begin{align*}
x_1&=\oint_{\mathfrak{a}_1}{d}u_1=
\int_{\gamma_1(\lambda_1,\lambda_2)}\frac{{d}z}{w}+
\int_{\gamma_2(\lambda_2,\lambda_1)}\frac{{d}z}{w}
=\int_{\lambda_1}\sp{\lambda_2}\frac{{d}z}{w}- \rho\sp2\,
\int_{\lambda_1}\sp{\lambda_2}\frac{{d}z}{w}\\
&=(1-\rho\sp2) \int_{\alpha}\sp{\rho\sp2\beta}\frac{{d}z}{w}
=(1-\rho\sp2)\left[-\mathcal{I}(\alpha)+
\int\limits_{0}^{\rho^2\beta}\frac{{d}z}{w}\right]
=(1-\rho\sp2)\left[-\mathcal{I}_1(\alpha)+
\rho^2\mathcal{J}_1(\alpha)\right]\\
&=-2\mathcal{I}_1(\alpha)-\mathcal{J}_1(\alpha)-\rho\left[
\mathcal{I}_1(\alpha)+2\mathcal{J}_1(\alpha)\right].
\end{align*}
Here we have used that on the second sheet $w_2=\rho w_1$ to
obtain the last expression of the first line, and also that
$1+\rho+\rho\sp2=0$ to obtain the final expression. Similarly we
find (upon dropping the $\alpha$ dependence from $\mathcal{I}_1$
and $\mathcal{J}_1$ when no confusion arises) that
\begin{equation}
\begin{array}{rlrl}
x_{1}&=-(2\mathcal{J}_1+\mathcal{I}_1)\rho -2\mathcal{I}_1
-\mathcal{J}_1, &x_{2}&=(\mathcal{J}_1- \mathcal{I}_1 )\rho+
\mathcal{I}_1+2\mathcal{J}_1,
\\
x_{3}&=(\mathcal{J}_1+2\mathcal{I}_1)\rho-\mathcal{J}_1+\mathcal{I}_1,
&x_{4}&=3(\mathcal{J}_1-\mathcal{I}_1)\rho+3\mathcal{J}_1.
\end{array}
\label{ij}
\end{equation}
 Note that
\begin{equation} x_2=\rho x_1,\quad x_3=\rho^2 x_1 .\label{ij1} \end{equation}

\subsection{The Ercolani-Sinha constraints}
We next reduce the Ercolani-Sinha constraints to a number
theoretic one. Using (\ref{expxx}) and (\ref{ij}) we may rewrite
the constraints as
\begin{equation}
    x_{i} = \xi( \epsilon_{i}n_{i} + \rho^{2}m_{i}) = (\alpha_{i}\mathcal{I}_1 +
    \beta_{i} \mathcal{J}_1) + (\gamma_{i}\mathcal{I}_1 + \delta_{i}
    \mathcal{J}_1)\rho .\label{esred}
\end{equation}
We may solve for the various $n_{i},m_{i}$ in terms of
$n_{1},m_{1}$ as follows. Set
$$C_{i}=\begin{pmatrix} \epsilon_{i} & -1 \\
    0 & -1 \end{pmatrix},\
D_{i}=\begin{pmatrix} \alpha_{i} & \beta_{i} \\
    \gamma_{i} &\delta_{i} \end{pmatrix},\
\hat{\mathcal{I}} = {\mathcal{I}_1}/{\xi},\ \hat{\mathcal{J}} =
{\mathcal{J}_1}/{\xi}.$$ Then (\ref{esred}) may be rewritten as
\begin{align*}
    C_{i}\, \begin{pmatrix} n_{i} \\ m_{i}
    \end{pmatrix}  = D_i \begin{pmatrix} \hat{\mathcal{I}} \\
    \hat{\mathcal{J}} \end{pmatrix}
\end{align*}
giving
\begin{align*}
\begin{pmatrix} n_{i} \\ m_{i}
    \end{pmatrix} & = C^{-1}_{i} D_{i} \begin{pmatrix}\hat{\mathcal{I}}
    \\ \hat{\mathcal{J}} \end{pmatrix} = C^{-1}_{i} D_{i} D^{-1}_{1} C_{1}
    \begin{pmatrix} n_{1} \\m_{1} \end{pmatrix}.
\end{align*}
This yields that the vectors $\boldsymbol{n}$, $\boldsymbol{m}$
are of the form
\begin{equation}\label{solvn}\boldsymbol{n}=
\begin{pmatrix} n_{1} \\ n_{2} \\ n_{3} \\n_{4}\end{pmatrix} =
\begin{pmatrix} n_{1} \\ m_{1}-n_{1} \\ -m_{1}
\\2n_{1}-m_{1}\end{pmatrix}, \qquad
\boldsymbol{m}=\begin{pmatrix} m_{1} \\ m_{2} \\ m_{3}
\\m_{4}\end{pmatrix} =
\begin{pmatrix} m_{1} \\ -n_{1} \\ n_{1}-m_{1}
\\-3n_{1}\end{pmatrix}.
\end{equation}
One may verify that for vectors of this form then
$(\boldsymbol{n},\boldsymbol{m})\mathcal{M}=-(\boldsymbol{n},\boldsymbol{m})$
as required by (\ref{cplxMT}). Recall further that
$(\boldsymbol{n},\boldsymbol{m})$ is to be a primitive vector:
that is one for which the greatest common divisor of the
components is 1, and hence a generator of $\mathbb{Z}\sp8$. We see
that $(\boldsymbol{n},\boldsymbol{m})$ is primitive if and only if
\begin{equation}(n_1,m_1)=1.\label{primcond}\end{equation}

From
\begin{equation*}
    \begin{pmatrix} \hat{\mathcal{I}} \\ \hat{\mathcal{J}} \end{pmatrix} = D^{-1}_{i}
    C_{i} \begin{pmatrix}n_{i} \\m_{i}\end{pmatrix} =
    \frac{1}{3}\begin{pmatrix}-2 & 1 \\ 1 & 1 \end{pmatrix}
    \begin{pmatrix}n_{1} \\ m_{1}\end{pmatrix}
\end{equation*}
we obtain
\begin{equation*}
\begin{matrix}
    \dfrac{\hat{\mathcal{I}}}{\hat{\mathcal{J}}}
    & = &\dfrac{\mathcal{I}}{\mathcal{J}} = \dfrac{m_{1} - 2n_{1}}{m_{1} +
    n_{1}}, \\
    \mathcal{I}_1 & =& \dfrac{m_{1} - 2n_{1}}{3} \, \xi \quad & = &- \dfrac{2
    \pi}{3\sqrt{3}} \ \alpha\ {_2F_1}(\frac{1}{3}, \frac{1}{3}; 1,
    -\alpha^{6}), \\
    \mathcal{J}_1 & =& \dfrac{m_{1} + n_{1}}{3} \, \xi \quad & = &\dfrac{2
    \pi}{3\sqrt{3}} \ \dfrac{1}{\alpha}\ {_2F_1}(\frac{1}{3}, \frac{1}{3}; 1,
    -\alpha^{-6}).
\end{matrix}
\end{equation*}
Now given (\ref{solvn})  we find that
\begin{equation*}
    \boldsymbol{n}.H \boldsymbol{n} - \boldsymbol{m} .
\boldsymbol{n} + \boldsymbol{m}.H\boldsymbol{m} = 2(m_{1} +
n_{1})(m_{1} -
    2n_{1})
\end{equation*}
and so the constraint (\ref{condition1}) is satisfied if
$$
    \boldsymbol{\bar x}\sp{T}H\boldsymbol{x}  = \xi^{2}[
    \boldsymbol{n}.H \boldsymbol{n} - \boldsymbol{m} .
\boldsymbol{n} + \boldsymbol{m}.H\boldsymbol{m}
    ]= 2\xi^{2}(m_{1} +
n_{1})(m_{1} -
    2n_{1})< 0.
    $$
This requires
\begin{equation}\label{esccnm}
(m_{1} + n_{1})(m_{1} -
    2n_{1})< 0.
\end{equation}
In particular we have from (\ref{esnu}) that
$$\xi=\frac{3\chi^{\frac{1}{3}}}{(n_{1} + m_{1})(m_{1} -
2n_{1})}.$$

Thus we have to solve
\begin{equation}
\begin{split}
    \mathcal{I}_1 &= \frac{\chi^{\frac{1}{3}}}{n_{1} + m_{1}}
= - \frac{2 \pi}{3
    \sqrt{3}}\ \alpha\ {_2F_1}(\frac{1}{3}, \frac{1}{3}; 1,
    -\alpha^{6}), \\
    \mathcal{J}_1 &= \frac{\chi^{\frac{1}{3}}}{m_{1} - 2n_{1}}
= \frac{2 \pi}{3
    \sqrt{3}}\ \frac{1}{\alpha}\ {_2F_1}(\frac{1}{3}, \frac{1}{3}; 1,
    -\alpha^{-6}).
\end{split}\label{escj}
\end{equation}

Using the identity
\begin{equation*}
{_2F_1}(\frac{1}{3}, \frac{1}{3}; 1,x)=(1-x)\sp{-{1}/{3}}\,
{_2F_1}(\frac{1}{3}, \frac{2}{3}; 1,\frac{x}{x-1})
\end{equation*}
we then seek solutions of
\begin{equation*}
\dfrac{\mathcal{I}_1}{\mathcal{J}_1} = \dfrac{m_{1} -
2n_{1}}{m_{1} +    n_{1}}=- \frac{{_2F_1}(\frac{1}{3},
\frac{2}{3}; 1,t)}{{_2F_1}(\frac{1}{3}, \frac{2}{3};
1,1-t)},\qquad t=\frac{\alpha\sp6}{1+\alpha\sp6}=
\frac{-b+\sqrt{b^2+4}}{2\sqrt{b^2+4}}.
\end{equation*}
From (\ref{esccnm}) the ratio of ${\mathcal{I}_1}/{\mathcal{J}_1}$
is negative. Consideration of the function
\[f(t)=\frac{{_2F_1}\left(\frac13,\frac23;1;t\right)}
{ {_2F_1}\left(\frac13,\frac23;1;1-t\right)} .
\]
(see Figure 2 for its plot) shows that there exists unique root
$t\in(0,1)$ for each value $f(t)\in(0,\infty)$ and correspondingly
a unique real positive $\alpha=\sqrt[6]{t/(1-t)}$.
\begin{figure}\label{figratio}
\epsfxsize=12cm \epsfysize=10cm \epsffile{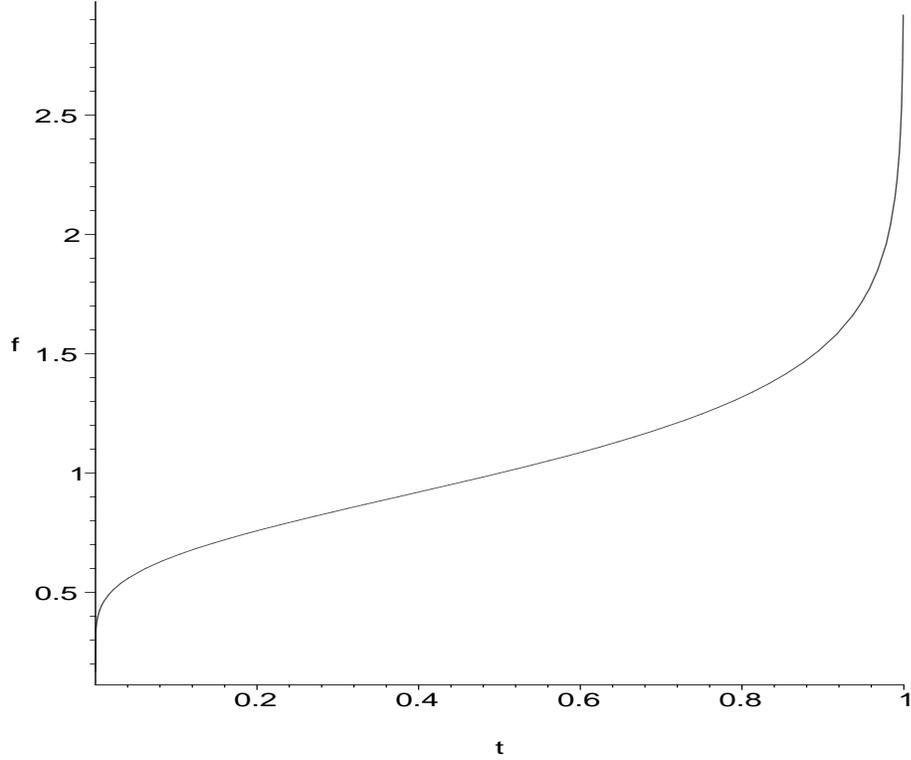} \caption{The
function $f(t)= \frac{{_2F_1}(\frac{1}{3}, \frac{2}{3};
1,t)}{{_2F_1}(\frac{1}{3}, \frac{2}{3}; 1,1-t)}$.}
\end{figure}

Bringing these results together we have established:
\begin{proposition}\label{propesnt}
To each pair of relatively prime integers
$(n_1,m_1)=1$ for which
$$(m_{1} + n_{1})(m_{1} -2n_{1})<0$$
we obtain a solution to the Ercolani-Sinha constraints for a curve
of the form (\ref{curve}) as follows. First we solve for $t$,
where
\begin{equation}\label{esct}
\dfrac{2n_{1}-m_{1}}{m_{1} + n_{1}}=\frac{{_2F_1}(\frac{1}{3},
\frac{2}{3}; 1,t)}{{_2F_1}(\frac{1}{3}, \frac{2}{3}; 1,1-t)}.
\end{equation}
Then
\begin{equation}
b=\frac{1-2t}{\sqrt{t(1-t)}},\qquad t=
\frac{-b+\sqrt{b^2+4}}{2\sqrt{b^2+4}},
\end{equation}
and we obtain $\chi$ from
\begin{equation}\label{esevch}
\chi^{\frac{1}{3}} = -(n_{1} + m_{1} )\, \frac{2 \pi}{3
    \sqrt{3}}\ \frac{\alpha}{(1+\alpha\sp6)\sp\frac13}\ {_2F_1}(\frac{1}{3}, \frac{2}{3}; 1,
    t)
\end{equation}
with $\alpha\sp6=t/(1-t)$.
\end{proposition}

\subsection{Ramanujan}
Thus far we have reduced the problem of finding an appropriate
monopole curve within the class (\ref{curve}) to that of solving
the transcendental equation (\ref{esct}) for which a unique
solution exists. Can this ever be solved apart from numerically?
Here we shall recount how a (recently proved) result of Ramanujan
enables us to find solutions.

Let $n$ be a natural number. A modular equation of degree $n$ and
signature $r$ ($r=2,3,4,6$) is a relation between $\alpha$,
$\beta$ of the form
\begin{equation}\label{modnr}
n\,\frac{_2F_1(\frac{1}{r},\frac{r-1}{r};1;1-\alpha)}
{_2F_1(\frac{1}{r},\frac{r-1}{r};1;\alpha)}=
\frac{_2F_1(\frac{1}{r},\frac{r-1}{r};1;1-\beta)}
{_2F_1(\frac{1}{r},\frac{r-1}{r};1;\beta)}. \end{equation} When
$r=2$ we have the complete elliptic integral
$\mathbf{K}(k)=\frac{\pi}{2}\, _2F_1(\frac12,\frac12;1;k\sp2)$ and
(\ref{modnr}) yields the usual modular relations. By interchanging
$\alpha\leftrightarrow\beta$ we may interchange $n\leftrightarrow
1/n$. This, together with iteration of these modular equations,
means we may obtain relations with $n$ being an arbitrary rational
number. Our equation (\ref{esct}) is precisely of this form for
signature $r=3$ and starting with say $\alpha=1/2$.

Ramanujan in his second notebook presents results pertaining to
these generalised modular equations and various theta function
identities. For example, if $n=2$ in signature $r=3$ then $\alpha$
and $\beta$ are related by
\begin{equation}\label{mod23}
    (\alpha\beta)\sp\frac{1}{3}+\left(
    (1-\alpha)(1-\beta)\right)\sp\frac{1}{3}=1.
\end{equation}
He also states that (for $0\le p<1$)
\begin{equation}\label{ramf23}
    (1+p+p^2)\, _2F_1\left(\frac12,\frac12;1,\frac{p^3(2+p)}{1+2p}\right)=
    \sqrt{1+2p}\;
    _2F_1\left(\frac13,\frac23;1,\frac{27p^2(1+p)^2}{4(1+p+p^2)^3}\right).
\end{equation}
Ramanujan's results were derived in \cite{bbg95} (see also
\cite{h98}), though some related to expansions of $1/\pi$ had been
obtained earlier by J.M. and P.B. Borwein \cite{BB87}. An account
of the history and the associated theory of these equations may be
found in the last volume dedicated to Ramanujan's notebooks
\cite{bv98}. The associated theory of these modular equations
presented in the accounts just cited is largely based on direct
verification that appropriate expressions of hypergeometric
functions satisfy the same differential equations and initial
conditions and so are equal: we shall present a more geometric
picture in due course.

Analogous expressions to (\ref{mod23}) are known for $n=3$, $5$,
$7$ and $11$ \cite[7.13, 7.17, 7.24, 2.28 respectively]{bv98}.
Thus by iteration we may solve (\ref{esct}) for rational numbers
whose numerator and denominator have these as their only factors.
We include some examples of these in the table below. Thus to get
the value $2$ for the ratio $(2n_1-m_1)/(m_1+n_1)$ we set
$\alpha=\frac12$ in (\ref{mod23}) and solve for
$$t\sp\frac13+(1-t)\sp\frac13=2\sp\frac13,$$
taking the larger value $t=\frac{1}{2}+\frac{5\sqrt{3}}{18}$ (the
smaller value yielding the ratio $\frac12$).
$$
\begin{array}{|c|c|c|c|c|} \hline
n_1&m_1&(2n_1-m_1)/(m_1+n_1)&t&b\\
\hline 2&1&1&\frac{1}{2}&0\\
\hline
1&0&{2}&\frac{1}{2}+\frac{5\sqrt{3}}{18}&5\sqrt{2}\\
\hline1&1&\frac{1}{2}&\frac{1}{2}-\frac{5\sqrt{3}}{18}&5\sqrt{2}\\
 \hline 4&-1&3&(63+171\sqrt[3]{2}-18\sqrt[3]{4})/250&
(44+38\sqrt[3]{2}+26\sqrt[3]{4})/3\\
\hline 5&-2&4&\frac{1}{2}+\frac{153\sqrt{3}-99\sqrt{2}}{250}&
9\sqrt{458+187\sqrt{6}}\\
\hline
\end{array}
$$

A theory exists then for solving (\ref{modnr}) and this has been
worked out for various low primes. These results enable us to
reduce the Ercolani-Sinha conditions (\ref{esct}) to solving an
algebraic equation.

\subsection{Covers of the sextic} We shall now describe some
geometry underlying our curves (\ref{curve}) which will lead to an
understanding of the results of the last section. We shall first
present a more computational approach, useful in actual
calculations, and then follow this with a more invariant
discussion. We begin with the observation that our curves each
cover four elliptic curves.

\begin{lemma}\label{coverlem}
The curve $\mathcal{C}:= \{(x,y)| y^3+x^6+bx^3-1=0\}$ with
arbitrary value of the parameter $b$ is a simultaneous covering of
the four elliptic curves $\mathcal{E}_{\pm}$, $\mathcal{E}_{1,2}$
as indicated in the diagram, where $\mathcal{C}\sp*$ is an
intermediate genus two curve:

\vskip 0.5cm
\begin{center}
\unitlength=1mm
\begin{picture}(60,30)(0,0)
\put(9,23){\makebox(0,0){${\mathcal{C}}=(x,y)$}}
\put(6,20){\vector(-4,-1){20}} \put(-8,20){\makebox(0,0){$\pi^*$}}
\put(-18,15){\makebox(0,0){$\mathcal{C}^*$}}
\put(-26,0){\makebox(0,0){$\mathcal{E}_+=(z_+,w_+)$}}
\put(-20,12){\vector(-1,-1){9}} \put(-16,12){\vector(1,-1){9}}
\put(0,0){\makebox(0,0){$\mathcal{E}_-=(z_-,w_-)$}}
\put(-9,10){\makebox(0,0){$\pi_-$}}
\put(-26,10){\makebox(0,0){$\pi_+$}}
\put(10,19){\vector(1,-1){16}} \put(16,10){\makebox(0,0){$\pi_1$}}
\put(28,0){\makebox(0,0){$\mathcal{E}_1=(z_1,w_1)$}}
\put(12,20){\vector(2,-1){34}} \put(38,10){\makebox(0,0){$\pi_2$}}
\put(55,0){\makebox(0,0){$\mathcal{E}_2=(z_2,w_2)$}}
\end{picture}
\end{center}
\vskip 1cm

The equations of the elliptic curves are
\begin{align}
\mathcal{E}_{\pm}&= \{ (z_{\pm},w_{\pm})|
w_{\pm}^2=z_{\pm}(1-z_{\pm})
(1-k_{\pm}^2z_{\pm})   \},\label{curvepm}\\
\mathcal{E}_1&= \{ (z_1,w_1)
|z_1^3+{w_1}^3+3z_1+b=0\},\label{curve1}\\
\mathcal{E}_2&= \{ (z_2,w_2) | {w_2}^3+{z_2}^2+bz_2-1 =0  \},
\label{curve2}
\end{align}
where the Jacobi moduli, $k_{\pm}$ are given by
\begin{equation}
k_{\pm}^2=-\frac{\rho(\rho M \pm 1)(\rho M \mp 1)^3  } {(M\pm
1)(M\mp 1 )^3}
\end{equation}
with
\begin{equation} M=\frac{K}{L},\quad K=(2\imath -b)^{\frac13},\quad
L=(b^2+4)^{\frac16}. \label{KLM}  \end{equation}

The covers $\pi_{\pm},\pi_{1,2}$ are given by
\begin{align}
\pi_{\pm}:\quad\begin{split} z_{\pm}&=-\frac{K^2-L^2}{K^2-\rho
L^2}\,
\frac{Kx-y}{\rho Kx-y}\,\frac{L^2x-Ky}{L^2x-K\rho y},\\
w_{\pm}&=\imath\sqrt{2+\rho}\sqrt{\frac{L\pm K}{L\mp K}}
\frac{K^2}{L} \frac{L^2-\rho K^2}{\rho L^2-K^2}\frac{(Lx\mp
y)(x^6+1)} {(\rho K x- y)^2(L^2x-\rho K y)^2  }
\end{split}\label{pipm}
\end{align}
and
\begin{align*}
\pi_1:&\quad z_1=x-\frac{1}{x},\quad w_1=\frac{y}{x},\\
\pi_2:&\quad z_2=x^3,\quad w_2=y.
\end{align*}

The elliptic curves $\mathcal{E}_{1,2}$ are equianharmonic
($g_2=0$) and consequently have vanishing j-invariant,
$j\left(\mathcal{E}_{1,2}\right)=0$.
\end{lemma}

\begin{proof}
The derivation of the covers $\pi_{1,2}$ and the underlying curves
is straightforward. The pullbacks $\pi_{1,2}\sp{-1}$ of these
covers are
\begin{align*}
\pi_1^{-1}:= \begin{cases}   x=(z_1
\pm\sqrt{{z_1}^2+4})/2\\
                            y=w_1(z_1
\pm\sqrt{{z_1}^2+4})/2\end{cases}\qquad
 \pi_2^{-1}:= \begin{cases}
x=\rho\sqrt[3]{z_2}\\
                               y=w_2\end{cases}
\end{align*}
showing that the degrees of the cover are 2 and 3 respectively. A
direct calculation putting these elliptic curves into Weierstrass
form shows $g_2=0$ and hence the  elliptic curves
$\mathcal{E}_{1,2}$ are equianharmonic. Their j-invariants are
therefore vanishing and $\mathcal{E}_{1,2}$ are birationally
equivalent.

To derive the covers $\pi_{\pm}$ we first note that the curve
$\mathcal{C}$ is a covering of the hyperelliptic  curve
$\mathcal{C}^{\ast}$ of genus two,
\begin{equation}
\mathcal{C}^{\ast}= \{ (\mu,\nu)| \nu^2=(\mu^3+b)^2+4   \}.
\label{hcurve}
\end{equation}
The cover of this curve is given by the formulae
\begin{equation}
\pi^{\ast}:\quad \mu=\frac{y}{x},\quad
\nu=-x^3-\frac{1}{x^3}.\label{piast}
\end{equation}

The curve $\mathcal{C}^{\ast}$ covers two-sheetedly the two
elliptic curves $\mathcal{E}_{\pm}$ given in (\ref{curvepm})
\begin{align}\begin{split}
  z_{\pm}&=\frac{K^2-L^2}{K^2-\rho L^2}\,
\frac{K-\mu}{\rho K-\mu}\,\frac{L^2-K\mu}{L^2-K\rho \mu},\\
w_{\pm}&=-\imath\sqrt{2+\rho}\sqrt{\frac{L\pm K}{L\mp K}}
\frac{K^2}{L} \frac{L^2-\rho K^2}{\rho L^2-K^2}\frac{\nu(L\mp
\mu)} {(\mu-\rho K)^2(L^2-\rho K \mu)^2
}.\end{split}\label{cover2a}
\end{align}
Composition of (\ref{piast})  and (\ref{cover2a}) leads to
(\ref{pipm}).
\end{proof}

Using these formulae direct calculation then yields
\begin{corollary}
The holomorphic differentials of $\mathcal{C}$ are mapped to
holomorphic differentials of $\mathcal{E}_{\pm}$,
$\mathcal{E}_{1,2}$ as follows
\begin{align}\begin{split}
\frac{{d} z_{\pm}}{w_{\pm}}&=\sqrt{1+2\rho}\frac{L}{K}
\sqrt{(L\pm K)(L\mp K)^3}\,\frac{Lx\pm y}{y^2}\,{d}x,\end{split}
\label{invdiffpma}\\
&=\sqrt{1+2\rho}\frac{L}{K} \sqrt{(L\pm K)(L\mp K)^3}\,
(L\pm \mu)\,\frac{{d} \mu}{\nu}  \nonumber\\
\frac{{d} z_1}
{w_1^2}&=\frac{x^2+1}{y^2}{d}x, \label{invdiff1}\\
\frac{{d}z_2 }{w_2^2}&=\frac{3x^2}{y^2}{d}x, \label{invdiff2}
\end{align}
where $L,K$ are given in (\ref{KLM}).
\end{corollary}

The absolute invariants $j_{\pm}$ of the curves
$\mathcal{E}_{\pm}$ are
\begin{equation}
j_{\pm}=108\,{\frac {{L}^{3} \left( 5\,{L}^{3}\mp 4\,b \right)
^{3}}{ \left( {L}^ {3}\pm b \right) ^{2} }} .\label{invariants}
\end{equation}
Evidently $j_{\pm}\neq 0$ in general, as well $j_+\neq j_-$;
therefore these elliptic curves are not birationally equivalent to
that one appearing in Hitchin's theory of the tetrahedral monopole
which is equianharmonic \cite{hmm95}. We observe that the
substitution
\[ M=\frac{1+2\rho+p}{1+2\rho-p}  \]
leads to the
parameterisation of Jacobi moduli being
\begin{equation} k_+^2=\frac{(p+1)^3(3-p)}{16p},\quad
                  k_-^2=\frac{(p+1)(3-p)^3}{16p^3},
\end{equation}
which Ramanujan used in his hypergeometric relations of signature
3, see e.g. \cite{bbg95}. The $\theta$-functional representation
of the moduli $k_{\pm}$ and parameter $p$ can be found in
\cite[Section 9.7]{lawd89},
\[  k_+=\frac{\vartheta_2^2(0|\tau) }{\vartheta_3^2(0|\tau)},\quad
   k_-=\frac{\vartheta_2^2(0|3\tau) }{\vartheta_3^2(0|3\tau)},\quad
  p=\frac{3\vartheta_3^2(0|3\tau) }{\vartheta_3^2(0|\tau)}.
  \]

We shall now describe the geometry of the covers we have just
presented explicitly. Our curve has several explicit symmetries
which lie behind the covers described. We will first describe
these symmetries acting on the field of functions $\mathfrak{k}$
of our curve as this field does not depend on whether we have a
singular or nonsingular model of the curve; we will subsequently
give a projective model for these, typically working in weighted
projective spaces where the curves will be nonsingular.

Viewing $\bar y=y/x$ and $x$ as functions on $\mathcal{C}$ we see
that
$${\bar y}^3=x^3+b-\frac{1}{x^3}$$
has symmetries ($\rho=e\sp{2\imath\pi /3}$)
\begin{align*}\mathrm{a}:\ &
x\rightarrow  x,\quad {\bar y}\rightarrow \rho{\bar y},\\
\mathrm{b}:\ &
x\rightarrow \rho x,\quad {\bar y}\rightarrow {\bar y},\\
\mathrm{c}:\ & x\rightarrow  -1/x,\quad {\bar y}\rightarrow {\bar
y}.
\end{align*}
Together these yield the group $G=C_3\times S_3$, with
$C_3=<\mathrm{a}|\,\mathrm{a}^3=1>$ and
$S_3=<\mathrm{b},\mathrm{c}|\,\mathrm{b}^3=1,\mathrm{c}^2=1,\mathrm{cbc}=\mathrm{b}^2>$.
When $b=5\sqrt{2}$, the dihedral symmetry $S_3$ is enlarged to
tetrahedral symmetry by
$$ \mathrm{t}: x\rightarrow \frac{\sqrt{2}-x}{1+\sqrt{2}x},\quad \bar y\rightarrow
\frac{3 x \bar y}{(1+\sqrt{2}x)(x-\sqrt{2})},\qquad
\mathrm{t}^2=1,$$ with $A_4$ being generated by $\mathrm{b}$ and
$\mathrm{t}$. Now to each subgroup $H\leq G$ we have the fixed
field $\mathfrak{k}\sp{H}$ associated to the quotient curve
$\mathcal{C}/H$.

The canonical curve of a non-hyperelliptic curve of genus 4 is
given by the intersection of an irreducible quadric and cubic
surface in $\mathbb{P}\sp3$. In our case the quadric is in fact a
cone and we may represent our curve $\mathcal{C}$  as the
nonsingular curve\footnote{Had we represented
$\mathcal{C}\subset\mathbb{P}\sp2$ as the plane curve given by the
vanishing of $z^6+b\, z^3 t^3 -t^6-w^3 t^3$  the curve is
singular. When $b$ is real the point $[z,t,w]=[0,0,1]$ is the only
singular point of $\mathcal{C}$ with delta invariant $6$ and
multiplicity $3$ yielding $g_\mathcal{C}=4$.} in the weighted
projective space $\mathbb{P}\sp{1,1,2}=\{[z,t,w]\,|\,[z,t,w]\sim
[\lambda z,\lambda t, \lambda\sp2 w]\}$ given by the vanishing of
$$f(z,t,w)=z^6+b\, z^3 t^3 -t^6-w^3 .$$
The group $G$ acts on this as ($x=z/t$, $\bar y=w/(zt)$)
\begin{align*}\mathrm{a}:\ &
[z,t,w]\rightarrow  [z,t,\rho w]\sim [\rho z,\rho t,w],\\
\mathrm{b}:\ & [z,t,w]\rightarrow [\rho z,t,\rho w]\sim [\rho\sp2 z,\rho t, w] ,\\
\mathrm{c}:\ & [z,t,w]\rightarrow [t,-z,-w]\sim [\imath t, -\imath
z,w].
\end{align*}
The fixed points of these actions on $\mathcal{C}$ and quotient
curves are as follows:
\begin{description}\item[$\mathrm{a}$]
There are 6 fixed points, $[1,\rho\sp{k}\,\alpha_\pm,0]$, where
$\alpha_\pm$ are the two roots of $\alpha\sp2-b\alpha-1=0$. For
other points we have a $3:1$ map
$\mathcal{C}\rightarrow\mathcal{C}/<\mathrm{a}>$. An application
of the Riemann-Hurwitz theorem shows the genus of
$\mathcal{C}/<\mathrm{a}>$ to be $g_{\mathcal{C}/<\mathrm{a}>}=0$.
\item[$\mathrm{b}$]
The has no fixed points and an application of the Riemann-Hurwitz
theorem shows the genus of $\mathcal{C}/<\mathrm{b}>$ to be
$g_{\mathcal{C}/<\mathrm{b}>}=2$.
\item[$\mathrm{c}$]There are 6 fixed points, $[1,\pm\imath,
\rho\sp{k}\,\beta_\pm]$, where $\beta_\pm$ is a root of
$\beta_\pm\sp3=2\pm\imath b$. Here the Riemann-Hurwitz theorem
shows the genus of $\mathcal{C}/<\mathrm{c}>$ to be
$g_{\mathcal{C}/<\mathrm{c}>}=1$.
\end{description}

By using the invariants of $H$ we may obtain nonsingular
projective models of $\mathfrak{k}\sp{H}$. Take for example
$H=<\mathrm{c}>$ with invariants $u=zt$, $v=z^2-t^2$ and $w$ (in
degree $2$). Then we obtain the quotient curve $w^3=v^3+3u^2 v+b
u^3$ in
$\mathbb{P}\sp{2,2,2}\sim\mathbb{P}\sp{1,1,1}=\{[u,v,w]\}$. The
genus of the quotient is seen to be $1$. We recognize this as the
curve $\mathcal{E}_1$. One verifies that
$$\mathrm{c}\sp*\left(\frac{x^2+1}{y^2}\,dx\right)=\frac{x^2+1}{y^2}\,dx$$
giving us the invariant differential (\ref{invdiff1}). Similarly,
by taking $H=<\mathrm{b}\mathrm{ c}>$ and
$H=<\mathrm{b}^2\mathrm{c}>$, we also obtain equianharmonic
elliptic curves. The invariants of the involution
$\mathrm{b}^2\mathrm{c}$ are again all in degree $2$ and now are
$u=zt$, $v=\rho\sp{1/2}z^2-\rho\sp{-1/2}\,t^2$ and $w$.

By taking $H=<\mathrm{a}^2\mathrm{b}>$ we may identify
$\mathcal{E}_2$. The invariant of
$<\mathrm{a}^2\mathrm{b}>:[z,t,w]\rightarrow[\rho z,t, w]$ is
$u=z^3$ and the curve $w^3=u^2-but^3+t^6$ in
$\mathbb{P}\sp{3,1,2}=\{[u,t,w]\}$.Using the formula for the genus
of a smooth curve of degree $d$ in $\mathbb{P}\sp{a_0,a_1,a_2}$,
$$g=\frac12\left(\frac{d\sp2}{a_0a_1a_2}-d\sum_{i<j}\frac{\textrm{gcd}(a_i,a_j)}{a_i a_j}+
\sum_{i=0}\sp{2}\frac{\textrm{gcd}(a_i,d)}{a_i}-1 \right),$$ the
genus is seen to be $1$. Now (\ref{invdiff2}) is the invariant
differential for this action. If we had taken $H=<\mathrm{a}>$
with invariants $u=z^3$, $v=t^3$ and $w$ we obtain the curve
$w^3=u^2+buv-v^2$ in $\mathbb{P}\sp{3,3,2}$ (which is equivalent
to $W=u^2+buv-v^2$ in $\mathbb{P}\sp{1,1,2}$). The genus of this
quotient is seen to be $0$.

We obtain the genus $2$ curve $\mathcal{C}^{\ast}$ as follows. The
invariants of $H=<\mathrm{b}>$ are $U=zt$, $V=z^3$, $T=t^3$ and
$w$, subject to the relation $U^3=VT$. The curve $\mathcal{C}$ may
be written $T^2=-w^3+bU^3+V^2$, and hence
$U^6=V^2T^2=V^2(-w^3+bU^3+V^2)$. This curve has genus 2 in
$\mathbb{P}\sp{2,3,2}=\{[U,V,w]\}$ and may be identified with
$\mathcal{C}^{\ast}$. Be setting $\nu=2V^2-(w^3-bU^3)$ this curve
takes the form $$\nu^2=(w^3-bU^3)^2 +4U^6$$ in
$\mathbb{P}\sp{1,3,1}=\{[U,\nu,w]\}$ and the identification with
$\mathcal{C}^{\ast}$ in the affine chart of earlier is given by
$\mu=-w$, $U=1$. In this latter form we find that the action of
$\mathrm{c}$ is given by $[U,\nu,w]\rightarrow [-U,\nu,-w]\sim
[U,-\nu,w]$ which is the hyperelliptic involution; further
quotienting yields a genus 0 curve.

The remaining genus 1 curves $\mathcal{E}_\pm$ are identified with
the quotients of $\mathcal{C}^{\ast}$ by $U\rightarrow \pm
w/\sqrt[6]{4+b^2}$, $w\rightarrow\pm\sqrt[6]{4+b^2}U$,
$\nu\rightarrow\nu$. This action has invariants $A=Uw$ (in degree
2), $B=w\pm\sqrt[6]{4+b^2}U$ (in degree 1), and $\nu$ (in degree
3). The resulting degree 6 curve is
$$\nu^2=B^6\mp 6 L A B^4+9 L^2 A^2 B^2\mp2L^3A^3-2b A^3,$$
where, as previously, $L=\sqrt[6]{4+b^2}$. These curves have genus
$1$ in $\mathbb{P}\sp{2,1,3}=\{[A,B,\nu]\}$. To complete the
identification with $\mathcal{E}_\pm$ we compute the j-invariants
of these curves. In the affine patch with $B\ne0$ which looks like
$\mathbb{C}\sp2$ (the other affine patches have orbifold
singularities and hence this choice) the curve takes the form
$$Y^2=1\mp 6 L X+9 L^2 X^2-2(b\pm L^3) X^3.$$
The j-invariants of these curves agree with (\ref{invariants}) and
hence the identifications as stated.

Both the differentials ${d}x/y$ and $x\,{d}x/y^2$ are invariant
under $\mathrm{b}$. These may be obtained by linear combinations
of ${d}z_\pm/w_\pm$ (\ref{invdiffpma}). The latter differentials
are those invariant under the symmetry of (\ref{hcurve})
$$\mu\rightarrow \frac{L\sp2}{\mu},\qquad \nu\rightarrow
\pm \frac{L\sp3 \nu}{\mu\sp3},
$$
which yield the quotients $\mathcal{E}_\pm$. A birational
transformation makes this symmetry more manifest\footnote{We thank
Chris Eilbeck for this observation.}. Let
$$T=\frac{L+\mu}{L-\mu},\ S=\frac{8\nu}{(L-\mu)\sp3},\qquad
\mu=L\,\frac{T-1}{T+1},\ \nu=\frac{L^3 S}{(T+1)^3}.$$ Then
(\ref{hcurve}) transforms to
$$S^2=(T-1)^6+2\frac{b}{L^3}(T^2-1)^3+(T+1)^6$$
which is manifestly invariant under $T\rightarrow-T$,
$S\rightarrow\mp S$. The substitution $W=T^2$ reduces the
canonical differentials ${d}T/S$ and $T\,{d}T/S^2$ to the
canonical differentials the elliptic curves
\begin{align*}
\mathcal{E}_+:&\quad
S^2=2(1+\frac{b}{L^3})W^3+6(5-\frac{b}{L^3})W^2+6(5+\frac{b}{L^3})W+
2(1-\frac{b}{L^3})
,\\
\mathcal{E}_-:&\quad
S^2=2(1+\frac{b}{L^3})W^4+6(5-\frac{b}{L^3})W^3+6(5+\frac{b}{L^3})W^2+
2(1-\frac{b}{L^3})W,
\end{align*}
which correspond to our earlier parameterizations.

\subsection{Role of the higher Goursat hypergeometric identities}
We have seen that complete Abelian integrals of the curve
$\mathcal{C}$ (\ref{bren03}) are given by hypergeometric
functions. The same is true for the various curves given in lemma
\ref{coverlem} covered by $\mathcal{C}$. Relating the periods of
$\mathcal{C}$ and the curves it covers leads to various relations
between hypergeometric functions, and this underlies the higher
hypergeometric identities of Goursat \cite{goursat81}. Goursat
gave detailed tables of transformations of hypergeometric
functions up to order four that will be enough for our purposes.

The simplest example of this is the cover
$\pi:\mathcal{C}\rightarrow\mathcal{C}\sp*$ for which
$\pi\sp*(\mu\,d\mu/\nu)=dx/y$ and $\pi\sp*(d\mu/\nu)=x\,dx/y^2$.
One then finds for example that
\begin{equation}\label{gour1}
\int_0\sp\alpha\frac{dx}{y}=\int_0\sp\infty\frac{\mu\,d\mu}{\nu},
\end{equation}
where both $y$ and $\nu$ are evaluated on the first sheet. A
change of variable shows that
$$\int_0\sp\infty\frac{\mu\,d\mu}{\nu}=\frac{2\pi}{3\sqrt{3}}\,
(b-2\imath)\sp{-1/3}\,{_2F_1}\left(\frac12,\frac13;1;\frac{4\imath}{2\imath-b}\right).$$
Now the left-hand side of equation (\ref{gour1}) is
$-\mathcal{I}_1$ (the minus sign arising when we go to Wellstein
variables $y\rightarrow -w$) and this has been evaluated in
(\ref{integralsij}). Comparison of these two representations
yields the hypergeometric equality
\[ F\left(\frac12,\frac13;1;\frac{4\imath}{2\imath-b}\right)=
      \left( \frac{2(b-2\imath)}{b+\sqrt{b^2+4}}   \right)^{\frac13}
F\left( \frac13,\frac13;1;\frac{b-\sqrt{b^2+4}}{b+\sqrt{b^2+4}}
\right) ,  \] which is one of Goursat's quadratic equalities
\cite{goursat81}; see also \cite[Sect. 2.11, Eq. (31)]{ba55}.
Further identities ensue from the coverings
$\mathcal{C}\rightarrow\mathcal{E}_\pm$ and we shall describe
these as needed below.

We remark that the curve (\ref{hcurve}) already appeared in
Hutchinson's study \cite{hut02} of automorphic functions
associated with singular, genus two, trigonal curves in which he
developed earlier investigations of Burkhardt \cite{bur93}.  These
results were employed by Grava and one of the authors \cite{eg04}
to solve the Riemann-Hilbert problem and associated Schlesinger
system for certain class of curves with $Z_N$-symmetry.

\subsection{Weierstrass reduction}
It is possible for the theta functions associated to a period
matrix $\tau$ to simplify (or admit reduction) and be expressible
in terms of lower dimensional theta functions. Such happens when
the curve covers a curve of lower genus, but it may also occur
without there being a covering. Reduction may be described purely
in terms of the Riemann matrix of periods (see \cite{ma92a}; for
more recent expositions and applications see
\cite{be02a},\cite{be02b}).
A $2g\times g$ Riemann matrix $\Pi=\left(%
\begin{array}{c}
  \mathcal{A} \\
  \mathcal{B} \\
\end{array}%
\right)$ is said to admit \emph{reduction} if there exists a
$g\times g_1$ matrix of complex numbers $\lambda$ of maximal rank,
a $2g_1\times g_1$ matrix of complex numbers $\Pi_1$ and a
$2g\times 2g_1$ matrix of integers $M$ also of maximal rank such
that
\begin{equation} \Pi \lambda  = \Pi_1 M, \end{equation}
where  $1\leq g_1 < g$. When a Riemann matrix admits reduction the
corresponding period matrix may be put in the form
\begin{equation}\label{redpergen}
    \tau=\left(%
\begin{array}{cc}
  \tau_1 & Q\\
  Q\sp{T} & \tau\sp{\#} \\
\end{array}%
\right),
\end{equation}
where $Q$ is a $g_1\times(g-g_1)$ matrix with rational entries and
the matrices $\tau_1$ and $\tau\sp{\#}$ have the properties of
period matrices. Because $Q$ here has rational entries there
exists a diagonal $(g-g_1)\times(g-g_1)$ matrix
$D=\diag(d_1,\ldots,d_{g-g_1})$ with positive integer entries for
which $(QD)_{jk}\in\mathbb{Z}$. With
$(z,w)=(z_1,\ldots,z_{g_1},w_1,\ldots,w_{g-g_1})$ the theta
function associated with $\tau$ may then be expressed in terms of
lower dimensional theta functions as
\begin{equation}\label{redthetagen}
\theta((z,w);\tau)=\sum_{\substack{
\mathbf{m}=(m_1,\ldots,m_{g-g_1})\\
0\le m_i\le d_i-1}}
\theta(z+Q\mathbf{m};\tau_1)\,\theta\left[%
\begin{array}{c}
  D\sp{-1}\mathbf{m} \\
  0 \\
\end{array}%
\right](Dw;D\tau\sp{\#}D).
\end{equation}

Our curve admits many reductions. Of itself this just means that
the theta functions may be reduced to theta functions of fewer
variables. It is only when the Ercolani-Sinha vector
correspondingly reduces that we obtain real simplification. In the
remainder of this section we shall describe these reductions and
later see how dramatic simplifications occur.

First let us describe the Riemann matrix of periods. We may
evaluate the remaining period integrals as follows. Let
\begin{align*}
\int\limits_0^{\alpha}{d}u_i=\mathcal{I}_i(\alpha),\quad
\int\limits_0^{\beta}{d}u_i=\mathcal{J}_i(\alpha),\quad
i=1,\ldots,4.
\end{align*}
Then for $k=1,2$ we have that
\begin{align*}
\int\limits_0^{\rho^k\alpha}{d}u_{1,2}&=\rho^k\mathcal{I}_{1,2}(\alpha),\quad
\int\limits_0^{\rho^k\beta}{d}u_{1,2}=\rho^k\mathcal{J}_{1,2}
(\alpha),\\
\int\limits_0^{\rho^k\alpha}{d}u_{3}&=\rho^{2k}\mathcal{I}_{3}(\alpha),\quad
\int\limits_0^{\rho^k\beta}{d}u_{3}=\rho^{2k}\mathcal{J}_{3}
(\alpha),\\
 \int\limits_0^{\rho^k\alpha}{d}u_{4}&=\mathcal{I}_{4}(\alpha),\quad\quad
\int\limits_0^{\rho^k\beta}{d}u_{4}=\mathcal{J}_{4} (\alpha),
\end{align*}
where it is again supposed that the integrals  $\mathcal{I}_*$ and
$\mathcal{J}_*$ are computed on the first sheet. We have already
computed $\mathcal{I}_1(\alpha)$ and $\mathcal{J}_1(\alpha)$. The
integrals $\mathcal{I}_*$ and $\mathcal{J}_*$ are found to be
\begin{align*}
\mathcal{I}_1(\alpha)&=
-\frac{2\pi\alpha}{3\sqrt{3}}\,{_2F_1}\left(\frac13,\frac13;1;-\alpha^6\right)
=-\frac{2\pi}{3\sqrt{3}}\,\frac{\alpha}{(1+\alpha^6)\sp\frac13}
\ {_2F_1}(\frac{1}{3}, \frac{2}{3}; 1,t),\\
\mathcal{J}_1(\alpha)&=
\frac{2\pi}{3\sqrt{3}\alpha}\,{_2F_1}\left(\frac13,\frac13;1;-\frac{1}{\alpha^{6}}
\right)=\frac{2\pi}{3\sqrt{3}}\,\frac{\alpha}{(1+\alpha^6)\sp\frac13}
\ {_2F_1}(\frac{1}{3}, \frac{2}{3}; 1,1-t),\\
\mathcal{I}_2(\alpha)&= \frac{4\pi^2}{9\Gamma\left(
\frac23\right)^3}\frac{\alpha}{(1+\alpha^6)^{\frac13}},\\
\mathcal{J}_2(\alpha)&=-\frac{4\pi^2}{9\Gamma\left(
\frac23\right)^3}\frac{\alpha}{(1+\alpha^6)^{\frac13}},\\
\mathcal{I}_3(\alpha)&=
\frac{2\pi\alpha^2}{3\sqrt{3}}\,{_2F_1}\left(\frac23,\frac23;1;-\alpha^6\right)
=\frac{2\pi}{3\sqrt{3}}\,\frac{\alpha^2}{(1+\alpha^6)\sp\frac23}
\ {_2F_1}(\frac{1}{3}, \frac{2}{3}; 1,t),\\
\mathcal{J}_3(\alpha)&=
\frac{2\pi}{3\sqrt{3}\alpha^2}{_2F_1}\left(\frac23,\frac23;1;-\frac{1}{\alpha^{6}}
\right)=\frac{2\pi}{3\sqrt{3}}\,\frac{\alpha^2}{(1+\alpha^6)\sp\frac23}
\ {_2F_1}(\frac{1}{3}, \frac{2}{3}; 1,1-t),\\
\mathcal{I}_4(\alpha)&= \alpha^3\, {_2F_1}\left(\frac23,1;\frac43;
-\alpha^6\right),\\
\mathcal{J}_4(\alpha)&=-\frac{1}{\alpha^3}\,
{_2F_1}\left(\frac23,1;\frac43;-\frac{1}{\alpha^{6}} \right),
\end{align*}
with $t=\alpha^6/(1+\alpha^6)$.

We observe that the relations
\begin{equation}\mathcal{R} \equiv\frac{\mathcal{I}_1(\alpha)}{\mathcal{J}_1(\alpha) }
=-\frac{\mathcal{I}_3(\alpha)}{\mathcal{J}_3(\alpha) }, \qquad
\mathcal{I}_2(\alpha)+\mathcal{J}_2(\alpha)=0, \qquad
\mathcal{I}_4(\alpha)-\mathcal{J}_4(\alpha)=\mathcal{I}_2(\alpha),
\label{relation13}\end{equation} follow from the above formulae.

The vectors $\boldsymbol{x},\ldots,\boldsymbol{d}$ are
\begin{equation}
\begin{split}
\boldsymbol{x}&=\left(\begin{array}{c}
 - (2\mathcal{J}_1+\mathcal{I}_1)\rho-2\mathcal{I}_1-\mathcal{J}_1\\
  (\mathcal{J}_1-\mathcal{I}_1)\rho+\mathcal{I}_1+2\mathcal{J}_1\\
 (\mathcal{J}_1+2\mathcal{I}_1)\rho+\mathcal{I}_1-\mathcal{J}_1\\
  3(\mathcal{J}_1-\mathcal{I}_1)\rho+3\mathcal{J}_1
\end{array}\right),\quad \boldsymbol{b}=\mathcal{I}_2
\left(\begin{array}{c} 1+2\rho\\-2-\rho\\1-\rho\\0
\end{array}\right),\\
\boldsymbol{c}&=\left(\begin{array}{c}
 (\mathcal{I}_3+2\mathcal{J}_3)\rho+\mathcal{J}_3-\mathcal{I}_3\\
 (\mathcal{I}_3-\mathcal{J}_3)\rho+\mathcal{J}_3+2\mathcal{I}_3\\
 - (2\mathcal{I}_3+\mathcal{J}_3)\rho-2\mathcal{J}_3-\mathcal{I}_3\\
 3(\mathcal{I}_3-\mathcal{J}_3)\rho+3\mathcal{I}_3
\end{array}    \right),\quad \boldsymbol{d}
=(\rho-1)\mathcal{I}_2 \left(\begin{array}{c} 1\\1\\1\\0
\end{array}\right).
\end{split}
\end{equation}

One may can easily check that
\[ \boldsymbol{x}\sp{T}H\boldsymbol{b}= \boldsymbol{x}\sp{T} H\boldsymbol{c}=
   \boldsymbol{x}\sp{T}H\boldsymbol{d}=0 .  \]
We then have that
\begin{align}\begin{split}
\mathcal{A}&= \left(\begin{array}{cccc}
-1-2\rho-(2+\rho)\mathcal{R}&1+2\rho&1+2\rho+(1-\rho)\mathcal{R}&-1+\rho\\
2+\rho+(1-\rho)\mathcal{R}&-2-\rho&1-\rho-(2+\rho)\mathcal{R}&-1+\rho\\
-1+\rho+(1+2\rho)\mathcal{R}&1-\rho&-2-\rho+(1+2\rho)\mathcal{R}&-1+\rho\\
3+3\rho-3\rho\mathcal{R}&0&-3\rho-3(1+\rho)\mathcal{R}&0
\end{array}\right)\left(\begin{array}{cccc} \mathcal{J}_1&&&\\
                           &\mathcal{I}_2\\
                           &&\mathcal{J}_3\\
                           &&&\mathcal{I}_2
\end{array}\right),\\
\mathcal{B}&=\left(\begin{array}{cccc}
2+\rho+(1-\rho)\mathcal{R}&1-\rho&1-\rho-(2+\rho)\mathcal{R}&2+\rho\\
-1+\rho+(1+2\rho)\mathcal{R}&1+2\rho&-2-\rho+(1+2\rho)\mathcal{R}&2+\rho\\
-1-2\rho-(2+\rho)\mathcal{R}&-2-\rho&1+2\rho+(1-\rho)\mathcal{R}&2+\rho\\
3-3(1+\rho)\mathcal{R}&0&3-3\rho\mathcal{R}&0
\end{array}\right)\left(\begin{array}{cccc} \mathcal{J}_1&&&\\
                           &\mathcal{I}_2\\
                           &&\mathcal{J}_3\\
                           &&&\mathcal{I}_2
\end{array}\right). \end{split}\label{ABsym}
\end{align}
The Ercolani-Sinha conditions,
$\boldsymbol{n}^T\mathcal{A}+\boldsymbol{m}\sp{T}\mathcal{B} =6
\chi\sp{\frac13} (1,0,0,0)$ written for the vectors
\begin{equation}
\boldsymbol{n}=\left(\begin{array}{c}n_1\\m_1-n_1\\-m_1\\2n_1-m_1
\end{array}\right), \qquad
\boldsymbol{m}=\left(\begin{array}{c}m_1\\-n_1\\n_1-m_1\\-3n_1
\end{array}\right)
\end{equation}
lead to the equations
\begin{equation}
\mathcal{R}=-\frac{2n_1-m_1}{m_1+n_1},\qquad
\mathcal{J}_1=\frac{\chi\sp{\frac13}}{m_1-2n_1},
\end{equation}
which were obtained earlier. A calculation also shows that the
relation (\ref{cplxMT})
$$\mathcal{M}\left(%
\begin{array}{c}
  \mathcal{A} \\
  \mathcal{B} \\
\end{array}%
\right) =\overline{
\left(%
\begin{array}{c}
  \mathcal{A} \\
  \mathcal{B} \\
\end{array}%
\right)}\cdot T$$ yielding a nontrivial check of our procedure.

The integrals between infinities may be reduced to our standard
integrals by writing
$$\int_{\infty_i}^{\infty_j} d\boldsymbol{u}
=\int_{\tau(0_{\tau(i)})}^{\tau(0_{\tau(j)})} d\boldsymbol{u}
=\int_{0_{\tau(i)}}^{0_{\tau(j)}} \tau\sp*(d\boldsymbol{u})
=\overline{\int_{0_{\tau(i)}}^{0_{\tau(j)}}d\boldsymbol{u}}\cdot T
=\left( \overline{\int_{0_{\tau(i)}}^{\lambda_*}d\boldsymbol{u}}-
\overline{\int_{0_{\tau(j)}}^{\lambda_*}d\boldsymbol{u}}\right)\cdot
T,
$$
where we write $\tau(\infty_i)=0_{\tau(i)}$ and $\lambda_*$ is any
of the branch points. These are then calculated to be
\begin{equation}
\int_{\infty_1}^{\infty_2} d\boldsymbol{u}=\left(
\begin{array}{c}
(\rho-1)\mathcal{J}_1\\
-(\rho^2-1)\mathcal{J}_4\\
(\rho^2-1)\mathcal{J}_3\\
-(\rho^2-1)\mathcal{J}_2
\end{array} \right),\
\int_{\infty_1}^{\infty_3} d\boldsymbol{u} =\left(
\begin{array}{c}
(\rho^2-1)\mathcal{J}_1\\
-(\rho-1)\mathcal{J}_4\\
(\rho-1)\mathcal{J}_3\\
-(\rho-1)\mathcal{J}_2
\end{array} \right),\
\int_{\infty_2}^{\infty_3}
 d\boldsymbol{u}=\left(
\begin{array}{c}
(\rho^2-\rho)\mathcal{J}_1\\
-(\rho-\rho^2)\mathcal{J}_4\\
(\rho-\rho^2)\mathcal{J}_3\\
-(\rho-\rho^2)\mathcal{J}_2
\end{array} \right).\label{qintegrals}
\end{equation}

Our Riemann matrix admits a reduction with respect to any of it
columns. We will exemplify this with the first column, a result we
will use next; similar considerations apply to the other columns.
Now from the above and (\ref{expxx}) it follows that
\begin{align}
\Pi\lambda=\left(%
\begin{array}{c}
  \mathcal{A} \\
  \mathcal{B} \\
\end{array}%
\right) \,\left(%
\begin{array}{c}
  1 \\
  0 \\
  0 \\
  0 \\
\end{array}%
\right)=\left(\begin{array}{c}\boldsymbol{x}\\\boldsymbol{y}
\end{array}\right)=
\left(\begin{array}{c}\oint_{\mathfrak{a}_i}{d} u_1
\\ \oint_{\mathfrak{b}_i}{d} u_1
\end{array}\right)=
\left(\begin{array}{c}\xi(H\boldsymbol{n}+\rho^2\boldsymbol{m})\\
\xi(\rho \boldsymbol{n}+H\boldsymbol{m})
\end{array}\right)=\xi\,M\left(\begin{array}{c}1\\\rho
\end{array}\right),\label{escond2}
\end{align}
where $M$ is the $2g\times2$ integral matrix
\begin{equation} \label{escond3}
M^T=\left(\begin{array}{cccccccc}n_1-m_1&n_2-m_2&n_3-m_3&-n_4-m_4
&m_1&m_2&m_3&-m_4      \\ -m_1&-m_2&-m_3&-m_4&n_1&n_2&n_3&n_4
\end{array}\right).
\end{equation}
Then to every two Ercolani-Sinha vectors $\boldsymbol{n}$,
$\boldsymbol{m}$ we have that
\begin{equation}
M^TJM=d\left(\begin{array}{cc}0&1\\-1&0\end{array}\right),\quad
d=\boldsymbol{n}.H\boldsymbol{n}- \boldsymbol{m}. \boldsymbol{n}
+\boldsymbol{m}.H\boldsymbol{m}= \sum_{j=1}^4(\varepsilon_j
n_j^2-n_jm_j+\varepsilon_j m_j^2).
\end{equation}
The number $d$ here is often called the Hopf number. In particular
for $d\ne0$ then $M$ is of maximal rank and consequently our
Riemann matrix admits reduction.

Let us now focus on the consequences of reduction for symmetric
monopoles.

\begin{theorem}\label{symwp} For the
symmetric monopole we may reduce by the first column using the
vector (\ref{escond2}) whose elements are related by
(\ref{solvn}), with $(n_1,m_1)=1$. Then
$$d=2(n_1+m_1)(m_1-2n_1)$$ and for $d\ne0$ there exists an element $\sigma$ of the
symplectic group $\mathrm{Sp}_{2g}(\mathbb{Z})$ such that
\begin{align}
\tau'_{\mathfrak{b}}=\sigma\circ\tau_{\mathfrak{b}}=
\left(\begin{array}{ccccc}
(\rho+2)/d&{\alpha}/{d}&0&\ldots&0\\
{\alpha}/{d}&&&&\\
0&&{\tau}^{\#}&&\\
\vdots&&&&\\ 0&&&&
\end{array}\right).\label{taureduced}
\end{align}
Letting $p\, m_1+q\, n_1=1$ then
\begin{equation}\label{alphareduced}
\alpha=\gcd(m_1+4n_1-q\,[m_1-2n_1],n_1-2m_1-p\,[m_1-2n_1]).
\end{equation}
When $\alpha=1$ a further symplectic transformation allows the
simplification $\tau'_{11}=\rho/d$.

Under $\sigma$ the Ercolani-Sinha vector transforms as
\begin{equation}\label{essymptrans}
  \sigma\circ{\boldsymbol U}=  \sigma\circ({\boldsymbol m}\sp{T}+{\boldsymbol
    n}\sp{T}\tau_{\mathfrak{b}})=(1/2,0,0,0).
\end{equation}
\end{theorem}
The proof of the theorem is constructive using work of Krazer,
Weierstrass and Kowalewski. Martens \cite{ma92,ma92a} has given an
algorithm for constructing $\sigma$ which we have implemented
using $Maple$. Because $\sigma$ depends on number theoretic
properties of $n_1$ and $m_1$ the form is rather unilluminating
and we simply record the result (though an explicit example will
be given in the following section). What is remarkable however is
the simple universal form the Ercolani-Sinha vector takes under
this transformation. This has great significance for us as we next
describe.

Using (\ref{redthetagen}),(\ref{taureduced}) and say
$D=\diag(d,1,1)$ we have that\footnote{When $\gcd(\alpha,d)\ne1$ a
smaller multiple than $d_1=d$ would suffice here with
correspondingly fewer terms in the sums $0\le m\le d_1-1$.}
\begin{align*}\theta((z,w);\tau'_\mathfrak{b})&=\sum_{m=0}\sp{d-1}
\theta(z+\frac{m\alpha}{d};\frac{\rho+2}{d})\,\theta\left[%
\begin{array}{ccc}
  \frac{m}{d} & 0 & 0 \\
  0 & 0 & 0 \\
\end{array}%
\right](Dw;D{\tau}^{\#}D)\\
&=\sum_{m=0}\sp{d-1} \theta \left[%
\begin{array}{c}\frac{m}{d}\\0\end{array}%
\right](dz;d(\rho+2))\,\theta((w_1+\frac{m\alpha}{d},w_2,w_3);{\tau}^{\#}),
\end{align*}
where we have genus one and three theta functions on the right
hand-side here. Comparison of (\ref{ourq0}) and
(\ref{essymptrans}) then reveals that the theta function
dependence of $Q_0(z)$ is given wholly by the genus one theta
functions. Further simplifications ensue from the identity
$$\theta \left[%
\begin{array}{c}\frac{\epsilon}{d}\\ \epsilon'\end{array}%
\right](dz;d\tau)=\mu(\tau)\,\prod_{l=0}\sp{d-1}
\theta \left[%
\begin{array}{c}\frac{\epsilon}{d}\\
\frac{ \epsilon'}{d}+\frac{d-(1+2l)}{2d}\end{array}%
\right](z;\tau),
$$
where $\mu(\tau)$ is a constant. We then have
\begin{theorem}\label{symel} For symmetric monopoles the theta function $z$-dependence of
$Q_0(z)$is expressible in terms of elliptic functions.
\end{theorem}

Thus far we have not discussed the final Hitchin constraint for
symmetric monopoles. This theorem reduces the problem to one of
the zeros of elliptic functions.
\begin{figure}
\centering
\begin{minipage}[c]{0.5\textwidth}
\includegraphics[width=7cm]{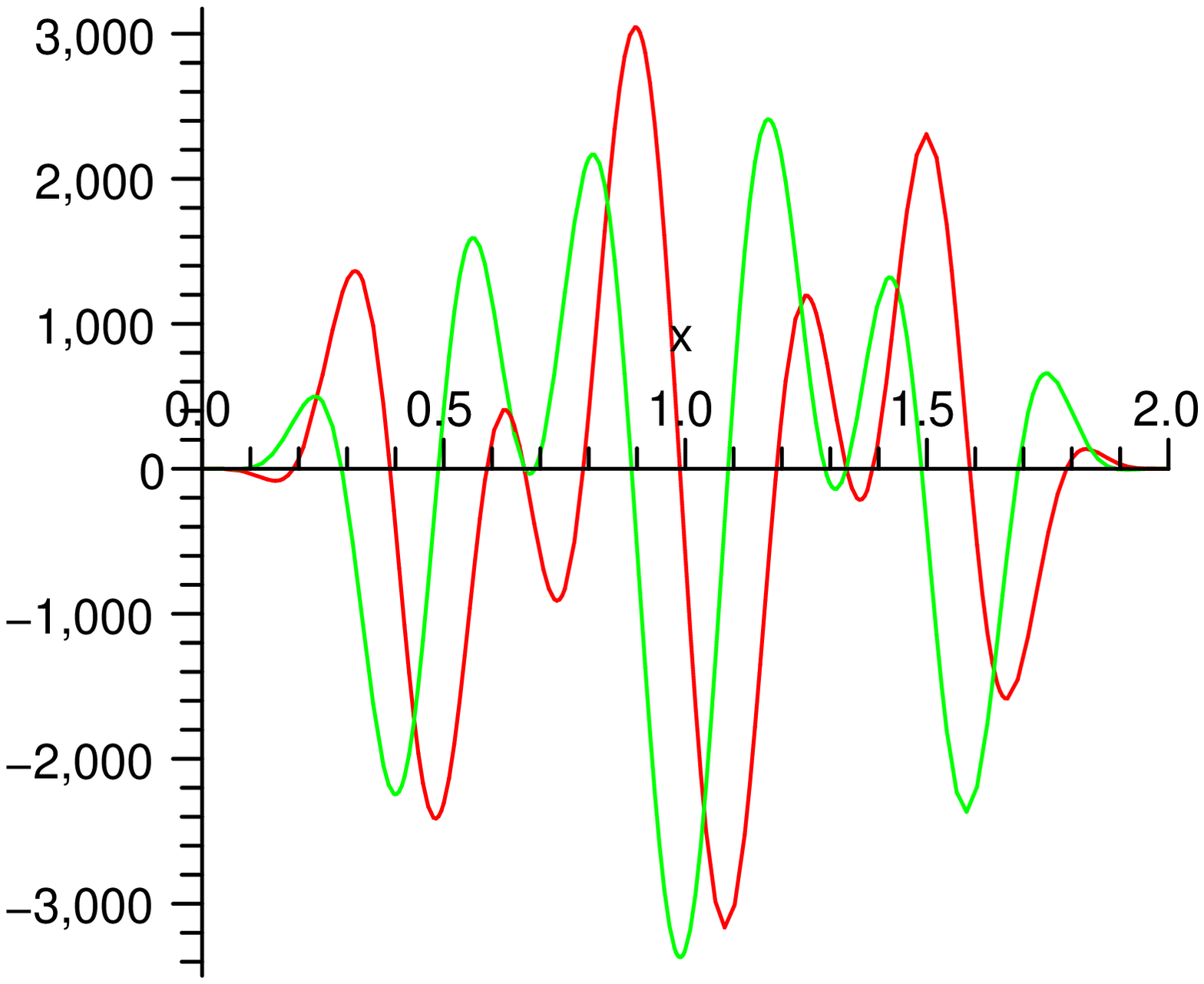} \caption{$n_1=2$, $m_1=1$. }
\label{fig:21}
\end{minipage}%
\begin{minipage}[c]{0.5\textwidth}
\includegraphics[width=7cm]{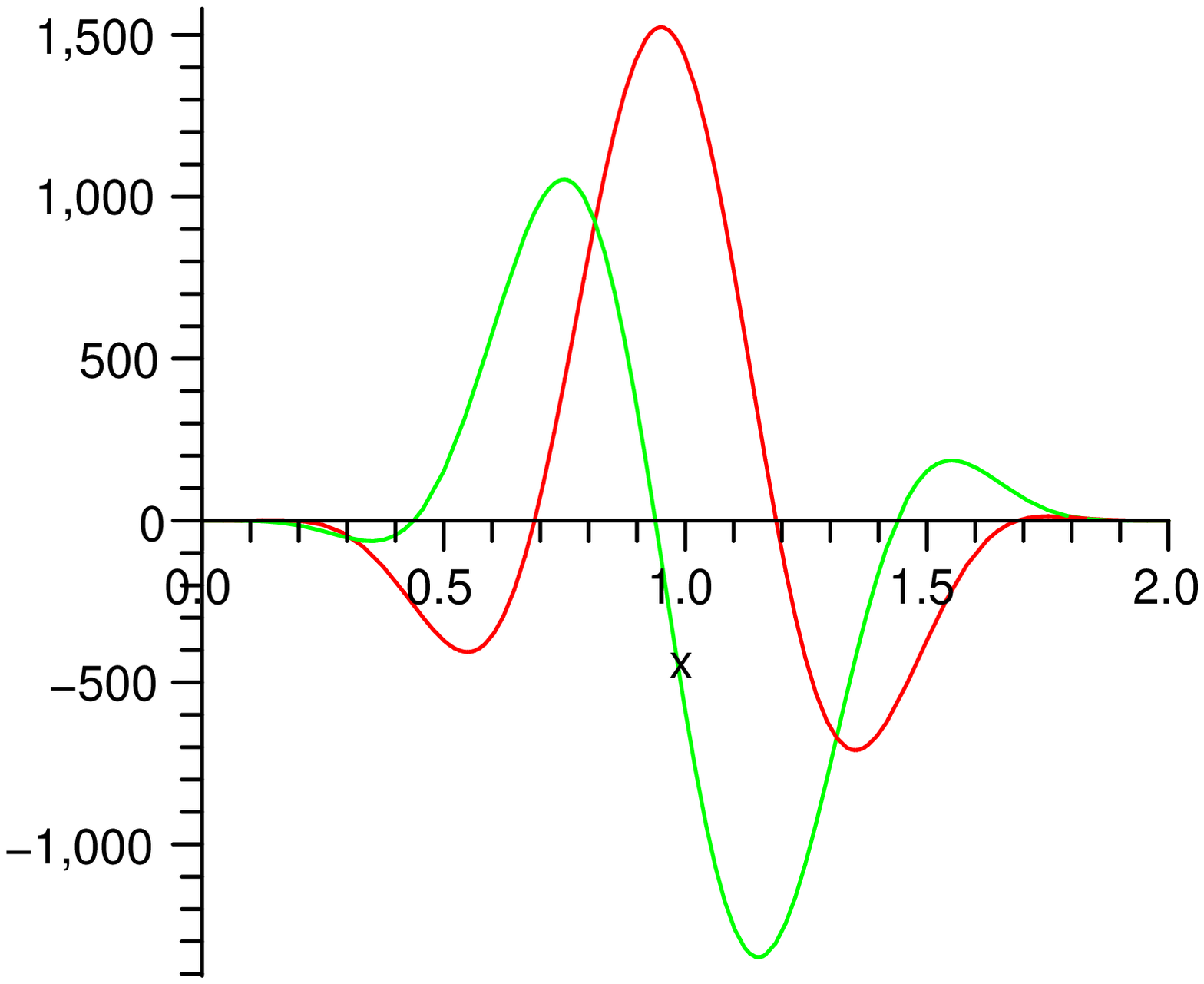} \caption{$n_1=1$, $m_1=1$. }
\label{fig:11}
\end{minipage}
\end{figure}
The graph in Figure~\ref{fig:21} shows the real and imaginary
parts of the theta function denominator of $Q_0(z)$ for the
$n_1=2$, $m_1=1$ symmetric monopole, the $b=0$ Ramanjuan case.
These vanish at $z=0$ and $z=2$ as desired, but additionally one
finds vanishing at $z=2/3$ and $z=4/3$. Calculating the theta
function with shifted argument in the numerator shows that there
is no corresponding vanishing and consequently $Q_0(z)$ yields
unwanted poles in $z\in(0,2)$. Thus the $n_1=2$, $m_1=1$ curve
does not yield a monopole.

A similar evaluation of the relevant $n_1=4$, $m_1=-1$ and
$n_1=5$, $m_1=-2$ theta functions also reveals unwanted zeros and
of the those cases from our table of symmetric 3-monopoles only
the tetrahedrally symmetric case has the required vanishing. As
yet we don't know whether there are further symmetric 3-monopoles
with the required vanishing for a genuine monopole curve. Before
turning to a more detailed examination of the tetrahedrally
symmetric case in our next section we first describe how to
calculate the remaining quantities appearing in our formula
(\ref{ourq0}) for $Q_0(z)$.

\subsection{Calculating $\nu_i-\nu_j$} Here we follow section
\S5.4. We calculate the $\mathfrak{a}$-periods of the differential
$dr_1$ in a manner similar to the period integrals already
calculated. Introduce integrals on the first sheet
\begin{align}
\mathcal{K}_1(\alpha)=\int_{0}^{\alpha}\frac{z^4 dz}{3w^2},\qquad
\mathcal{L}_1(\beta)&=\int_{0}^{\beta}\frac{z^4 dz}{3w^2}, \qquad
\beta=-\frac{1}{\alpha}.
\end{align}
Evidently
$\mathcal{K}_1(\rho^k\alpha)=\rho^{2k}\mathcal{K}(\alpha) $ and
$\mathcal{L}_1(\rho^k\beta)=\rho^{2k}\mathcal{L}(\beta)$ and one
finds that
\begin{equation}
\mathcal{K}_1=-\frac{4\sqrt{3}\pi}{27} \alpha^5\,
_2F_1\left(\frac23,\frac53;2; -\alpha^6   \right),\qquad
\mathcal{L}_1=\frac{4\sqrt{3}\pi}{27} \frac{1}{\alpha^5}\,
_2F_1\left(\frac23,\frac53;2; -\frac{1}{\alpha^6}
\right).\label{kl}
\end{equation}

We find, as before in the case of holomorphic differentials, that
\begin{align*}
y_1&=(\mathcal{K}_1+2\mathcal{L}_1)\rho-\mathcal{K}_1+\mathcal{L}_1,
&y_2&=(\mathcal{K}_1-\mathcal{L}_1)\rho+2\mathcal{K}_1+\mathcal{L}_1\\
y_3&=-(2\mathcal{K}_1+\mathcal{L}_1
)\rho-\mathcal{K}_1-2\mathcal{L}_1,
&y_4&=3(\mathcal{K}_1-\mathcal{L}_1)\rho +3\mathcal{K}_1.
\end{align*}
The Legendre relation (\ref{legndre}) gives a non trivial
consistency check of our calculations. This may be written in the
form of the following hypergeometric equality
\begin{align*}\frac{27}{4\sqrt{3}\pi}=
\alpha^4\,
_2F_1\left(\frac13,\frac13;1;-\frac{1}{\alpha^6}\right)\, _2F_1
\left(\frac23,\frac53;2;-\alpha^6 \right)+\frac{1}{\alpha^4}\,
_2F_1\left(\frac13,\frac13;1;-\alpha^6\right)\, _2F_1
\left(\frac23,\frac53;2;-\frac{1}{\alpha^6} \right)  \label{new}
\end{align*}
and this may be established by standard means.

To calculating $\nu_i-\nu_j$ using (\ref{nuijgpgr}) introduce the
differential of the second kind,
\begin{equation}s= d\left( \frac{w}{z}\right)(P)  -3 d r_1(P)\equiv
\frac{dz}{z^2 w^2},
\end{equation}
with second order pole at $0$ on all sheets,
\begin{align*}
\left.\frac{dz}{z^2 w^2}\right|_{P=0_k}&= \left\{
\frac{1}{w(0_k)^2}
 \frac{1}{\xi^2}  +\frac{2b}{3}\xi+\ldots  \right\} d\xi
= \left\{-
 \frac{w(0_k)}{\xi^2}  +\frac{2b}{3}\xi+\ldots  \right\} d\xi.
 \end{align*}
(Here we took into account $w(0_k)^3=-1$ for $k=1,2,3$.) Then
\begin{align}
\nu_i-\nu_j&=3 \boldsymbol{y}.\int_{\infty_j}^{\infty_j}
\boldsymbol{v}+\int_{\infty_j}^{\infty_j} \frac{dz}{z^2 w^2}
\label{nuij1s}
\end{align}
The last integral in (\ref{nuij1s}) may also be expressed in terms
of hypergeometric functions as follows. First we remark that
\[ \int_{\infty_i}^{\infty_j} \frac{dz}{z^2 w^2}
= (\rho_i-\rho_j)\int_{\alpha}^{\infty_1} \frac{dz}{z^2 w^2}, \]
where $\rho_i=\rho^{i-1}$. Next, for the integrals on the first
sheet we have
\begin{align*} \int_{\alpha}^{\infty} \frac{dz}{z^2 w^2}
=\frac{4\sqrt{3}\pi}{27} \frac{1}{\alpha^5}
F\left(\frac23,\frac53;2;
-\frac{1}{\alpha^6}   \right)=\mathcal{L}_1,\\
\int_{-\frac{1}{\alpha}}^{\infty} \frac{dz}{z^2 w^2}
=-\frac{4\sqrt{3}\pi}{27} \alpha^5 F\left(\frac23,\frac53;2;
-\alpha^6   \right)=\mathcal{K}_1.
\end{align*}

\section{The tetrahedral 3-monopole}
The curve of the tetrahedrally symmetric monopole is of the form
\begin{equation} \eta^3+\chi(\zeta^6+5\sqrt{2} \zeta^3-1)=0.
\label{thetracurve}
\end{equation}
In this case we may take
$$
t=\frac12-\frac{5\sqrt{3}}{18},\ \alpha
=\frac{\sqrt{3}-1}{\sqrt{2}},\
\mathcal{J}_1(\alpha)=-2\mathcal{I}_1(\alpha).
$$
For these values we may explicitly evaluate the various
hypergeometric functions. Using Ramanujan's identity
(\ref{ramf23}) together with the standard quadratic transformation
of the hypergeometric function
$$_2F_1\left(\frac12,\frac12,1,z\right)=(1+\sqrt{z})\sp{-1}\,
 _2F_1\left(\frac12,\frac12,1,\frac{4\sqrt{z}}{(1+\sqrt{z})\sp2}\right),
$$
(valid for $|z|<1$, $\arg{z}<\pi$) we find that
$$
_2F_1\left(\frac13,\frac23,1,t\right)=\frac{3\sp{\frac{5}{4}}}{4}\,
 _2F_1\left(\frac12,\frac12,1,\frac{2-\sqrt{3}}{4}\right).
$$
(In verifying this we note that
$p=4+3\sqrt{3}-2\sqrt{6}-3\sqrt{2}$ is the relevant value leading
to our $t$ in (\ref{ramf23}).) Now this last hypergeometric
function is related to an elliptic integral we may evaluate
\cite[p 86]{lawd89},
$$
K\left(\frac{\sqrt{3}-1}{2\sqrt{2}}\right)=\frac{\pi}{2}\,
_2F_1\left(\frac12,\frac12,1,\frac{2-\sqrt{3}}{4}\right)
=\frac{\Gamma(\frac16)\Gamma(\frac13)}{3\sp{\frac14}\, 4
\sqrt{\pi}}.
$$
Bringing these results together we finally obtain
\begin{equation}\label{evalhyp}
_2F_1\left(\frac13,\frac23,1,t\right)=
\frac{3\,\Gamma(\frac16)\Gamma(\frac13)}{8\pi\sp{\frac32}}\, .
\end{equation}
Then from (\ref{esevch}) we obtain that
\begin{equation}\label{tetchi}
\chi^{\frac{1}{3}} = -2\, \frac{2 \pi}{3
    \sqrt{3}}\ \frac{\alpha}{(1-\alpha\sp6)\sp\frac13}\ {_2F_1}(\frac{1}{3}, \frac{2}{3}; 1,
    t)=-\frac{1}{2\sp\frac16 \,\sqrt{3}}\,
    \frac{\Gamma(\frac16)\Gamma(\frac13)}{2\sqrt{3}\pi\sp{\frac12}}.
\end{equation}
This agrees with the result of \cite{hmr99}. We also note that
upon using Goursat's identity \cite[(39)]{goursat81}
$$
_2F_1\left(\frac13,\frac23,1,x\right)= (1-2x)\sp{\frac{-1}{3}}\,
_2F_1\left(\frac16,\frac23,1,\frac{4x(x-1)}{(2x-1)\sp2}\right),
$$
we may establish the result of \cite{hmr99} based on numerical
evaluation, that
$$
_2F_1\left(\frac16,\frac23,1,-\frac{2}{25}\right)=
\frac{5\sp{\frac13}\,3}{8}\,
\frac{\Gamma(\frac16)\Gamma(\frac13)}{\sqrt{3}\,\pi\sp{\frac32}}.
$$

Using these results and those of the previous section we have,
\begin{theorem}
The tetrahedral 3-monopole for which $b=5\sqrt{2}$ admits the
$\tau$-matrix of the form
\begin{equation}\label{tetratau}
\begin{split}\tau&=
{\frac {1}{98}}\, \left( \begin {array}{cccc} -73+51\,\imath\sqrt
{3}&9-13
\,i\sqrt {3}&15+11\,\imath\sqrt {3}&42-28\,\imath\sqrt {3}\\
\noalign{\medskip}9- 13\,\imath\sqrt {3}&-34+60\,\imath\sqrt
{3}&2\,\imath\sqrt {3}-24&21+35\, \imath\sqrt {3}
\\\noalign{\medskip}15+11\,\imath\sqrt {3}&2\,\imath\sqrt {3}-24&-40+36\,
\imath\sqrt {3}&-63-7\,\imath\sqrt
{3}\\\noalign{\medskip}42-28\,\imath\sqrt {3}&21+35\, \imath \sqrt
{3}&-63-7\,\imath\sqrt {3}&49+49\,\imath\sqrt {3}\end {array}
\right) \\
&=\left( \begin {array}{cccc} -{\frac {11}{49}}+{\frac
{51}{49}}\,\rho& -{\frac {2}{49}}-{\frac {13}{49}}\,\rho&{\frac
{13}{49}}+{\frac {11}{
49}}\,\rho&\frac17-\frac47\,\rho\\\noalign{\medskip}-{\frac
{2}{49}}-{\frac { 13}{49}}\,\rho&{\frac {13}{49}}+{\frac
{60}{49}}\,\rho&-{\frac {11}{49 }}+{\frac
{2}{49}}\,\rho&\frac47+\frac57\,\rho\\\noalign{\medskip}{\frac
{13}{ 49}}+{\frac {11}{49}}\,\rho&-{\frac {11}{49}}+{\frac
{2}{49}}\,\rho&-{ \frac {2}{49}}+{\frac
{36}{49}}\,\rho&-\frac57-\frac17\,\rho
\\\noalign{\medskip}\frac17-\frac47\,\rho&\frac47+\frac57\,\rho&-\frac57-\frac17\,\rho&1+\rho
\end {array} \right).
\end{split}
\end{equation}
\end{theorem}

We have already seen that the symmetric monopole curve
$\mathcal{C}$ covers two equianharmonic torii $\mathcal{E}_{1,2}$.
For the value of the parameter $b=5\sqrt{2}$ the curve covers
three further equianharmonic elliptic curves. These may be
described as follows.  For $i=3$, $4$, $5$ let $\pi_i:\,
\mathcal{C}\rightarrow\mathcal{E}_i$ be defined by the formulae
\begin{align}
\mu_3&=-\frac{\imath}{2^{\frac43}}\frac{(1+z\alpha)^4
+(z-\alpha)^4}{\alpha^2w^2},& \nu_3&=\frac{1+\imath}{\sqrt{2}}
\frac{(1+\alpha^2)(z^2+1)}{(z-\alpha)(z\alpha+1)},\nonumber \\
\mu_4&=-2^{\frac23}\frac{(1+z\alpha)^4 +(z-\alpha)^4}{2\alpha w
(1+z\alpha)(z-\alpha) },& \nu_4&=2^{\frac23}\alpha w \frac{
(1+z\alpha)^4 -(z-\alpha)^4
}{(z-\alpha)^3(z\alpha+1)^3},\label{cover345}\\
\mu_5&=-\sqrt{3}\imath\frac{(z^2+1)(z^2-2\sqrt{2}z-1)}
{(z^2+\sqrt{2}z-1)^2},& \nu_5&=-4\sqrt{6}\imath \frac{w
(z^4-\sqrt{2}z^3+3z^2 +\sqrt{2}z+1 )}
{(z^2+\sqrt{2}z-1)^3}.\nonumber
\end{align}
Then
\begin{align}
\mathcal{E}_3:&\quad \{(\nu_3,\mu_3)\,|\, {\nu_3}^2-{\mu_3}^3-2\imath=0 \},\nonumber\\
 \mathcal{E}_4:&\quad \{(\nu_4,\mu_4)\,|\,
  {\nu_4}^2-  \mu_4({\mu_4}^3+4)=0\}, \label{curves345} \\
\mathcal{E}_5:&\quad \{(\nu_5,\mu_5)\,|\,
{\nu_5}^3+24\sqrt{6}\imath ({\mu_5}^2-1)^2=0 \},\nonumber
 \end{align}
and we have the following relations between holomorphic
differentials
\begin{align}
{d}u_2&=\frac{z\,{d}z}{w^2}=\frac{1}{2^{\frac53}\sqrt{3}}\left\{
 (\imath-1)\pi_3\sp*\left(\frac{{d}\mu_3 }{\nu_3}\right) +
\pi_4\sp*\left(\frac{{d}\mu_4 }{\nu_4}\right) \right\} \label{differentials34} \\
{d}u_1&=\frac{{d} z}{w}
=\pi_5\sp*\left(\frac{{d}\mu_5}{\nu_5}\right).
\label{differentials5}
 \end{align}
The final of these rational maps was introduced by \cite{hmr99}
and has the following significance.

\begin{proposition}
Let $\boldsymbol{x}$ and $\boldsymbol{y}$ be the $\mathfrak{a}$
and $\mathfrak{b}$-periods of the differential ${d}u_1$ and denote
by $X$, $Y$ the $\mathfrak{a}$ and $\mathfrak{b}$-periods of the
elliptic differential ${d}\mu_5/\nu_5$. Then
\begin{equation}
\left(
\begin{array}{c}
  \boldsymbol{x} \\
  \boldsymbol{y} \\
\end{array}
\right) = M_5\left(%
\begin{array}{c}
  X \\
  Y \\
\end{array}
\right),\label{transfhom}
\end{equation}
where $M_5$ is the matrix
\begin{align}
M_5\sp{T}= \left(\begin{array}{rrrrrrrr}-1 &1 &0  &3 &1  &0 &-1  &1\\
                                 0 &1 &-1 &2 &1 & -1 &0  &3
  \end{array}\right)\label{mmatrix}
\end{align}
satisfying the condition
\begin{equation}
M_5\sp{T}\left(\begin{array}{cc}0_4&1_4\\
                        -1_4&0_4 \end{array}\right)M_5=
4\left(\begin{array}{cc}
 0&1\\-1&0  \end{array}\right).
\label{hopf1}\end{equation}
\end{proposition}

\begin{proof}
Introduce the homology basis for the elliptic curve as shown in
Figure~\ref{fig:Elldiffbasis} and
\begin{figure}
\centering
\begin{minipage}[c]{0.8\textwidth}
\includegraphics[width=6cm]{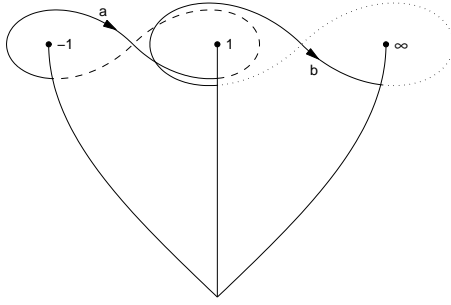} \caption{The elliptic curve homology basis }
\label{fig:Elldiffbasis}
\end{minipage}%
\end{figure}
set
\[ \mathcal{K}(\alpha)
=\int\limits_{-1}^{1}\frac{{d}\mu_5}{\nu_5}.\] Then
\begin{equation}
X=(2+\rho)\mathcal{K}(\alpha), \quad
Y=-(2\rho+1)\mathcal{K}(\alpha).\label{XY}
\end{equation}
From the reduction formula (\ref{differentials5}) we next conclude
that
\begin{equation}
\int\limits_{\alpha\rho}^{\alpha\rho^2}\frac{{d}z}{w}
=\mathcal{K}(\alpha)
\end{equation}
and therefore have that
\begin{align}\begin{split}
&2\mathcal{I}_1(\alpha)+\rho\mathcal{I}_1(\alpha)=\rho\mathcal{K}(\alpha),\\
&-2\rho\mathcal{I}_1(\alpha)-\mathcal{I}_1(\alpha)=\mathcal{K}(\alpha).
\end{split}\label{XY1}
\end{align}
Equations (\ref{XY}) and (\ref{XY1}) permit us to express
\begin{equation} \mathcal{I}_1(\alpha)=-\frac{Y}{3}   ,\qquad
\rho\mathcal{I}_1(\alpha)=-\frac{X}{3}
\end{equation}
and comparison with (\ref{ABsym}) yields the given $M_5$. The
condition (\ref{hopf1}) is checked directly. The number 4
appearing in (\ref{hopf1}) means that the cover $\pi_5$ given in
(\ref{cover345}) be of degree 4.

\end{proof}
We remark that the matrix $M_5$ of the proposition is obtained
from the $M$ of (\ref{escond3}) by 
$M_5=-\left(%
\begin{array}{cc}
  0 & 1 \\
  1 & 0 \\
\end{array}%
\right)M$, which simply reflects our choice of homology basis.
Thus we are discussing the reduction of the previous section.
Indeed with
$$\sigma=
\left[ \begin {array}{cccccccc}
1&0&0&0&0&0&0&0\\\noalign{\medskip}0&0
&0&0&0&1&0&0\\\noalign{\medskip}0&0&-1&0&0&1&0&0\\\noalign{\medskip}2&0
&-1&1&0&0&0&-1\\\noalign{\medskip}1&0&-1&1&1&-1&0&-3
\\\noalign{\medskip}5&-1&0&3&0&-1&1&-2\\\noalign{\medskip}-6&0&0&-3&0&0
&-1&2\\\noalign{\medskip}7&0&0&3&0&0&0&-2\end {array} \right]
\equiv \left(%
\begin{array}{cc}
  a & b \\
  c & d \\
\end{array}%
\right)
$$
we find that the $\tau$ matrix (\ref{tetratau}) transforms to
$$
\tau'=\sigma\circ\tau=(a\tau+b)(c\tau+d)\sp{-1}=\left[ \begin
{array}{cccc} \rho/4&1/4&0&0\\\noalign{\medskip}1/4&
5\rho/4&\rho&0\\\noalign{\medskip}0&\rho&2\,\rho&\rho
\\\noalign{\medskip}0&0&\rho&2/7+6\rho/7 \end {array} \right].
$$

Combined with Theorem \ref{symwp} we may reduce our expression for
the tetrahedral monopoles $Q_0(z)$ to one built out of Jacobi
elliptic theta functions. To compare with the Nahm data of
\cite{hmm95} we must solve for $C(z)$. This will be done
elsewhere.

\part{End Matters}
\section{Conclusions}
Although monopoles have been studied now for many years and from
various perspectives, relatively few analytic solutions are known.
This paper has sought to make effective the connection with
integrable systems to construct such solutions. It is nevertheless
only early steps upon this road.

The paper had two thrusts: an examination of the general
construction and then a focus on a (new) class of charge three
monopoles. In our general considerations we gave a further
constraint on the Ercolani-Sinha vector (Lemma \ref{ueventheta})
and presented a new solution to the matrix $Q_0$ (Theorem
\ref{ourq0}), from which the Nahm data is reconstructed by solving
a first order matrix differential equation. This latter step will
be considered elsewhere. Our construction of the matrix $Q_0$ has
been cast solely in terms of data built out of the spectral curve.
Previous expressions for this matrix in terms of Baker-Akhiezer
functions involve the choice of a non-special divisor which we
relate to a gauge choice. Our analysis clearly identifies each of
the ingredients necessary for the construction of this matrix and
we showed how the fundamental bi-differential may be used in
calculating this. Nearly all of the ingredients hinge on being
able to integrate explicitly on the curve.

To apply our general construction beyond the known case of charge
two we considered the restricted class of charge three monopoles
(\ref{welstein99}) which includes the tetrahedrally symmetric
monopole. This family of curves has many arithmetic properties
that facilitates analytic integration. In particular the period
matrix may be explicitly expressed in terms of just four
integrals. Using this we were able to explicitly solve the
Ercolani-Sinha constraints that are equivalent to Hitchin's
transcendental condition (A2) of the triviality of a certain line
bundle over the spectral curve (Proposition \ref{ourh2}). Our
approach reduces the problem to that of determining certain
rationality properties of the (four) relevant periods. (Our result
also admits another approach to seeking monopole curves: we may
solve the Ercolani-Sinha constraints and then seek to impose
Hitchin's reality conditions on the resulting curves. Results from
this approach will be explored elsewhere.) To proceed further in
this rather uncharted territory we further restricted our
attention to what we have referred to as ``symmetric 3-monopoles"
whose spectral curve has the form (\ref{bren03}). This reduced the
required independent integrals from four to two, each of which
were hypergeometric in form, and the rationality requirement is
now for the ratio of these (Proposition \ref{propesnt}).
Extensions of work by Ramanujan mean this latter question may be
replaced by number theory and of seeking solutions of various
algebraic equations (depending on the primes involved in the
rational ratio). Examples of such solutions were given (again
including the tetrahedral case). We further examined the
symmetries and coverings of these symmetric curves and their
relation to higher Goursat hypergeometric identities. Having at
hand now many putative spectral curves we proceeded to evaluate
the remaining integrals needed in our construction. Remarkably we
discovered that application of Weierstrass reduction theory showed
that the Ercolani-Sinha vector transformed to a universal form and
that all of the theta function $z$-dependence for symmetric
3-monopoles was expressible in terms of elliptic functions
(Theorems \ref{symwp},\ref{symel}). The final selection of
permissible spectral curves at last reduced to the question of
zeros of these elliptic functions. Unfortunately, of the symmetric
3-monopoles we have examined only the tetrahedral monopole has the
required zeros. Further investigation is required to ascertain
whether this is a general result.

Our final section then was devoted to the charge three
tetrahedrally symmetric monopole. Here we were able to
substantially simplify known expressions for the period matrix of
the spectral curve as well as prove a conjectured identity of
earlier workers. Again an explicit map was given and we have been
able to reduce entirely to elliptic functions. The final
comparison with the Nahm data of \cite{hmm95} requires the next
stage of the reconstruction, solving for the matrix $C(z)$. This
and other matters will be left for a subsequent work.

\section*{Acknowledgements}
This paper was conceived at the EPSRC funded Newton Institute
programme ``Integrable Systems'' in 2001 and had early
encouragement from Hermann Flaschka and Nick Ercolani. The many
technical hurdles encountered in this work meant growth has come
in spurts rather than continuous progress and it has occupied our
thoughts for much of this time. We are grateful to an EPSRC small
grant enabling both authors to come together to draw the work into
its present form. Over the intervening years we have benefited
from discussions and correspondence with many colleagues and we
wish to thank: Nigel Hitchin, Conor Houghton and Paul Sutcliffe
for their remarks on monopoles; David Calderbank, Miles Reid,
Richard Thomas and Armando Treibich for their geometric and
algebro-geometric advice; Raimundas Vidunas and Adri Olde Daalhuis
for references on matters hypergeometric; Mike Eggar and Keiji
Matsumoto for help with topological aspects of our curve; and
Chris Eilbeck, Tamara Grava, Yuri Fedorov, John McKay and Stan Richardson
for discussions pertaining to Riemann surfaces. We have enjoyed a
problem touching on so many aspects of mathematics.

\appendix
\section{Theta Functions}
 For $r\in \mathbb{N}$ the canonical Riemann
 $\theta$-function is given by
\begin{equation}
\theta(\boldsymbol{z};\tau) =\sum_{\boldsymbol{n}\in \mathbb{Z}^r}
\exp(\imath\pi \boldsymbol{n}^T \tau\boldsymbol{n}+2 \imath\pi
\boldsymbol{z}^T \boldsymbol{n}).
\end{equation}
The $\theta$-function is holomorphic on
$\mathbb{C}^r\times\mathbb{S}^r$ and satisfies
\begin{equation}\theta(\boldsymbol{z}+\boldsymbol{p}\, ;\tau)=
\theta(\boldsymbol{z};\tau),\quad
\theta(\boldsymbol{z}+\boldsymbol{p}\tau;\tau)=
\mathrm{exp}\{-\imath\pi(\boldsymbol{p}^T\tau \boldsymbol{p}
+2\boldsymbol{z}^T\boldsymbol{p})\}\, \theta(\boldsymbol{z};\tau),
\label{transformation}
\end{equation}
where $\boldsymbol{p}\in\mathbb{Z}^r$.

The Riemann $\theta$-function
$\theta_{\boldsymbol{a},\boldsymbol{b}}(\boldsymbol{z};\tau)$ with
characteristics $\boldsymbol{a},\boldsymbol{b}\in\mathbb{Q}$ is
defined by
\begin{align*}
\theta_{\boldsymbol{a},\boldsymbol{b}}(\boldsymbol{z};\tau)
&=\mathrm{exp} \left\{ \imath\pi
(\boldsymbol{a}^T\tau\boldsymbol{a}
+2\boldsymbol{a}^T(\boldsymbol{z}+\boldsymbol{b})))\right\}
\theta(\boldsymbol{z}+\tau\boldsymbol{a}+\boldsymbol{b};\tau)\\
& =\sum_{\boldsymbol{n}\in\mathbb{Z}^r}\mathrm{exp}
\left\{\imath\pi(\boldsymbol{n}+\boldsymbol{a})^T\tau
            (\boldsymbol{n}+\boldsymbol{a})
+2\imath\pi
(\boldsymbol{n}+\boldsymbol{a})^T(\boldsymbol{z}+\boldsymbol{b})
\right\},
\end{align*}
where $\boldsymbol{a},\boldsymbol{b}\in \mathbb{Q}^r$. This is also
written as
$$\theta_{\boldsymbol{a},\boldsymbol{b}}(\boldsymbol{z};\tau)=
\theta\left[\begin{matrix}\boldsymbol{a}
\\ \boldsymbol{b}\end{matrix}\right](\boldsymbol{z};\tau).
$$
For arbitrary $\boldsymbol{a},\boldsymbol{b}\in \mathbb{Q}^r$ and
$\boldsymbol{a}',\boldsymbol{b}'\in \mathbb{Q}^r$ the following
formula is valid
\begin{align}
\theta_{\boldsymbol{a},\boldsymbol{b}}
(\boldsymbol{z}+\boldsymbol{a}'\tau+\boldsymbol{b}';\tau)&=\mathrm{exp}\left\{
-\imath\pi
{\boldsymbol{a}'}^T\tau{\boldsymbol{a}'}-2\imath\pi{\boldsymbol{a}'}^T\boldsymbol{z}
 -2\imath\pi (\boldsymbol{b}+\boldsymbol{b}')^T{\boldsymbol{a}'}   \right\}
 \times
\theta_{\boldsymbol{a}+\boldsymbol{a}',\boldsymbol{b}+\boldsymbol{b}'}
(\boldsymbol{z};\tau).\label{thetatransf}
\end{align}

The function $\theta_{\boldsymbol{a},\boldsymbol{b}}(\tau)=
\theta_{\boldsymbol{a},\boldsymbol{b}}(\boldsymbol{0};\tau) $ is
called the $\theta$-constant with characteristic
$\boldsymbol{a},\boldsymbol{b}$. We have
\begin{align*} \
&\theta_{-\boldsymbol{a},-\boldsymbol{b}}(\boldsymbol{z};\tau)=
\theta_{\boldsymbol{a},\boldsymbol{b}}(-\boldsymbol{z};\tau)\\
&\theta_{\boldsymbol{a}+\boldsymbol{p},\boldsymbol{b}+\boldsymbol{q}}
(\boldsymbol{z};\tau)= \mathrm{exp}(2\pi\imath
\boldsymbol{a}^T\boldsymbol{q})
\theta_{\boldsymbol{a},\boldsymbol{b}}(\boldsymbol{z};\tau)
\end{align*}

The following transformation formula is given in \cite[p85,
p176]{igusa72}.
\begin{proposition}
For any $\mathfrak{g}=\left(\begin{array}{cc}A&B\\C&D
\end{array}\right)\in\mathrm{Sp}(2g,\mathbb{Z})$ and
$(\boldsymbol{a},\boldsymbol{b})\in\mathbb{Q}^{2g}$ we put
\begin{align*}
\mathfrak{g}\cdot(\boldsymbol{a},\boldsymbol{b})&=
(\boldsymbol{a},\boldsymbol{b})\mathfrak{g}^{-1}
+\frac12(\mathrm{diag}(CD^T),\mathrm{diag}(AB^T) )\\
\boldsymbol{\phi}_{\boldsymbol{a},\boldsymbol{b}}(\mathfrak{g})&=-\frac12
(\boldsymbol{a}D^TB\boldsymbol{a}^T
-2\boldsymbol{a}B^TC\boldsymbol{b}^T+
\boldsymbol{b}C^TA\boldsymbol{b}^T)
+\frac12(\boldsymbol{a}D^T-\boldsymbol{b}C^T)^T\mathrm{diag}(AB^T)
,\end{align*} where $\mathrm{diag}(A)$ is the row vector
consisting of the diagonal components of $A$. Then for every
$\mathfrak{g}\in\mathrm{Sp}(2g,\mathbb{Z})$ we have
\begin{align}\begin{split}
&\theta_{\mathfrak{g}\cdot(\boldsymbol{a},
\boldsymbol{b})}(0;(A\tau_\mathfrak{b}+B)(C\tau_\mathfrak{b}+D)^{-1})
=\kappa(\mathfrak{g})\mathrm{exp}(2\pi\imath
\boldsymbol{\phi}_{\boldsymbol{a},\boldsymbol{b}}(\mathfrak{g})   )\,
\mathrm{det}(C\tau_\mathfrak{b}+D)^{\frac12}
\theta_{(\boldsymbol{a},\boldsymbol{b})}(0;\tau_\mathfrak{b})
\end{split}\label{igusa}
\end{align}
in which $\kappa(\mathfrak{g})^2$ is a $4$-th root of unity
depending only on $\mathfrak{g}$ while
\begin{equation} \begin{split}
\theta_{\mathfrak{g}\cdot(\boldsymbol{a},
\boldsymbol{b})}(z(C\tau_\mathfrak{b}+D)^{-1};(A\tau_\mathfrak{b}+B)(C\tau_\mathfrak{b}+D)^{-1})
&=\mu\,\exp\left(i\pi z(C\tau_\mathfrak{b}+D)^{-1}Cz\sp{T}\right)
\,
\mathrm{det}(C\tau_\mathfrak{b}+D)^{\frac12}\\
&\qquad
\times\theta_{(\boldsymbol{a},\boldsymbol{b})}(z;\tau_\mathfrak{b})
\end{split} \label{igusab}
\end{equation}
and $\mu$ is a complex number independent of $\tau$ and $z$ such
that $|\mu|=1$.
\end{proposition}

\subsection{The Vector of Riemann Constants} The convention we
adopt for our vector of Riemann constants is
$$\theta\left(\boldsymbol{\phi}(P)-\boldsymbol{\phi}(\sum_{i=1}\sp{g}Q_i) -K\right)=0$$
in the Jacobi inversion. This is the convention used by Farkas and
Kra and the negative of that of Mumford; the choice of signs
appears in the actual construction of $K$, such as (2.4.1) of
Farkas and Kra. Then \begin{align}\label{vecR}\begin{split} (K_Q)_j&=\frac12
\tau_{jj}
-\sum_{k}\oint_{\mathfrak{a}_k}\omega_k(P)\int_Q\sp{P}\omega_j,\\
 & = \frac{1}{2}\left( \tau_{jj} + 1\right) -
    \sum_{k\ne j}\oint_{\mathbf{\mathfrak{a}}_{k}}
    \omega_{k}(P)\int^{P}_{Q}\omega_{j}.\end{split}
\end{align}
The vector of Riemann constants depends on the homology basis and
base point $Q$. If we change base points of the Abel map
$\boldsymbol{\phi}_{Q}\rightarrow \boldsymbol{\phi}_{ Q'}$ then $K_{Q}=K_{ Q'}+\boldsymbol{\phi}_{
Q'}({Q}\sp{g-1})$. With this convention
\begin{equation}\label{KKC}
\boldsymbol{\phi}_Q(\div(K_\mathcal{C}))=-2K_Q.
\end{equation}

\subsection{Theta Characteristics}
\def\Pic{\mathop{\rm Pic}\nolimits}
\def\div{\mathop{\rm div}\nolimits}
The set $\Sigma$ of divisor classes $D$ such that
$2D=K_\mathcal{C}$, the canonical class, is called the set of
\emph{theta characteristics} of $\mathcal{C}$. The set $\Sigma$ is a
principal homogeneous space for the group $J_2$, the group of
$2$-torsion points of the group $\Pic\sp0(\mathcal{C})$ of degree
zero line bundles on $\mathcal{C}$. Equivalently this may be viewed
as the $2$-torsion points of the Jacobian, $J_2=\frac12
\Lambda/\Lambda$. Geometrically if $\xi$ is a holomorphic line
bundle on $\mathcal{C}$ such that $\xi\sp2$ is holomorphically
equivalent to $K_\mathcal{C}$ then the divisor of $\xi$ is a theta
characteristic. If $L$ is a holomorphic line bundle of order $2$,
that is $L\sp2$ is holomorphically trivial, then the divisor of
$\xi\otimes L$ is also a theta characteristic. Thus there are
$|J_2|=2\sp{2g}$ theta characteristics.

We may view $J_2=\{v\in \Pic\sp0(\mathcal{C})|2v=0\}$ as a vector
space of dimension $2g$ over $\mathbb{F}_2$. This vector space has a
nondegenerate symplectic (and hence symmetric as the field is
$\mathbb{F}_2$) form defined by the Weil pairing. If $D$ and $E$ are
divisors with disjoint support in the classes of $u$ and $v$
respectively, and $2D=\div(f)$, $2E=\div(g)$ then the Weil Pairing
is
$$\lambda_2:J_2\times J_2\rightarrow \mathbb{F}_2,\qquad
\lambda_2(u,v)=\frac{g(D)}{f(E)},$$ where if $D=\sum_j n_j x_j$ then
$g(D)=\prod_j g(x_j)\sp{n_j}$. Mumford identifies $\mathbb{F}_2$
with $\pm1$ by sending $0$ to $1$ and $1$ to $-1$. (In general we
may consider $J_r$, the $r$-torsion points of
$\Pic\sp0(\mathcal{C})$, and the Weil pairing gives us a
nondegenerate antisymmetric map $\lambda_2:J_r\times J_r\rightarrow
\mu_r$ where $\mu_r$ are $r$-th roots of unity.) The $\mathbb{F}_2$
vector space $J_2$ may be identified with
$H\sp1(\mathcal{C},\mathbb{F}_2)$ and with this identification
$\lambda_2$ is simply the cup product.

Define $\omega_\xi:J_2\rightarrow \mathbb{F}_2$ by
\begin{equation}\label{quadform}\omega_\xi(u)=\dim H\sp0(\mathcal{C},\xi\otimes u)- \dim
H\sp0(\mathcal{C},\xi)\quad (\mod 2),\end{equation}
 where $u=L_D$ is
the line bundle with divisor $D$. Then
$$\lambda_2(u,v)=\omega_\xi(u\otimes v)-\omega_\xi(u)-
\omega_\xi(v).$$ Any function $\omega_\xi$ satisfying this identity
is known as an Arf function, and any Arf function is given by
$\omega_\xi$ for some theta characteristic with corresponding line
bundle $\xi$. Thus the space of theta characteristics may be
identified with the space of quadratic forms (\ref{quadform}).

\section{Integrals between branch points}
We shall now describe how to integrate holomorphic differentials
between branch points.  We use the fact that for non-invariant
holomorphic differentials (as we have)
\begin{equation*}
    \sum^{3}_{i=1} \int_{\gamma_{i}(\lambda_{A},\lambda_{B})} \omega =
    \int^{\lambda_{B}}_{\lambda_{A}}(\omega + R_{*}\omega + R^{2}_{*}\omega) =
    0.
\end{equation*}
Indeed, if $\omega$ is any holomorphic differential on a compact
Riemann surface which is an $N$-fold branched cover of
$\mathbb{CP}\sp1$ then $\sum_{j=1}\sp{N}\omega(P\sp{(j)})=0$,
where $P\sp{(j)}$ are the preimages of $P\in\mathbb{CP}\sp1$.
 Then
\begin{align*}
   \oint_{ \mathfrak{a}_{1} - \mathfrak{b}_{1} }\omega= 3
   \int_{ \gamma_{1}(\lambda_{1},\lambda_{2})}\omega,\qquad
    \oint_{\mathfrak{a}_{2} - \mathfrak{b}_{2}}\omega = 3
    \int_{\gamma_{1}(\lambda_{3},\lambda_{4})}\omega,\qquad
    \oint_{\mathfrak{a}_{3} - \mathfrak{b}_{3}}\omega = 3
    \int_{\gamma_{1}(\lambda_{5},\lambda_{6})}\omega,
\end{align*}
and consequently
\begin{align*}
    \int_{\gamma_{1}(\lambda_{1}, \lambda_{2})}\omega & = \frac{1}{3}
    \oint_{\mathfrak{a}_{1} - \mathfrak{b}_{1}} \omega, \\
    \int_{\gamma_{2}(\lambda_{1}, \lambda_{2})}\omega & =
    \int_{\gamma_{1}(\lambda_{1}, \lambda_{2})-\mathfrak{a}_{1}}\omega =
    \frac{1}{3} \oint_{-2\mathfrak{a}_{1} - \mathfrak{b}_{1}} \omega, \\
    \int_{\gamma_{3}(\lambda_{1}, \lambda_{2})}\omega & = \frac{1}{3}
    \oint_{2\mathfrak{b}_{1} + \mathfrak{a}_{1}} \omega, \\
\end{align*}
with similar expressions obtained for $\gamma_{i}(\lambda_{3},
\lambda_{4})$ and $\gamma_{i}(\lambda_{5},\lambda_{6})$.

Further utilising $
    \gamma_{1}(\lambda_{2},\lambda_{6})  = \gamma_{1}(\lambda_{2},\lambda_{1}) +
    \gamma_{1}(\lambda_{1},\lambda_{6})$ and $\gamma_{2}(\lambda_{5},\lambda_{1})
    = \gamma_{2}(\lambda_{5},\lambda_{6}) +
    \gamma_{2}(\lambda_{6},\lambda_{1})$ we may write
\begin{align*}
    \mathfrak{a}_{4} & =  \mathfrak{b}_{1}
    -\mathfrak{b}_{3}-\mathfrak{a}_{3}+
    \gamma_{1}(\lambda_{1},\lambda_{6})+
     \gamma_{2}(\lambda_{6},\lambda_{1}), \\
    \mathfrak{b}_{4} & = \mathfrak{a}_{1} +
    \mathfrak{b}_{1}-\mathfrak{a}_{3}+ \gamma_{1}(\lambda_{1},\lambda_{6})+
     \gamma_{3}(\lambda_{6},\lambda_{1}) .
\end{align*}
Appropriate linear combinations of these yield
$\int_{\gamma_{i}(\lambda_{1},\lambda_{6})}\omega$ for $i=1,2,3$.
For example
$$\int_{\gamma_{1}(\lambda_{1},\lambda_{6})}\omega=
\frac{1}{3} \oint_{2\mathfrak{a}_{3} -
2\mathfrak{b}_{1}-\mathfrak{a}_{1}+\mathfrak{b}_{3}+\mathfrak{a}_{4}+\mathfrak{b}_{4}}
\omega .$$

In order to be able to integrate a holomorphic differential
between any branch point we must show how we may integrate such
between $\lambda_4$ and $\lambda_5$ on any branch. Now we use that
there exist meromorphic functions $f=w/{(z - \lambda_{1})^{2}}$
and $g=(z - \lambda_{i})/(z - \lambda_{j})$ (for each $i$, $j$)
with (respective) divisors
\begin{align*}
    (f) = \lambda_{2} + \lambda_{3} + \lambda_{4} + \lambda_{5} +
    \lambda_{6} - 5\lambda_{1},\qquad (g)=3(\lambda_{i} - \lambda_{j}).
\end{align*}
Thus for any normalized holomorphic differential $\boldsymbol{v}$
\begin{align*}
    \Lambda & \ni \int^{\lambda_{2}}_{\lambda_{1}}\boldsymbol{v}+
    \int^{\lambda_{3}}_{\lambda_{1}}\boldsymbol{v} +
    \int^{\lambda_{4}}_{\lambda_{1}}\boldsymbol{v} +
    \int^{\lambda_{5}}_{\lambda_{1}}\boldsymbol{v} +
    \int^{\lambda_{6}}_{\lambda_{1}}\boldsymbol{v}
     = 4\int^{\lambda_{2}}_{\lambda_{1}}\boldsymbol{v} +
    3\int^{\lambda_{3}}_{\lambda_{2}}\boldsymbol{v} +
    2\int^{\lambda_{4}}_{\lambda_{3}}\boldsymbol{v} +
    \int^{\lambda_{5}}_{\lambda_{4}}\boldsymbol{v} +
    \int^{\lambda_{6}}_{\lambda_{1}}\boldsymbol{v},
\end{align*}
and $  3\int^{\lambda_{i}}_{\lambda_{j}}\boldsymbol{v}
\in\Lambda$, where $\Lambda$ is the period lattice. These
equalities hold (modulo a lattice vector) for a path of
integration on any branch and so, for example,
$$\int_{\gamma_{1}(\lambda_{4},\lambda_{5})}\boldsymbol{v}\equiv\int_{\gamma_{1}(\lambda_{3},\lambda_{4})}\boldsymbol{v}
-\int_{\gamma_{1}(\lambda_{1},\lambda_{2})}\boldsymbol{v}
-\int_{\gamma_{1}(\lambda_{1},\lambda_{6})}\boldsymbol{v}\qquad\mod\Lambda.$$

\section{M\"obius Transformations}
We wish to determine when there is a M\"obius transformation
between the sets $H = \{\alpha_{1}, -{1}/{\overline{\alpha}_{1}},
\alpha_{2}, -{1}/{\overline{\alpha}_{2}}, \alpha_{3},
-{1}/{\overline{\alpha}_{3}}\}$ and $S = \{0, 1, \infty,
\Lambda_{1}, \Lambda_{2}, \Lambda_{3}\}$.  The former corresponds
to reality constraints on our data arising from $(H1)$ while the
latter may be constructed from the period matrix of the curve in
terms of various theta constants.  If we have a period matrix
satisfying
$(H2)$ then we must satisfy $(H1)$. \\
At the outset we note that the M\"obius transformation $M$ sending
$a \rightarrow 0, b \rightarrow 1, c \rightarrow \infty$ and its
inverse $M^{-1}$
\begin{equation*}
    \begin{matrix}
        M(a) = 0 & & M^{-1}(0) = a \\
        M(b) = 1 & & M^{-1}(1) = b \\
        M(c) = \infty & & M^{-1}(\infty) = c
    \end{matrix}
\end{equation*}
are given by
\begin{equation}\label{mob1}
    M(z) = \frac{b - c}{b - a} \frac{z-a}{z-c} \qquad M^{-1}(z) =
    \frac{z\,c(b - a) - a(b - c)}{z(b - a) - (b - c)}.
\end{equation}
The transformation
\begin{equation*}
    M(z) = \lambda \cdot \frac{z - a}{z - c} \qquad = \frac{\alpha z
    + \beta}{\gamma z + \delta}
\end{equation*}
may be represented by the $SL(2, \mathbb{Z})$ matrix
\begin{equation}\label{mob2}
    \begin{pmatrix}
        \alpha & \beta \\ \gamma & \delta
    \end{pmatrix} =
    \begin{pmatrix}
        \frac{i \sqrt{\lambda}}{\sqrt{a + c}}
        & - \frac{ia \sqrt{\lambda}}{\sqrt{a + c}} \\
        \frac{i}{\sqrt{\lambda}\sqrt{a + c}} &
        -\frac{ic}{\sqrt{\lambda}\sqrt{a + c}}
    \end{pmatrix}
\end{equation}
and upon setting $\lambda = {(b - c)}/{(b - a)}$ we may determine
a $SL(2, \mathbb{Z})$ representation of (\ref{mob1}).

A M\"obius transformation is conjugate to a rotation if and only
if it is of the  form $M(z) = \dfrac{(\alpha z +
\beta)}{(-\overline{\beta}Z + \overline{\alpha})}$.  In terms of
(\ref{mob2}) this means
\begin{equation*}
    a \overline{c} = -1 \qquad \text{and} \qquad \lambda
    \overline{\lambda} = \frac{1}{a \overline{a}}.
\end{equation*}
Then $M(0) \overline{M(\infty)} = -1$.

The rotation $\begin{pmatrix}\dfrac{
\overline{\alpha}_{1}}{\sqrt{1
+ |\alpha_{1}|^{2}}} & \dfrac{1}{\sqrt{1 + | \alpha_{1}|^{2}}} \\
\dfrac{-1}{\sqrt{1
+ |\alpha_{1}|^{2}}} & \dfrac{\alpha_{1}}{\sqrt{1 + | \alpha_{1}|^{2}}} \\
\end{pmatrix}$ transforms the set $H$ to one of the form $\{0, \infty,
\tilde{\alpha}_{2}, -{1}/{\tilde{\alpha}_{2}}, \tilde{\alpha}_{3},
-{1}/{\tilde{\alpha}_{3}} \}$ where $\tilde{\alpha}_{r} =
M(\alpha_{r}) = ({1 + \overline{\alpha}_{1}
\alpha_{r}})/({\alpha_{1} - \alpha_{r}})$ ($r = 2,3$).  Upon
setting $\tilde{\alpha}_2 = ae^{i \theta}$, $ a =
|\tilde{\alpha}_{2}|$ the rotation $\begin{pmatrix} e^{i
{\theta}/{2}} & 0 \\ 0 & e^{-i {\theta}/{2}}\end{pmatrix}$ will
transform the latter set to one of the form $\{0, \infty, a,
{-1}/{a}, w, {-1}/{\overline{w}}\}$.  Finally the scaling $z
\rightarrow {z}/{a}$ given by $\begin{pmatrix}\dfrac{1}{\sqrt{a}}
& 0 \\ 0 & \sqrt{a}\end{pmatrix}$ transforms $H$ to $H_{s} = \{0,
1, \infty,{-1}/{a^{2}}, {w}/{a}, {-1}/{(a \overline{w})}\}$ . Such
a set is of the desired form $S$ and is characterised by 3 (real)
parameters.  With $\Lambda_{1} = {-1}/{a^{2}}$, $\Lambda_{2} =
{w}/{a}$, $\Lambda_{3} = {-1}/{a \overline{w}}$ we see we have
$\Lambda_{1} \in \mathbb{R}$, $\Lambda_{1} < 0$,
$\Lambda_{2}\overline{\Lambda}_{3} = \Lambda_{1}$. From a set
$H_{s}$ and a choice of $\theta$ and $\alpha_{1}$ (equivalently, a
rotation) we may reconstruct $H$.

More generally, let us consider images $M(H)$ under M\"obius
transformations.  Up to a relabelling of roots we have four
possibilities of those roots we map to $\{0, 1, \infty \}$:
\begin{equation*}
\begin{matrix}
    \text{a.} & \alpha_{1} \rightarrow 0 & \alpha_{2} \rightarrow 1
    & {-1}/{\overline{\alpha}_{1}} \rightarrow \infty, \\
    \text{b.} & \alpha_{1} \rightarrow 0 & {-1}/{\overline{\alpha}_{1}}
\rightarrow 1
    & \alpha_{2} \rightarrow \infty, \\
    \text{c.} & \alpha_{1} \rightarrow 0 & {-1}/{\overline{\alpha}_{2}}
\rightarrow 1
    & \alpha_{2} \rightarrow \infty, \\
    \text{d.} & \alpha_{1} \rightarrow 0 & \alpha_{3} \rightarrow 1
    & \alpha_{2} \rightarrow \infty.
\end{matrix}
\end{equation*}
We have already considered $(a)$ in the previous paragraph.  For
completeness let us give $\Lambda_{1}, \Lambda_{2}, \Lambda_{3}$
for
the various cases and the various restrictions arising\\
a.
\begin{align}
    \Lambda_{1} = M(\frac{-1}{\overline{\alpha}_{2}}) & = - \frac{(1
    + \overline{\alpha}_{1} \alpha_{2})(1 +
    \overline{\alpha}_{2}\alpha_{1} )}{(\alpha_{1} -
    \alpha_{2}){(\overline{\alpha}_{1} - \overline{\alpha}_{2})}} <
    0 ,\nonumber \\
    \Lambda_{2} = M(\alpha_{3}) & = \frac{\alpha_{1} - \alpha_{3}}{\alpha_{1} -
    \alpha_{2}}\frac{1 + \overline{\alpha}_{1} \alpha_{2}}{1 + \overline{\alpha}_{1}
    \alpha_{3}}, \nonumber\\
    \Lambda_{3} = M(\frac{-1}{\overline{\alpha}_{3}}) & = - \frac{1
    + \alpha_{1} \overline{\alpha}_{3}}{\overline{\alpha}_{1}- \overline{\alpha}_{3}}\frac{1 + \alpha_{2}
    \overline{\alpha}_{1}}{\alpha_{1} - \alpha_{2}},\nonumber \\
    \Lambda_{2} \overline{\Lambda}_{3} & = \Lambda_{1};
    \label{mobca}
\end{align}
b.
\begin{align}
    \Lambda_{1} = M(\frac{-1}{\overline{\alpha}_{2}}) & = - \frac{1
    + \overline{\alpha}_{1} \alpha_{2}}{1 + \alpha_{1}
\overline{\alpha}_{1}}\cdot
    \frac{1 + \alpha_{1} \overline{\alpha}_{2}}
    {1 + \alpha_{2}\overline{\alpha}_{2}} \in \mathbb{R}
\quad 0 <\Lambda_{1} < 1 ,\nonumber\\
    \Lambda_{2} = M(\alpha_{3}) & = \frac{1
+ \overline{\alpha}_{1} \alpha_{2}}{1 + \alpha_{1}
    \overline{\alpha}_{1}}\frac{\alpha_{3} - \alpha_{1}}{\alpha_{3}
- \alpha_{2}}, \nonumber\\
    \Lambda_{3} = M(\frac{-1}{\overline{\alpha}_{3}}) & = \frac{1
    + \overline{\alpha}_{1} \alpha_{2}}{1
+ \alpha_{1} \overline{\alpha}_{1}} \frac{1 + \alpha_{1}
    \overline{\alpha}_{3}}{1 + \alpha_{2} \overline{\alpha}_{3}}, \nonumber\\
    \frac{\Lambda_{2}}{\Lambda_{2} - 1}
\overline{\left(\frac{\Lambda_{3}}{\Lambda_{3} - 1}\right)} & =
    \frac{\Lambda_{1}}{\Lambda_{1} - 1}; \label{mobcb}
\end{align}
c.
\begin{align}
    \Lambda_{1} = M(\frac{-1}{\overline{\alpha}_{1}}) & = - \frac{1
    + \alpha_{2} \overline{\alpha}_{2}}{1 + \alpha_{1}
\overline{\alpha}_{2}}\cdot
    \frac{1 + \alpha_{1} \overline{\alpha}_{1}}
    {1 + \alpha_{2}\overline{\alpha}_{1}} \in \mathbb{R} \quad 1 <\Lambda_{1} < \infty\nonumber, \\
    \Lambda_{2} = M(\alpha_{3}) & = \frac{1 + \alpha_{2} \overline{\alpha}_{2}}{1 + \alpha_{1}
    \overline{\alpha}_{2}}\frac{\alpha_{3} - \alpha_{1}}{\alpha_{3} - \alpha_{2}} \nonumber,\\
    \Lambda_{3} = M(\frac{-1}{\overline{\alpha}_{3}}) & = \frac{1
    + \alpha_{2} \overline{\alpha}_{2}}{1 + \alpha_{1} \overline{\alpha}_{2}} \frac{1 + \alpha_{1}
    \overline{\alpha}_{3}}{1 + \alpha_{2} \overline{\alpha}_{3}}, \nonumber\\
   (1 - \Lambda_{2}) \overline{(1 - \Lambda_{3})} & = 1 - \Lambda_{1}
 \label{mobcc};
\end{align}
d.
\begin{align}
    \Lambda_{r} = M(-\frac{1}{\overline{\alpha}_{r}}) & =
    \frac{\alpha_{3} - \alpha_{2}}{\alpha_{3} - \alpha_{1}} \frac{1 +
    \alpha_{1}\overline{\alpha}_{r}}{1 + \alpha_{2}
    \overline{\alpha}_{r}}, \qquad r = 1, 2, 3, \nonumber\\
    0 < \Lambda_{1} \overline{\Lambda}_{2} &\in \mathbb{R},\
    1 < \frac{\Lambda_{1}}{\Lambda_{2}} \in \mathbb{R}, \qquad
    \Lambda_{3} = \Lambda_{2} \frac{(1 -
    \overline{\Lambda}_{1})}{1 - \overline{\Lambda}_{2}}.
   \label{mobcd}
\end{align}

The constraints $(\ref{mobcb})$ for case $(b)$ may be obtained as
follows. Further composing the M\"obius transformation leading to
$(b)$ with that giving $0 \rightarrow 0$, $1 \rightarrow \infty$,
 $\infty \rightarrow 1$ gives us case $(a)$ for which we know the
constraint.  This second M\"obius transformation is given by $M(z)
= M^{-1}(z) = {z}/{(z - 1)}$ and we may transfer the constraint of
$(a)$ to $(b)$.  Similarly composing $(c)$ with $M(z) = 1-z$
yields case $(a)$ up to a relabelling of roots. Geometrically
cases $(a)$, $(b)$, $(c)$ consist of the following.  A circle
passes through $\{ \alpha_{1}, {-1}/{\overline{\alpha}_{1}},
\alpha_{2},{-1}/{\overline{\alpha}_{2}} \}$.  Under a M\"obius
transformation to the set $\{ 0, 1, \infty, \mu \}$ the circle
becomes the real axis and so $\mu \in \mathbb{R}$. This is the
real parameter appearing in each of these cases.  A similar
argument composing $(d)$ with $M(z) = {z}/{(z - \Lambda_{1})}$
will give the constraints (\ref{mobcd}).

In each case, given $\alpha$, and a choice of $\theta$ (a
rotation)
we can construct $S$ from $H$.

\providecommand{\bysame}{\leavevmode\hbox
to3em{\hrulefill}\thinspace}
\bibliographystyle{amsalpha}

\end{document}